\documentclass[a4paper,fleqn]{cas-sc}

\usepackage[sort&compress, numbers]{natbib}
\usepackage{amssymb}
\usepackage{amsmath}
\usepackage{graphicx}
\usepackage{color}
\usepackage{braket}
\usepackage{siunitx}
\usepackage{subcaption}
\usepackage{comment}
\usepackage{eqparbox}
\usepackage{url}
\usepackage{caption}
\usepackage{cleveref}
\usepackage{rsfso}
\usepackage{upgreek}
\usepackage{hyperref}
\usepackage{subcaption}
\usepackage{orcidlink}

\def\tsc#1{\csdef{#1}{\textsc{\lowercase{#1}}\xspace}}
\tsc{WGM}
\tsc{QE}
\begin{document}
\let\WriteBookmarks\relax
\def\floatpagepagefraction{1}
\def\textpagefraction{.001}

\shorttitle{The Electron-Gamma Coincidence Setup DAGOBERT}

\shortauthors{B.~Hesbacher, G.~Steinhilber, J.~Isaak, N.~Pietralla et~al.}

\title [mode = title]{The Electron-Gamma Coincidence Setup DAGOBERT}                      

\author[1]{B.~Hesbacher\,\href{https://orcid.org/0009-0001-9246-0607}{\includegraphics[height=0.68em]{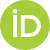}}}[orcid = 0009-0001-9246-0607]
\cormark[1]
\ead{bhesbacher@ikp.tu-darmstadt.de}
\credit{Methodology, Software, Validation, Formal analysis, Data Curation, Writing - Original Draft, Visualization}
	
\author[1]{G.~Steinhilber\,\href{https://orcid.org/0000-0001-5109-8990}{\includegraphics[height=0.68em]{ORCID-iD_icon_vector.pdf}}}[orcid = 0000-0001-5109-8990]
\credit{Methodology, Software, Validation, Formal analysis, Investigation, Data Curation, Writing - Review \& Editing, Visualization}
	
\author[1]{J.~Isaak\,\href{https://orcid.org/0000-0002-4735-8320}{\includegraphics[height=0.68em]{ORCID-iD_icon_vector.pdf}}}[orcid = 0000-0002-4735-8320]
%\cormark[2]
\ead{jisaak@ikp.tu-darmstadt.de}
\credit{Methodology, Validation, Investigation, Resources, Data Curation, Writing - Original Draft, Supervision}
	
\author[1]{N.~Pietralla\,\href{https://orcid.org/0000-0002-4797-3032}{\includegraphics[height=0.68em]{ORCID-iD_icon_vector.pdf}}}[orcid = 0000-0002-4797-3032]
\credit{Conceptualization, Methodology, Resources, Writing - Review \& Editing, Supervision, Project administration, Funding acquisition}
	
% Additional authors
\author[1]{J.~Birkhan\,\href{https://orcid.org/0000-0002-4366-7492}{\includegraphics[height=0.68em]{ORCID-iD_icon_vector.pdf}}}[orcid = 0000-0002-4366-7492]
\credit{Investigation, Writing - Review \& Editing, Supervision}
	
\author[1]{I.~Brandherm\,\href{https://orcid.org/0009-0006-3357-9487}{\includegraphics[height=0.68em]{ORCID-iD_icon_vector.pdf}}}[orcid = 0009-0006-3357-9487]
\credit{Investigation, Writing - Review \& Editing}
	
\author[1, 2]{M.~L.~Cort\'es\,\href{https://orcid.org/0009-0002-7497-6527}{\includegraphics[height=0.68em]{ORCID-iD_icon_vector.pdf}}}[orcid = 0009-0002-7497-6527]
\credit{Investigation, Writing - Review \& Editing}
	
\author[3]{D.~H.~Jakubassa-Amundsen\,\href{https://orcid.org/0000-0002-9782-3796}{\includegraphics[height=0.68em]{ORCID-iD_icon_vector.pdf}}}[orcid = 0000-0002-9782-3796]
\credit{Formal analysis, Writing - Review \& Editing, Visualization}
	
\author[1]{I.~Jurosevic\,\href{https://orcid.org/0000-0002-5847-7747}{\includegraphics[height=0.68em]{ORCID-iD_icon_vector.pdf}}}[orcid = 0000-0002-5847-7747]
\credit{Investigation, Writing - Review \& Editing}
	
\author[1]{P.~von~Neumann-Cosel\,\href{https://orcid.org/0000-0002-0256-5940}{\includegraphics[height=0.68em]{ORCID-iD_icon_vector.pdf}}}[orcid = 0000-0002-0256-5940]
\credit{Writing - Review \& Editing}
	
\author[1]{M.~Rech\,\href{https://orcid.org/0000-0002-2639-3688}{\includegraphics[height=0.68em]{ORCID-iD_icon_vector.pdf}}}[orcid = 0000-0002-2639-3688]
\credit{Investigation, Writing - Review \& Editing}
	
\author[4, 5, 6, 7]{X.~Roca-Maza\,\href{https://orcid.org/0000-0002-2100-6407}{\includegraphics[height=0.68em]{ORCID-iD_icon_vector.pdf}}}[orcid = 0000-0002-2100-6407]
\credit{Formal analysis, Writing - Review \& Editing}
	
\author[1]{D.~Schneider\,\href{https://orcid.org/0009-0002-8154-7472}{\includegraphics[height=0.68em]{ORCID-iD_icon_vector.pdf}}}[orcid = 0009-0002-8154-7472]
\credit{Resources, Writing - Review \& Editing}

\cortext[cor1]{Corresponding author}
%\cortext[cor2]{Corresponding author}

% Address affiliation
\affiliation[1]{organization={Institut f{\"u}r Kernphysik, Technische Universit{\"a}t Darmstadt},
    addressline={Schlossgartenstr. 9}, 
    city={Darmstadt},
    % citysep={}, % Uncomment if no comma needed between city and postcode
    postcode={64289}, 
    country={Germany}}
\affiliation[2]{organization={RIKEN Nishina Center},
    addressline={2-1 Hirosawa, Wako}, 
    city={Saitama},
    % citysep={}, % Uncomment if no comma needed between city and postcode
    postcode={351-0198}, 
    country={Japan}}
\affiliation[3]{organization={Mathematisches Institut, Ludwig-Maximilians Universit{\"a}t M{\"u}nchen},
    addressline={Theresienstr. 39}, 
    city={Munich},
    % citysep={}, % Uncomment if no comma needed between city and postcode
    postcode={80333}, 
    country={Germany}}
\affiliation[4]{organization={Departament de F\'isica Qu\`{a}ntica i Astrof\'{i}sica, Universitat de Barcelona},
    addressline={Mart\'{i} i Franqu\'{e}s, 1}, 
    city={Barcelona},
    % citysep={}, % Uncomment if no comma needed between city and postcode
    postcode={08028}, 
    country={Spain}}
\affiliation[5]{organization={Institut de Ci\`{e}ncies del Cosmos, Universitat de Barcelona},
    addressline={Mart\'{i} i Franqu\'{e}s, 1}, 
    city={Barcelona},
    % citysep={}, % Uncomment if no comma needed between city and postcode
    postcode={08028}, 
    country={Spain}}
\affiliation[6]{organization={Dipartimento di Fisica ''Aldo Pontremoli'', Universit\`{a} degli Studi di Milano},
    addressline={Via Celoria 16}, 
    city={Milano},
    % citysep={}, % Uncomment if no comma needed between city and postcode
    postcode={20133}, 
    country={Italy}}
\affiliation[7]{organization={INFN, Sezione di Milano},
    addressline={Via Celoria 16}, 
    city={Milano},
    % citysep={}, % Uncomment if no comma needed between city and postcode
    postcode={20133}, 
    country={Italy}}

\let\printorcid\relax % Remove ORCID footnote

\begin{abstract}
The QCLAM electron spectrometer at the S-DALINAC electron accelerator at Technische Universit\"at Darmstadt has been extended by the DAGOBERT $\gamma$-detector array consisting of fast timing and high efficiency LaBr\textsubscript{3}:Ce detectors to perform electron-gamma coincidence measurements. 
The functionality of the setup and data acquisition system was demonstrated in a commissioning measurement on $^{12}\textrm{C}$ observing the $4.44\,$MeV and $15.11\,$MeV states. 
A medium-heavy nucleus, $^{96}\textrm{Ru}$, has been studied for the first time up to excitation energies of $15\,$MeV using the $(e,e'\gamma)$ reaction. 
In particular, the angular distribution of the $2_1^+$ state and the $\gamma$-decay branching ratios of the mixed-symmetric $2_3^+$ state were observed.
DAGOBERT@QCLAM is a new and worldwide unique setup for nuclear structure studies of excitation and decay using purely electromagnetic probes, with a significantly improved sensitivity compared to previous experiments.
\end{abstract}

%% Research highlights
\begin{highlights}
\item DAGOBERT is worldwide unique for high-resolution $(e,e^\prime \gamma)$ experiments.
\item First $(e,e^\prime \gamma)$ experiment of an $A \geq 90$ nucleus conducted.
\item Techniques for subtraction of indistinguishable coincident bremsstrahlung developed.
\item First $(e,e^\prime \gamma)$ data on an off-yrast state analysed.
\item Investigated $\gamma$ decays in $(e,e^\prime n\gamma)$ reaction.
\end{highlights}

% Keywords

\maketitle

\section{Introduction}
Electron scattering is an established tool for nuclear structure studies since 70 years~\cite{Hofstadter.1956}. 
Elastic electron scattering has been used to study the size and shape of the nuclear ground-state charge distribution by measuring electromagnetic form factors~\cite{Hofstadter.1956, BeatHahn.1956, Hofstadter.1964, Donnelly.1984}. 
It has also been used to explore the ground-state quadrupole moments of deformed nuclei~\cite{BeatHahn.1956,Downs.1957,Pal.1960}, the nuclear magnetic dipole moments~\cite{Goldemberg.1966, Griffy.1965}, and the structure of individual nucleons~\cite{Hofstadter.1962}.
Inelastic electron scattering allows for the measurement of nuclear resonance excitation energies, multipolarities of nuclear transitions, parity quantum numbers of nuclear excited states, and nuclear dynamics by measuring transition densities~\cite{Barber.1962, Pitthan.1971, Heisenberg.1981-short}.

An interesting special case is an exclusive electron-gamma coincidence measurement, where the electromagnetic interaction is involved in both the excitation and the decay channel. A pioneering $(e,e^\prime \gamma)$ measurement was performed on the $2_1^+$ state of $^{12}\textrm{C}$ at the MUSL-2 accelerator at the University of Illinois~\cite{Papanicolas.1985, C.N.Papanicolas.1985}. 
It was demonstrated that the reaction is capable of separating the longitudinal and transverse form factors including their relative sign in the measurement of the angular distribution of the emitted $\gamma$ rays.
Further measurements were conducted on light nuclei with low level densities such as $^{15}\textrm{N}$~\cite{Mueller.1993} and $^{16}\textrm{O}$~\cite{Deininger.1996}.
Despite the success of these early $(e,e^\prime \gamma)$ experiments, the potential of the technique has not been fully exploited. 
The main experimental challenges were insufficient sensitivity due to low resolution for $\gamma$-energy measurements and low timing resolution for $(e,e^\prime \gamma)$ coincidences, needed for discriminating the signals of interest from the inevitable bremsstrahlung and beam-induced background. 
Advances in $\gamma$-ray detector technology, digital signal processing and the precise tuning of the electron beam promise a significant improvement of the experimental conditions, in particular the $(e,e^\prime \gamma)$ coincidence resolution, permitting to address nuclei with much higher level densities. 
This development paves the way for studying virtually all stable nuclei with the $(e,e^\prime \gamma)$ coincidence reaction. 

This article reports on the design and commissioning of the new $(e,e^\prime \gamma)$ experimental setup at the Superconducting DArmstadt electron LINear ACcelerator (S-DALINAC) at the Institute for Nuclear Physics~\cite{Pietralla.2018} at the Technische Universit\"at Darmstadt. 
The accelerator delivers a 3-GHz continuous-wave electron beam to the experimental site of the large-acceptance Quadrupole CLAM shell magnetic spectrometer QCLAM~\cite{DAlessio2019}. The $\gamma$ rays emitted in the $(e,e^\prime \gamma)$ reaction are detected with the recently established Darmstadt Array for coincident Gamma-ray OBservation in Electron scattering ReacTions (DAGOBERT) equipped with fast-timing and high-efficiency LaBr\textsubscript{3}:Ce scintillators~\cite{Rasmussen.2004}.

The article is organized as follows. Chapter~\ref{chap::Formalism} describes the underlying formalism in more detail. 
Chapter~\ref{chap::S-DALINAC} focuses on the S-DALINAC accelerator. 
The following Chapter~\ref{chap::EEPG_Fac} deals with the QCLAM electron spectrometer and the newly established electron-$\gamma$ coincidence experimental facility, while Chapter~\ref{chap::Electronics_DAQ} describes the electronics and data acquisition for $(e,e^\prime \gamma)$ experiments. 
The data analysis steps are discussed in Chapter~\ref{chap::Data_Analysis}. Chapter~\ref{chap::Results} points out the results and offers a discussion. 
Chapter~\ref{chap::Summary_Outlook} provides a summary and an outlook on future experiments.

\section{Formalism} \label{chap::Formalism}
An advantage of using electrons for scattering experiments is their purely electromagnetic interaction through the exchange of virtual photons described with high accuracy within the framework of quantum electrodynamics (QED)~\cite{Dirac}. 
Since the electromagnetic interaction is relatively weak, measurements on atomic nuclei are dominated by one-step processes, while multiple scattering processes can be neglected for momentum transfers relevant in nuclear structure studies~\cite{Forest.1966}. 
This fact allows for a precise quantitative description of the reactions and a clean separation from the desired information on nuclear structure.

\subsection{Inclusive Electron Scattering}
In inclusive $(e,e^\prime)$ electron scattering experiments, the incoming electron is scattered off a target and the scattered electron is detected.
The measured inclusive cross sections contain information about the structure of the nucleus, which is accessible via a comparison of the experimentally determined differential cross sections $\mathrm{d}\sigma_{\mathrm{exp}}/\mathrm{d}\Omega$ with the one of a theoretically calculated point-like, inert, and infinitely massive nucleus $\mathrm{d}\sigma_{\mathrm{point}}/\mathrm{d}\Omega$:
\begin{equation} \label{Eq:1}
    \frac{\mathrm{d}\sigma_{\mathrm{exp}}}{\mathrm{d}\Omega} = \frac{\mathrm{d}\sigma_{\mathrm{point}}}{\mathrm{d}\Omega}\cdot R(E_x,q).
\end{equation}
Here, $\Omega$ is the solid angle of the electron detection, $R$ denotes the nuclear response which depends on the excitation energy $E_x$ and the momentum transfer $q$ of the electron to the nucleus.

The nuclear response function is a linear combination of the longitudinal and transverse form factors $F_L$ and $F_T$~\cite{Uberall.1971, Suda.2017}:
\begin{equation} \label{Eq:2}
    R(E_x,q) = |F_L(q)|^2+\frac{1}{2}\left(1+2\tan^2\frac{\theta}{2}\right) |F_T(q)|^2.
\end{equation}
They contain information on the nuclear structure and describe the variation of the cross section as a function of the momentum transfer.
The terms longitudinal and transverse refer to the polarization of the virtual photon. 
Elastic scattering processes ($E_x = 0$) allow the investigation of ground-state properties such as charge radii. 
For a spinless nucleus $R(E_x = 0, q) = |F_L(q)|^2$, where $F_L$ represents the Fourier transformed charge distribution of the ground state. 
If the nucleus is excited ($E_x > 0$) during the scattering process, the scattered electron looses energy and, hence, this process is called inelastic scattering. In this case both form factors must be considered.
Inclusive electron scattering is an established method, in particular, to study isolated bound states of stable nuclei and their excitations from the ground state. 
Inclusive electron scattering does not provide information on all decay channels of excited states.

\subsection{Exclusive Electron Scattering}
Exclusive $(e,e^\prime x)$ electron scattering experiments are a way to expand the potential of inclusive electron measurements. 
One or several particle(s) $x$ (e.g., neutrons, protons, $\alpha$, $\gamma$) emitted during the deexcitation process of the nucleus are detected in coincidence to the scattered electron. 
This makes it possible to study the excitation and decay channels simultaneously.

The emitted particles are correlated with the range of excitation energies inferred from the energy loss of the coincidentally detected electrons. 
Moreover, decay multipole strengths can be determined by measuring the angular distribution of the emitted particle(s) $x$~\cite{vonNeumannCosel.1997,Diesener.1995}.
In the case of $\gamma$ emission, it is sensitive to the interference term between the transverse and longitudinal form factors leading to a rotation of the angular distribution of the coincident particle emission, with the relative sign of the form factors indicating the direction of the rotation~\cite{Papanicolas.1985}. 
The measurement of the rotation angle of the angular distribution represents an alternative to the Rosenbluth separation method for the simultaneous extraction of $F_L$ and $F_T$~\cite{C.N.Papanicolas.1985}. 
This was demonstrated, for example, in the study of the $E0$, $E1$ and $E2$ multipole strengths of $^{40}\textrm{Ca}$ using the $^{40}\textrm{Ca}(e,e^\prime p)^{39}\textrm{K}$ reaction~\cite{vonNeumannCosel.1994}. 
Furthermore, in exclusive electron scattering, the coincidence condition along with energy conservation eliminates the radiative tail of the electron scattering signals~\cite{Strauch.2000}, which is the main background source in the study of continuum excitations with electron scattering.

\subsection{Electron-$\gamma$ Coincidence Reaction}
\begin{figure}[htbp]
    \centering
    \begin{subfigure}{0.32\linewidth}
		\centering
		\includegraphics[width=\linewidth]{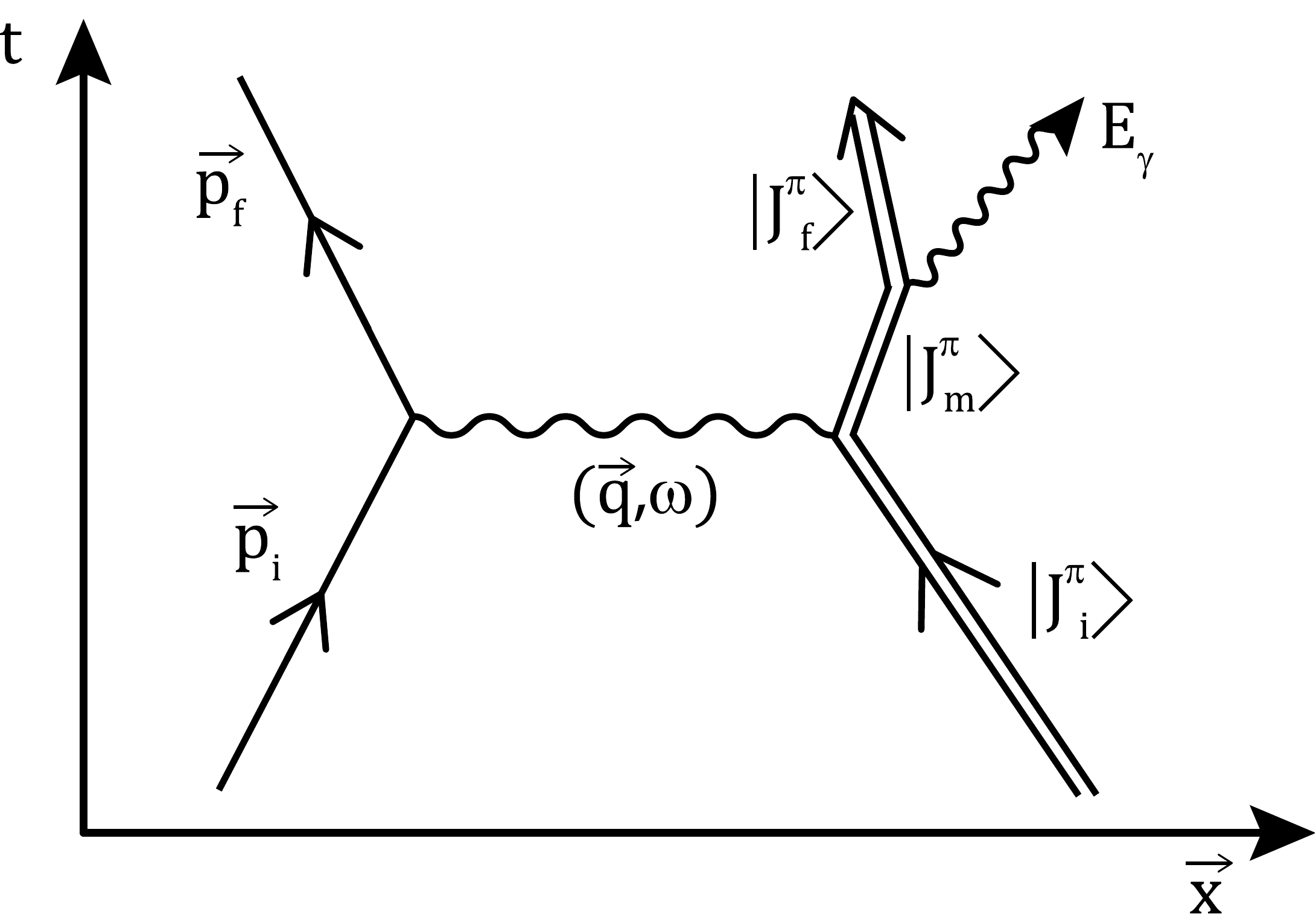}
        \caption{Inelastic scattering with excitation of the target nucleus and subsequent decay into a state with lower excitation energy or the ground state.}
        \label{fig::feynman:A}
	\end{subfigure}\hfill
	\begin{subfigure}{0.32\linewidth}
		\centering
		\includegraphics[width=\linewidth]{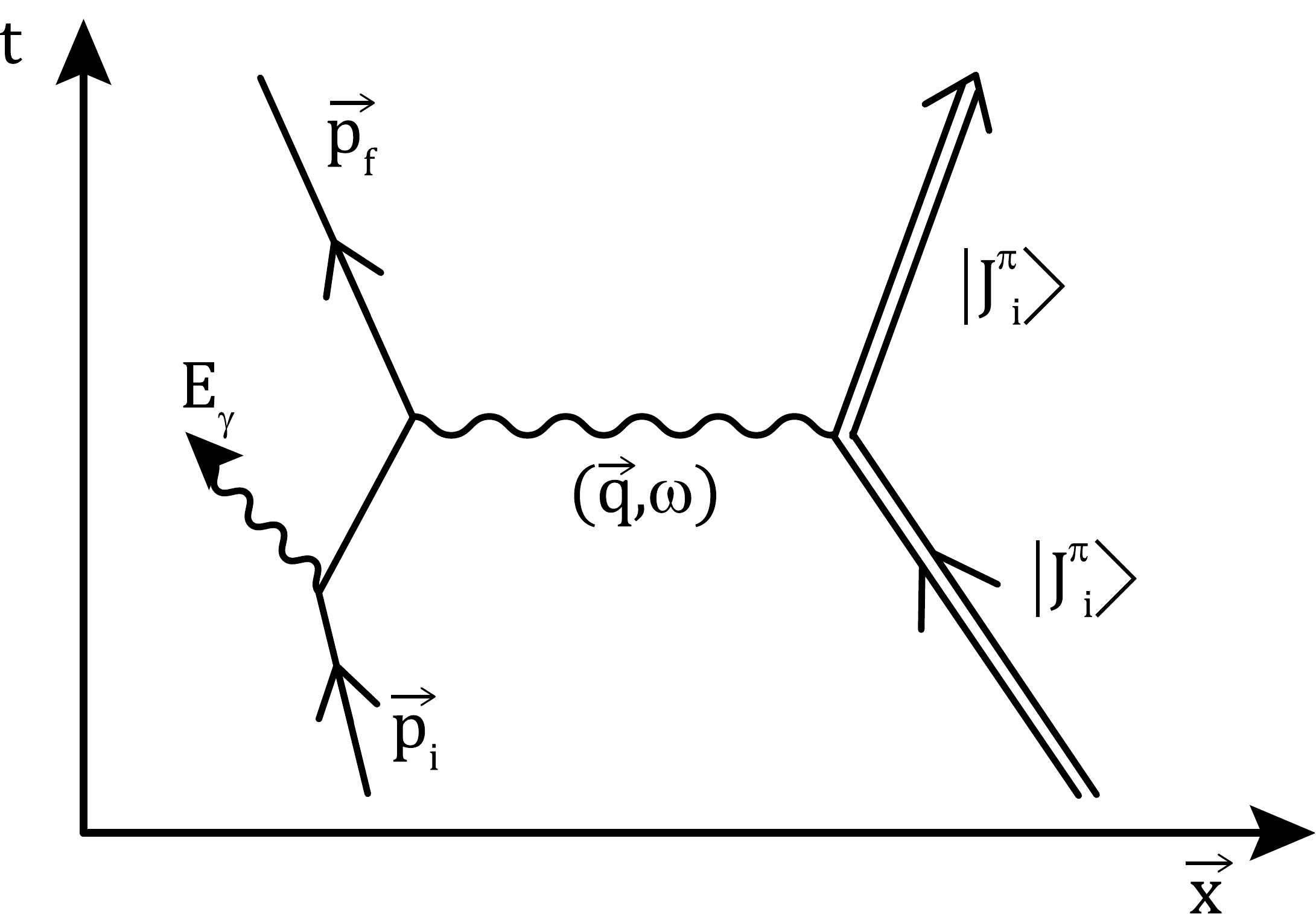}
        \caption{Bremsstrahlung process before scattering at the target nucleus.\vspace*{2\baselineskip}}
        \label{fig::feynman:B}
	\end{subfigure}\hfill
	\begin{subfigure}{0.32\linewidth}
		\centering
		\includegraphics[width=\linewidth]{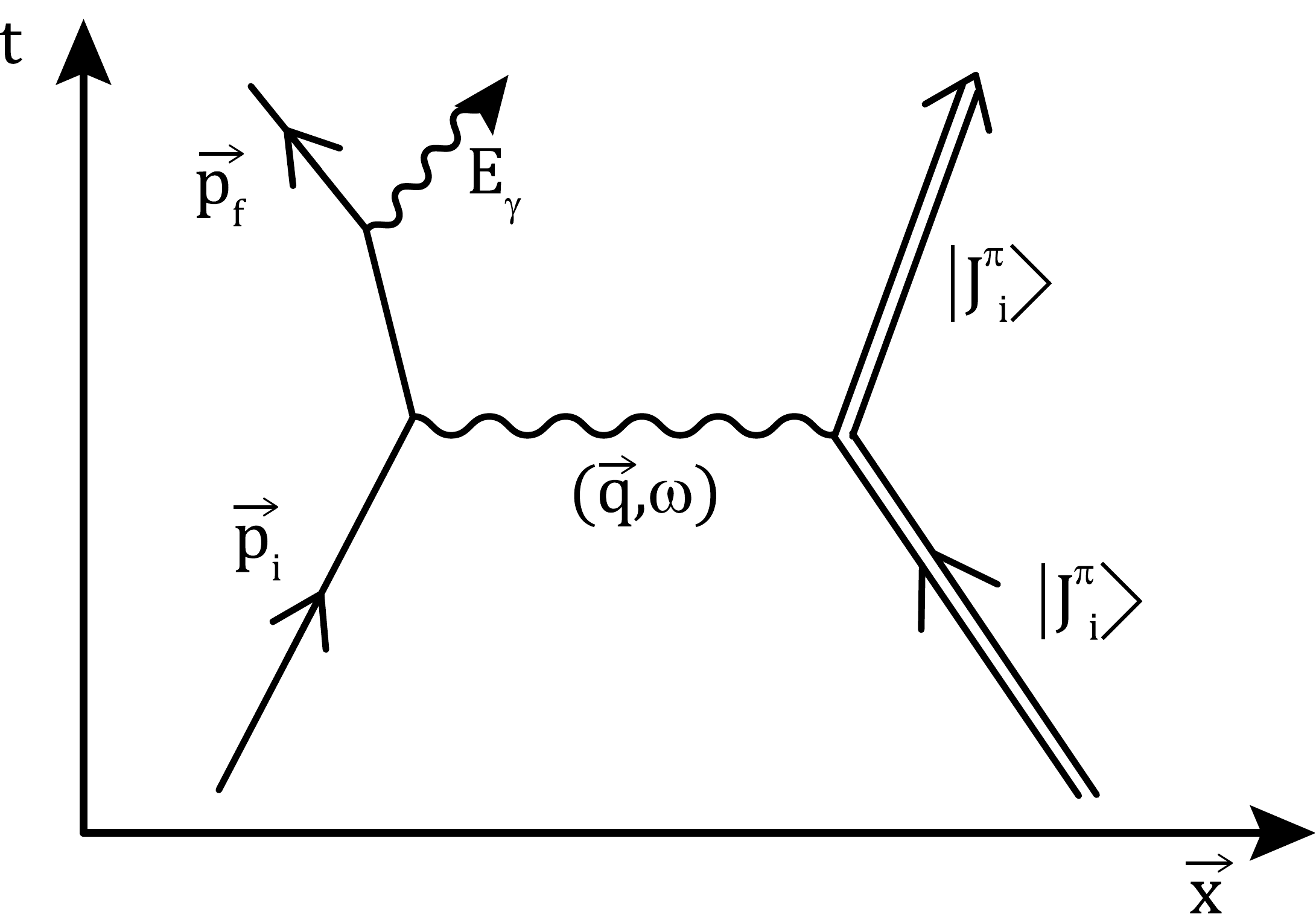}
        \caption{Bremsstrahlung process after scattering.\vspace*{2\baselineskip}}
        \label{fig::feynman:C}
	\end{subfigure}
    \caption{Feynman diagrams of the $(e,e^\prime \gamma)$ reaction.}
    \label{fig::feynman}
\end{figure}

Possible processes contributing to the $(e,e^\prime\gamma)$ reaction are illustrated using Feynman diagrams in Fig.~\ref{fig::feynman}.
\\
In inelastic electron scattering, a nucleus is excited from an initial state $J^{\pi}_i$ to a state $J^{\pi}_m$ (Fig.~\ref{fig::feynman:A}). By the emission of $\gamma$ radiation, the nucleus decays into a state $J^{\pi}_f$ with a lower excitation energy. 
In the case of a direct decay to the ground state, the energy of the emitted photon $E_{\gamma_0} = E_x$ corresponds to the energy loss of the electron during inelastic scattering. 
In competition to this quasi-prompt $\gamma$ radiation, bremsstrahlung may be generated by the elastic scattering reaction of an electron with the target material leading to the same energy loss of the electron and, hence, the same photon energy, as the $(e,e^\prime\gamma_0)$ reaction. 
Such a bremsstrahlung process can take place either before (Fig.~\ref{fig::feynman:B}) or after (Fig.~\ref{fig::feynman:C}) the elastic scattering of the electron off the target nucleus.  
The two processes, $\gamma$ decay to the ground state and bremsstrahlung leading to the same energy loss of the scattered electron, can only be distinguished by the angular distribution of the emitted $\gamma$ radiation.
Due to the indistinguishability of the emitted photons, the two process are coherent~\cite{Hubbard.1966}. 
Consequently, the total cross section of the $(e,e^\prime\gamma)$ reaction consists of three terms:
\begin{equation}
    \sigma_{(e,e^\prime \gamma)} = \sigma_{\mathrm{nucl}} + \sigma_{\mathrm{brems}} + \sigma_{\mathrm{inter}}.
\end{equation}
Here, $\sigma_{\mathrm{nucl}}$ describes the total cross section of the nuclear excitation with subsequent internal $\gamma$ decay, $\sigma_{\mathrm{brems}}$ the total cross section of the bremsstrahlung process, and $\sigma_{\mathrm{inter}}$ the interference term of the two previous processes. 

The differential $(e,e^\prime \gamma)$ cross section for a nuclear excitation of a spin-zero ground state with subsequent $\gamma$ decay is described in plane-wave Born approximation (PWBA) by~\cite{H.L.Acker.1967} (see Appendix~\hyperref[Sec:Appendix_A]{A} for details)
\begin{equation} \label{Eq:Acker-Rose}
    \begin{split}
        \frac{\mathrm{d}^3\sigma_{\mathrm{nucl}}}{\mathrm{d}\Omega_e \mathrm{d}\Omega_\gamma \mathrm{d}\omega} = 2 \left(\frac{Z\alpha \hbar c}{4\pi}\right)^2 (2J+1) \cdot \frac{p_f}{p_i} \frac{\Gamma_J^{\mathrm{rad}}}{\Gamma_J} \frac{\Gamma_J}{(E_x-\omega)^2+\frac{\Gamma_J^2}{4}} \cdot \left( V_LW_L+V_TW_T+V_IW_I+V_SW_S\right),
    \end{split}
\end{equation}
with the charge number of the excited nucleus $Z$, the spin of the excited state $J$, the initial electron momentum $p_i$, the momentum of the scattered electron $p_f$, the partial width for the coincident photon decay of the excited state $\Gamma_J^{\mathrm{rad}}$, the total width of the excited state $\Gamma_J$, the energy of the excited state $E_x$, the energy loss of the electron $\omega$ and a sum of products of kinematic factors $V$ and generalized form factors $W$~\cite{Donnelly.1985}. 
The indices $L$ and $T$ stand for longitudinal and transverse, respectively, $I$ for interference and $S$ for spin.
The generalized form factors $W$ are proportional to the form factors $F$ of the nucleus
\begin{align}
    W_L &\propto |F_L(q)|^2,\\
    W_T &\propto |F_T(q)|^2,\\
    W_I &\propto F_L(q)\cdot F_T(q),\\
    W_S &\propto |F_T(q)|^2,
\end{align}
where $F_L(q)$ and $F_T(q)$ are the longitudinal and transverse form factors for allowed values of transition angular momenta of the nuclear excitation under consideration~\cite{Donnelly.1984}.
Their relative sign is accessible through the interference term $F_L \cdot F_T$.
After integration over the photon angle, the well-known inclusive $(e,e^\prime)$ cross section from Eqs.~(\ref{Eq:1}) and~(\ref{Eq:2}) is obtained.
A full description of the formalism and explicit expressions for the kinematic factors $V$ and generalized form factors $W$ can be found in Appendix~\hyperref[Sec:Appendix_A]{A}.

The contribution of bremsstrahlung to the $(e,e^\prime \gamma)$ cross section can be calculated using PWBA with the Bethe-Heitler formula~\cite{Bethe.1934}:
\begin{equation} \label{Bethe-Heitler-Formel}
		\begin{split}
		  \frac{\mathrm{d}^3 \sigma_{\mathrm{brems}}}{\mathrm{d}\Omega_{\gamma}\mathrm{d}\Omega_{e}\mathrm{d}\omega} = {\left(\frac{Z \hbar c}{2\pi}\right)}^2 {\alpha}^3 \frac{p_f/p_i}{\omega {|\vec{q}-\vec{k}|}^4 {c}^4} \cdot \left(\frac{{|\vec{p}_f \times \vec{k}|}^2}{{\left(\tilde{p}_f \cdot \tilde{k}\right)}^2} \left(4{E}^2_i-{|\vec{q}-\vec{k}|}^2 {c}^2\right)\right. \left. + \frac{{|\vec{p}_i \times \vec{k}|}^2}{{\left(\tilde{p}_i \cdot \tilde{k}\right)}^2} \left(4{E}^2_f-{|\vec{q}-\vec{k}|}^2 {c}^2\right)\right. \\ \left. -2 \frac{\left(\vec{p}_f \times \vec{k}\right)\left(\vec{p}_i \times \vec{k}\right)}{\left(\tilde{p}_f \cdot \tilde{k}\right)\left(\tilde{p}_i \cdot \tilde{k}\right)}\left(4E_i E_f-{|\vec{q}-\vec{k}|}^2 {c}^2\right)\right. \left. + 2{\omega}^2 \frac{{|\vec{k} \times \vec{q}|}^2}{\left(\tilde{p}_f \cdot \tilde{k}\right)\left(\tilde{p}_i \cdot \tilde{k}\right)}\right).
		\end{split}
\end{equation}
Here, the energies $E_i$ and $E_f$ are the initial and final electron energy and $k$ represents the momentum of the photon.
A tilde indicates a four-vector.

\begin{figure}[htbp]
    \centering
    \includegraphics[width=0.53\linewidth]{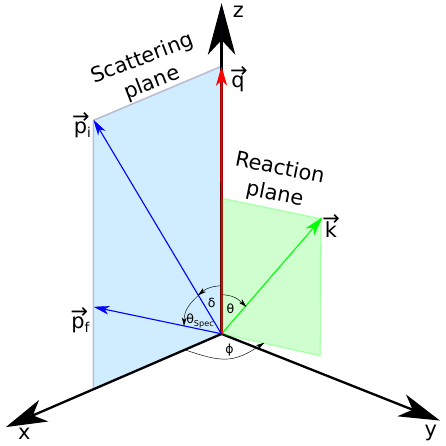}
    \caption{The relevant vectors and angles of the kinematics of an $(e,e^\prime \gamma)$ reaction are shown.}
    \label{fig:coordinate_system}
\end{figure}

The corresponding coordinate system can be seen in Fig.~\ref{fig:coordinate_system}.
\\
The scattering plane is defined by the vectors $\vec{p}_i$ and $\vec{p}_f$. 
The coordinate system is chosen such that the scattering plane coincides with the $x$-$z$-plane and the $z$-axis is defined by the momentum transfer $\vec{q} = \vec{p}_i-\vec{p}_f$. 
The vectors of momentum transfer $\vec{q}$ and momentum of the emitted photon $\vec{k}$ define the reaction plane. 
This coordinate system accounts for the symmetry of the PWBA with respect to the direction of $\vec{q}$. 

\begin{figure}[htbp]
    \centering
    \includegraphics[width=0.75\linewidth]{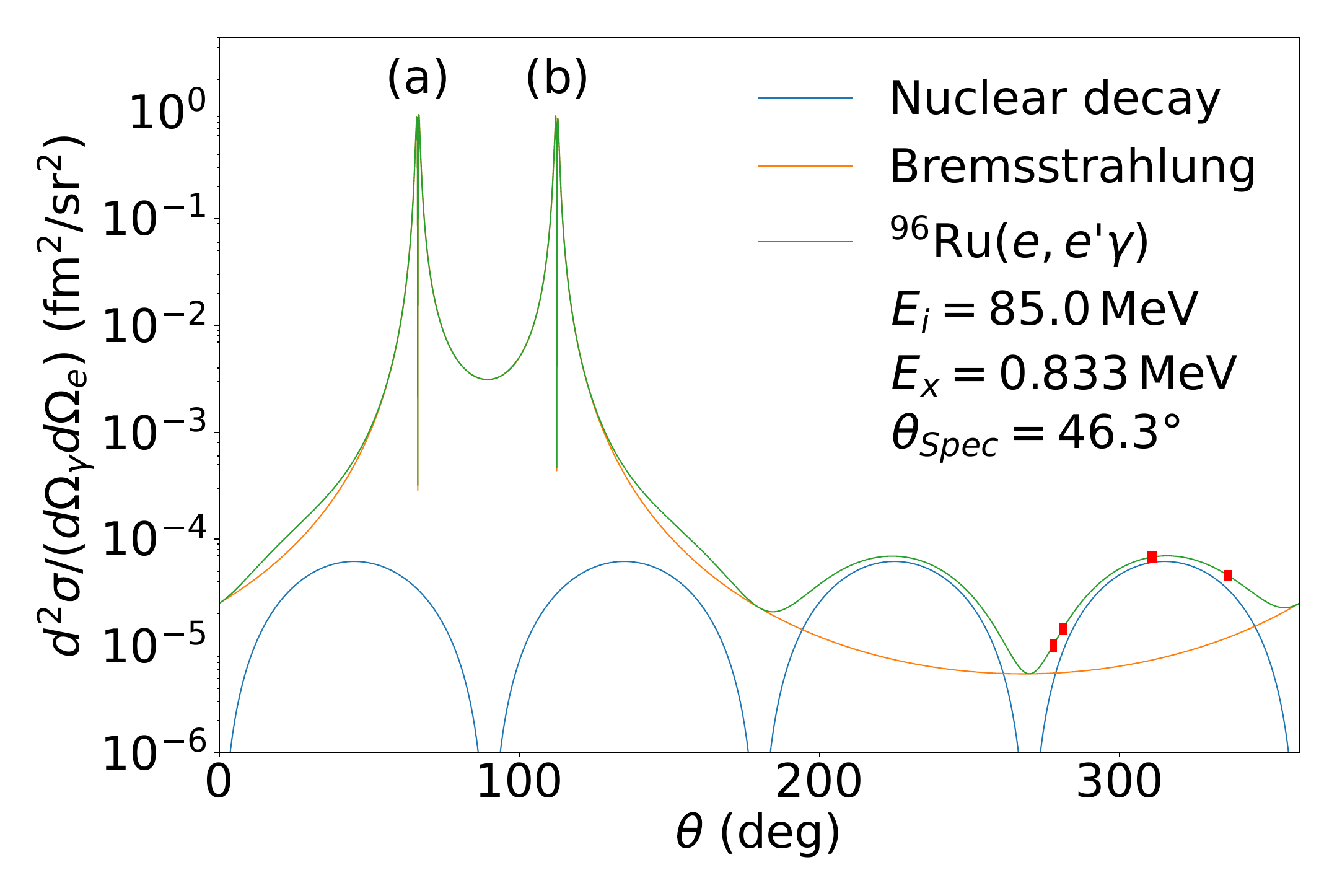}
    \caption{Double differential cross section of nuclear decay and bremsstrahlung contributions to the $(e,e^\prime \gamma)$ reaction in the $x$-$z$ plane for the $2_1^{+}$ state of $^{96}\textrm{Ru}$ and $E_i = 85\,$MeV. Positions of the $\gamma$ detectors in the experiment are shown as red rectangles. The peak marked with (a) represents the maximum of the bremsstrahlung cone in direction of the incoming electron beam and the peak marked with (b) the corresponding maximum in direction of the scattered electrons towards the QCLAM spectrometer.}
    \label{fig:nuclear_bremsstrahlung}
\end{figure}

An example of the double differential $(e,e^\prime \gamma)$ nuclear (blue line) and bremsstrahlung (orange line) cross sections is shown in Fig.~\ref{fig:nuclear_bremsstrahlung} for the population of the first $2_1^+$ state of $^{96}\textrm{Ru}$ as a function of the angle $\theta$, the angle of the emission direction of the photon $\vec{k}$ relative to the momentum transfer vector $\vec{q}$ for the nucleus (compare Fig.~\ref{fig:coordinate_system}).
\\
The cross sections from Eqs.~(\ref{Eq:Acker-Rose}) and~(\ref{Bethe-Heitler-Formel}) are averaged over the $\gamma$ detector resolution $\Delta \omega / \omega = 3\,\%$. 
In addition, due to the small line width, the integration over $\omega$ results in a contribution of the Breit-Wigner factor with the value $2\pi$ for Eq.~(\ref{Eq:Acker-Rose}).
The green line shows the coherent sum of both contributions.
Two conical maxima marked (a) and (b) are recognizable at angles of about $65.6^\circ$ and $111.9^\circ$, respectively. 
They correspond to the maximum of the bremsstrahlung emission in direction of the incoming electrons (a) and electrons scattered towards the QCLAM spectrometer (b). 
The nuclear cross section shows a periodic behavior as a function of $\theta$, which is characteristic for electric quadrupole transitions ($E2$).
The red rectangles represent the position of the six $\gamma$ detectors in the performed experiment, see Tab.~\ref{tab:measurement_detector_position} and discussion below.

\section{S-DALINAC} \label{chap::S-DALINAC}
\begin{figure}[htbp]
    \centering
    \includegraphics[width=1\linewidth]{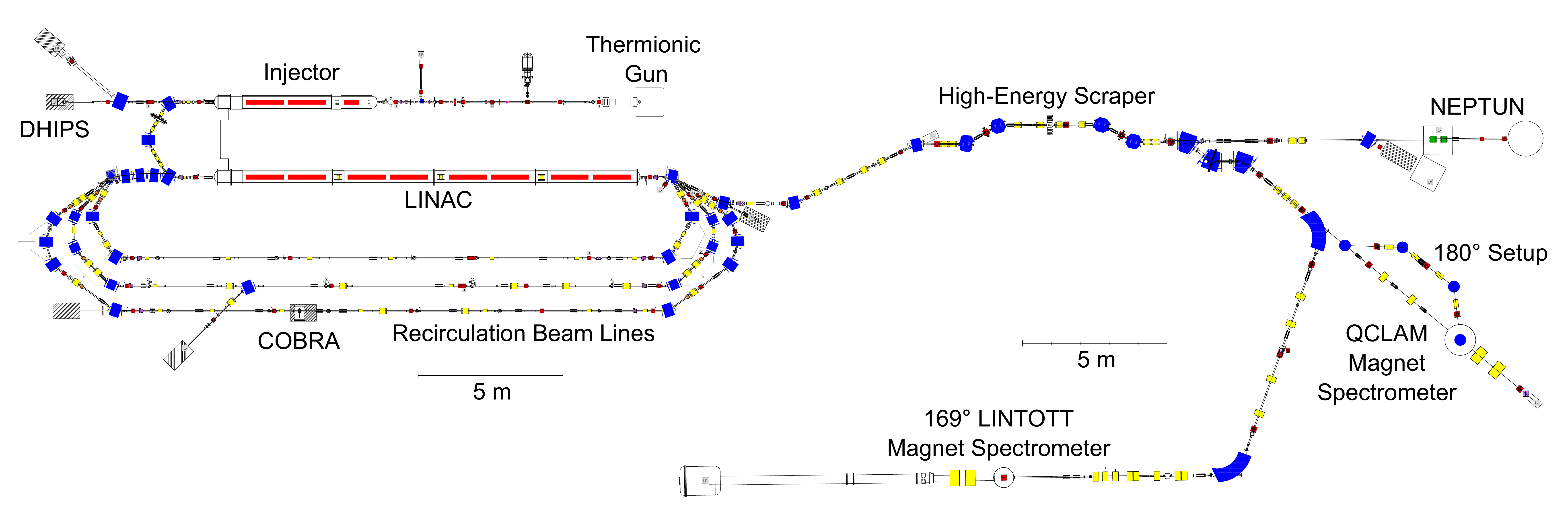}
    \caption{The S-DALINAC~\cite{Pietralla.2018} consisting of thermionic gun, injector and main accelerator (LINAC). The recirculations enable higher electron energies. Various experimental stations exist: DHIPS~\cite{Sonnabend.2011}, COBRA~\cite{Meier.2024}, NEPTUN~\cite{Savran.2010}, QCLAM~\cite{Knirsch.10.12.1991, Luttge.1994} and LINTOTT~\cite{Schuller.1978, Singer.Lintott.2021} spectrometer.}
    \label{fig:S-DALINAC}
\end{figure}

The Superconducting DArmstadt electron LINear ACcelerator S-DALINAC at the Institute for Nuclear Physics~\cite{Pietralla.2018} at Technische Universit\"at Darmstadt is a recirculating electron accelerator for nuclear physics studies, which is capable of high resolution electron scattering experiments at low momentum transfer. The floor plan of the accelerator is shown in Figure~\ref{fig:S-DALINAC}.

After emission from a thermionic electron source, the electrons are pre-accelerated by an electrostatic field and prepared for further acceleration in a chopper and pre-buncher section, imprinting a $2.997\,$GHz time structure on the continuous electron beam. 
The same frequency is applied to the superconducting cavities of the injector and the main linear accelerator (LINAC), which are operated at a temperature of $2\,$K. 
In the injector section, the electrons are accelerated to kinetic energies of up to $10\,$MeV. 
After exiting this section, they are guided into the main accelerator LINAC, where the electrons gain an additional up to $30\,$MeV in kinetic energy per passage.
Depending on the energy required, the main accelerator can be operated in single-pass, single-recirculation or thrice-recirculation mode, resulting in a maximum design electron energy of $130\,$MeV.
The electron beam can be transported to the Quadrupole CLAM shell magnetic spectrometer (QCLAM)~\cite{Knirsch.10.12.1991, Luttge.1994} with a large solid angle acceptance well suited for coincidence measurements.

For electron-$\gamma$ coincidence reactions it is crucial to minimize any source of beam-induced background. 
To improve the beam quality at the spectrometer, the electron beam is guided through a high-energy scraper system~\cite{Jurgensen.09.04.2018}. 
The high-energy scraper acts as a filter for fluctuations in beam energy and helps to detect and control slow energy fluctuations leading to an enhanced beam energy resolution and stability. Additionally, it removes beam halo contributions. The high-energy scraper significantly improves beam quality albeit at the expense of beam intensity.

\section{Electron-$\gamma$ Coincidence Setup} \label{chap::EEPG_Fac}
\begin{figure}[htbp]
	\centering
		\includegraphics[width=0.75\linewidth]{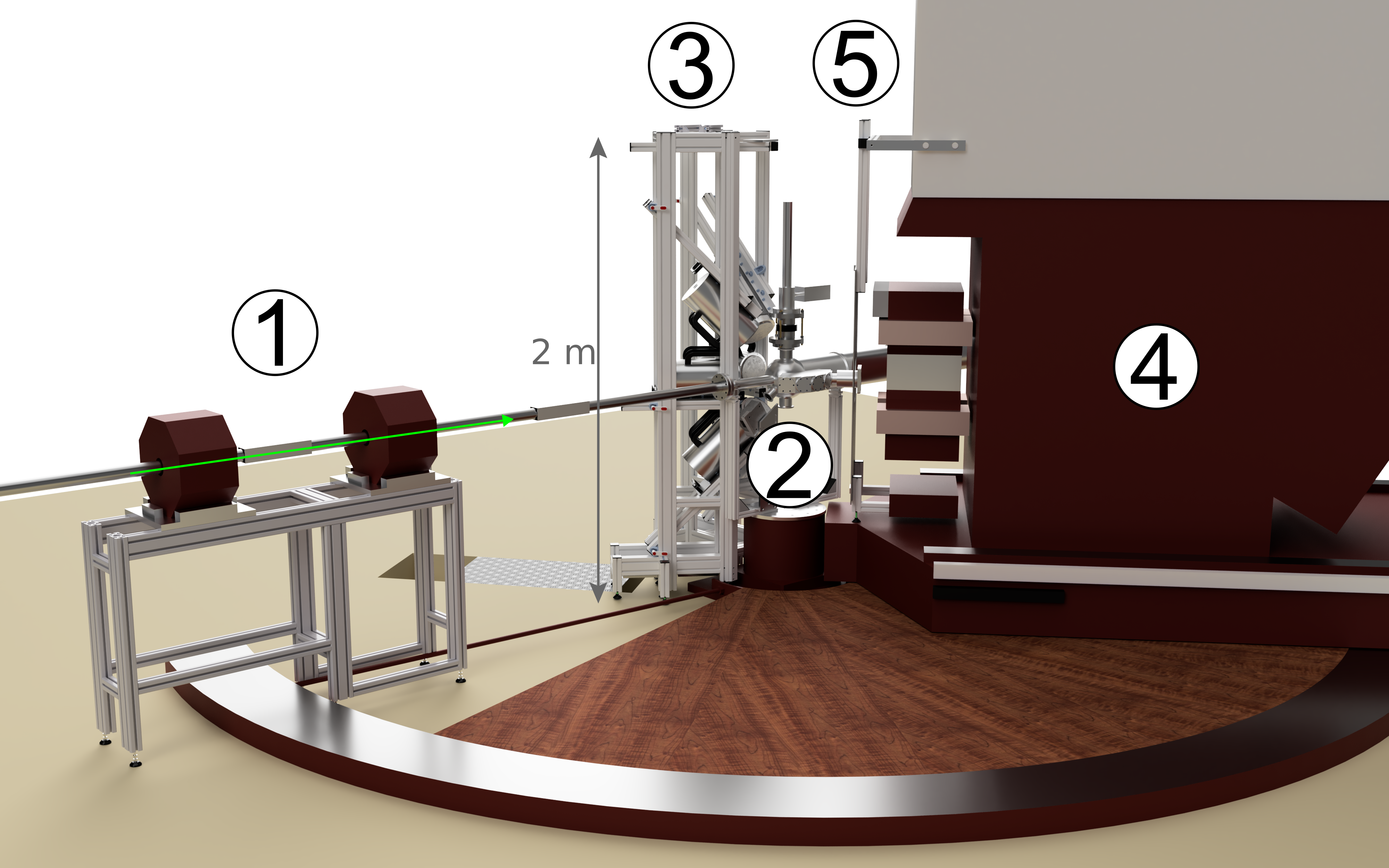}
	\caption{Overview of the new $(e,e^\prime \gamma)$ setup. \raisebox{.5pt}{\textcircled{\raisebox{-.9pt}{1}}}: Beam focusing elements, \raisebox{.5pt}{\textcircled{\raisebox{-.9pt}{2}}}: scattering chamber, \raisebox{.5pt}{\textcircled{\raisebox{-.9pt}{3}}}: DAGOBERT, \raisebox{.5pt}{\textcircled{\raisebox{-.9pt}{4}}}: QCLAM spectrometer, \raisebox{.5pt}{\textcircled{\raisebox{-.9pt}{5}}}: magnetic field shielding.}
	\label{fig:eepg_setup}
\end{figure}

Electron-$\gamma$ coincidence measurements are made possible by the unique combination of the QCLAM electron spectrometer and the $\gamma$-ray detector system DAGOBERT using the high-quality electron beam provided by the S-DALINAC. 
A CAD model of the experimental setup is shown in Fig.~\ref{fig:eepg_setup}. 
\\
The electron beam (green arrow) passes through a system of quadrupole magnets and steerer elements (1) that optimize the beam quality and the focusing on the target. 
A few meters downstream, the electron beam impinges the target located inside a vacuum scattering chamber (2). 
The $\gamma$ rays emitted in the $(e,e^\prime \gamma)$ reactions are detected with DAGOBERT (3), an array consisting of minimum six LaBr\textsubscript{3}:Ce scintillation detectors. 
The scattered electrons are observed with the QCLAM spectrometer (4). 
In order to protect DAGOBERT from the magnetic fields of the QCLAM spectrometer, a dedicated shielding is mounted (5). 
A detailed description of the individual components is provided in the following subsections.

\subsection{QCLAM Electron Spectrometer}
\begin{figure}[htbp]
    \centering
    \includegraphics[width=0.75\linewidth]{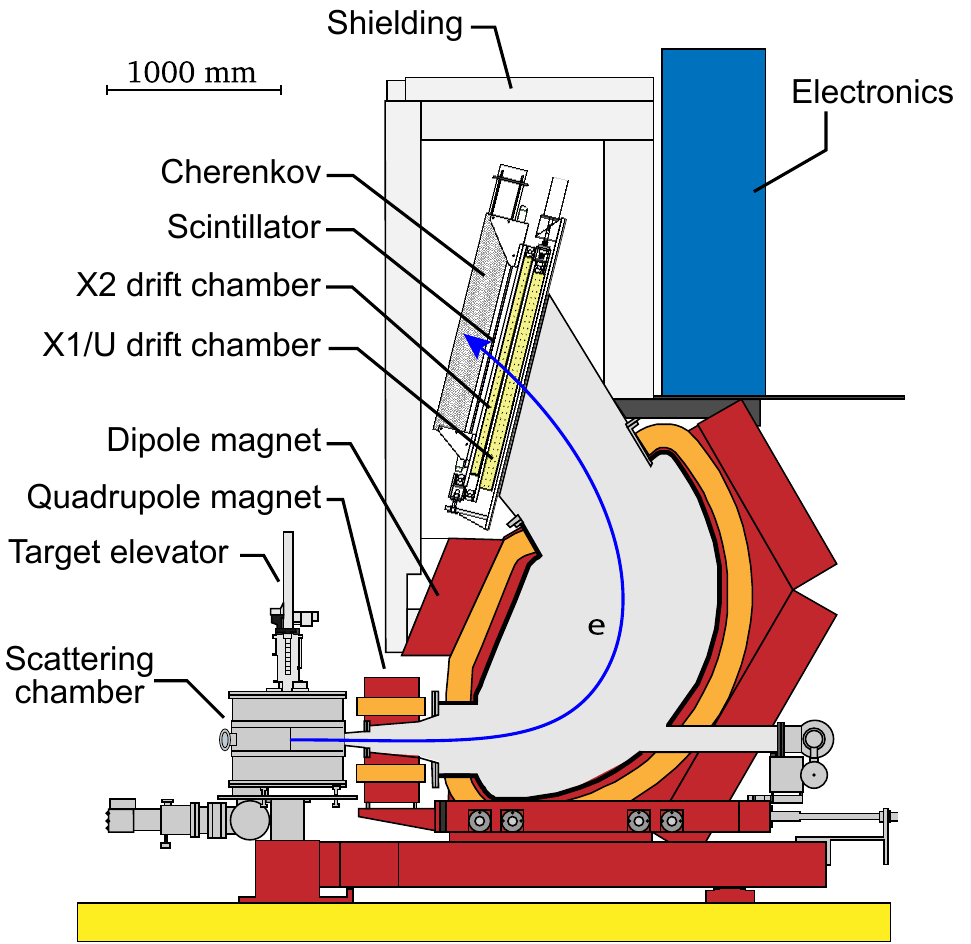}
    \caption{Cross-section through the QCLAM spectrometer. Electrons scattered off the target in the scattering chamber are focused horizontally in the quadrupole magnet and are bend by a dipole magnet towards the detector system.}
    \label{fig:qclam}
\end{figure}

\begin{figure}[htbp]
    \centering
    \begin{subfigure}[b]{0.75\textwidth}
        \includegraphics[width=\linewidth]{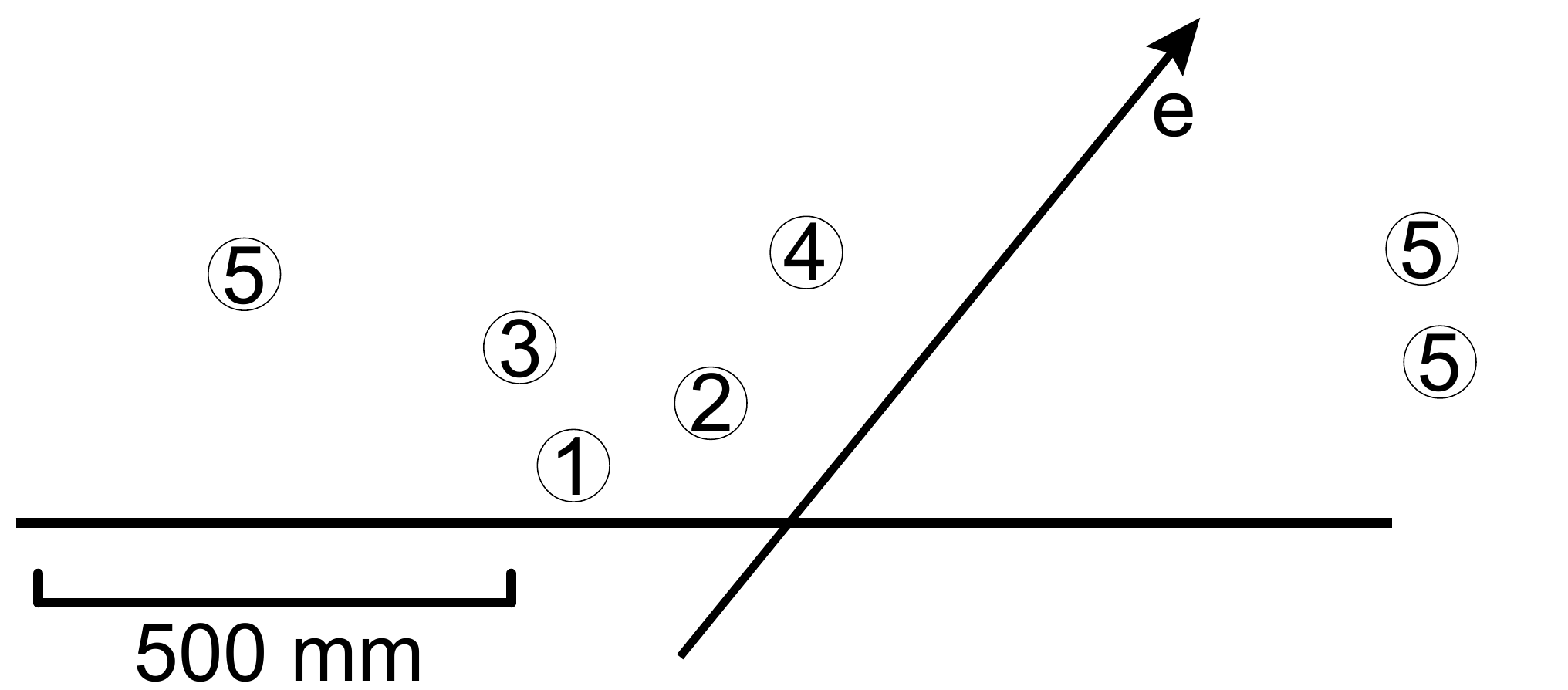}
        \caption{Cross section through the QCLAM spectrometer detector system consisting of three layers of multi wire drift chambers (MWDC), the X1/U double chamber {\large{\raisebox{-.6pt}{\textcircled{\raisebox{-.2pt}{{\small 1}}}}}} and the X2 chamber {\large{\raisebox{-.6pt}{\textcircled{\raisebox{-.2pt}{{\small 2}}}}}}, a plastic scintillation detector {\large{\raisebox{-.6pt}{\textcircled{\raisebox{-.2pt}{{\small 3}}}}}} and a Cerenkov detector {\large{\raisebox{-.6pt}{\textcircled{\raisebox{-.2pt}{{\small 4}}}}}}. Photo-multiplier tubes {\large{\raisebox{-.6pt}{\textcircled{\raisebox{-.2pt}{{\small 5}}}}}} are attached to the scintillation and Cerenkov detectors.}
        \label{fig:qclam_detector:A} 
    \end{subfigure}
    \begin{subfigure}[b]{0.75\textwidth}
        \includegraphics[width=\linewidth]{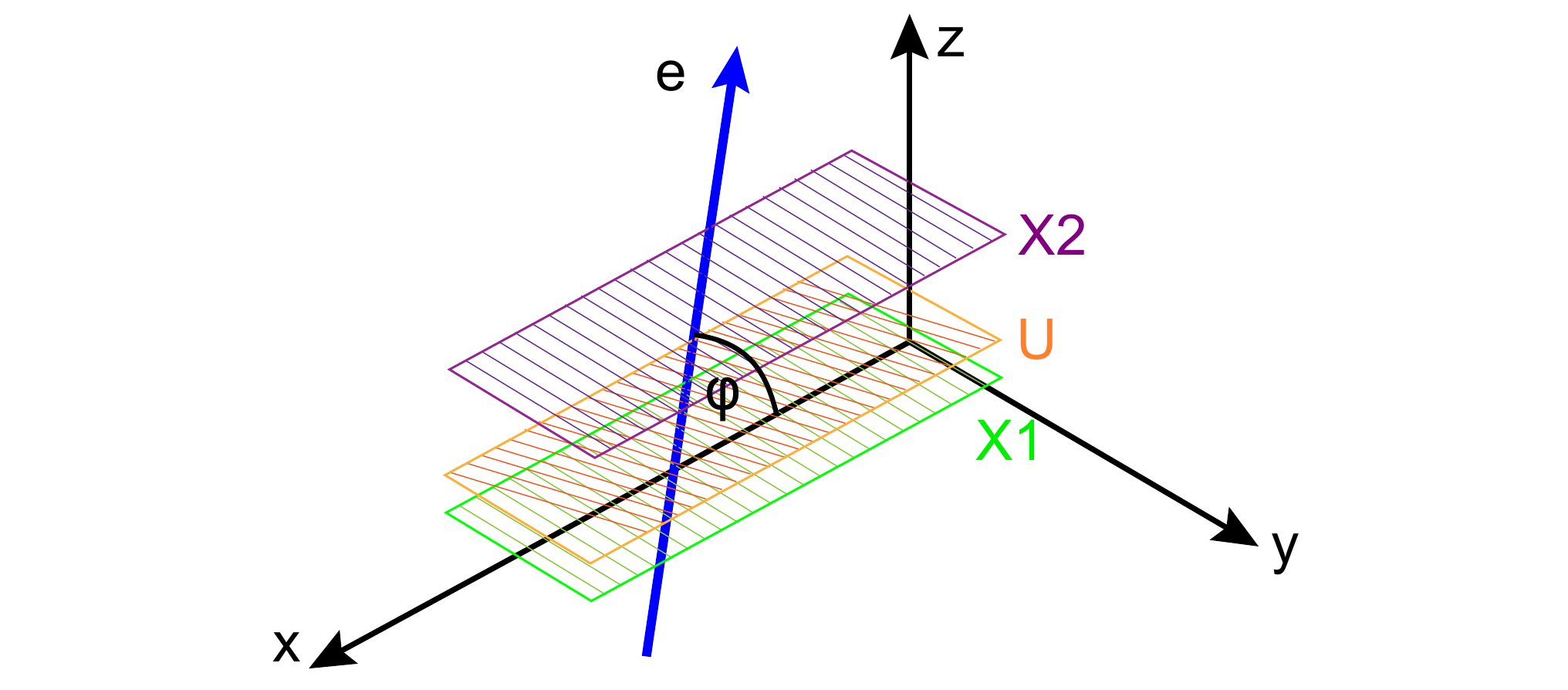}
        \caption{Coordinate system of the QCLAM spectrometer detector system.}
        \label{fig:qclam_detector:B}
    \end{subfigure}
    \caption{QCLAM spectrometer detector system.}
    \label{fig:qclam_detector}
\end{figure}

A cross section of the QCLAM spectrometer is shown in Fig.~\ref{fig:qclam}.
Electrons scattered into the QCLAM spectrometer first pass through a horizontally focusing quadrupole magnet.
The horizontal focus maintains the large solid angle acceptance despite the geometric limitation of the following dipole magnet in the form of a narrow gap between the pole shoes. 
In the dipole magnet, electrons are deflected onto a circular path allowing for a momentum analysis and thus energy measurement.
Imaging errors of the electron-optical system are reduced by a sextupole component which is created by a fifth, neutral pole of the quadrupole magnet and the dipole magnet's concave (convex) shape of the entrance (exit).
After passing through the magnet system, the electrons leave the vacuum through a mylar window and enter the detector system. 
Inside the detector system, shown in Fig.~\ref{fig:qclam_detector}, electrons first pass through three layers of multi wire drift chambers (MWDC), the X1/U double chamber and the X2 chamber, before entering a $2\,$cm thick plastic scintillation detector and a Cerenkov detector. 
Other arrangements of electron tracking and timing detectors could also fit the requirements of our measurements. This article addresses the functionality of the presently available setup.\\
Position information from the MWDC is used to reconstruct the trajectory for each electron through the spectrometer. 
This is required to calculate the intersection with its curved focal plane. 
By measuring electron energy spectra of known nuclei, the intersection point can be associated with an energy value.
The X1 and X2 chambers are used to calculate the dispersive coordinate and dispersive angle. 
The wires of the U chamber are rotated by $26.5^\circ$ with respect to the wires of the other two chambers. 
The non-dispersive coordinate and angle are calculated from the measured U coordinate and the information of the other two chambers~\cite{Hummel.20.10.1992}.
The scintillation detector is used to trigger the data acquisition and generate a position-independent timestamp.
The time signal is generated from two photo-multiplier tubes at the opposite ends of the scintillator. 
The Cerenkov detector is used to suppress background by discriminating between electrons and $\gamma$ radiation.

The spectrometer can be rotated about its pivot point allowing for the measurement of electrons scattered to angles between $25^\circ$ and $155^\circ$ relative to the incoming beam~\cite{Knirsch.10.12.1991}. 
The key design parameters of the QCLAM spectrometer are summarized in Tab.~\ref{tab:qclam_paras} and further details can be found in Refs.~\cite{Knirsch.10.12.1991, Neumeyer.1997, Singer.2019b, Steinhilber.2022}.
\begin{table}[htbp]
    \caption{Design parameters of the QCLAM spectrometer.} \label{tab:qclam_paras}
    \begin{tabular*}{\tblwidth}{@{}LL@{}}
        \toprule
        Maximum momentum & $220\,$MeV/c \\
        Momentum acceptance (relative) & $\pm 10\,\%$ \\
        Energy resolution (relative) & $10^{-4}$ \\
        Spectrometer angle range & $25^\circ$ to $155^\circ$ \\
        Horizontal/Vertical angle acceptance & $\pm 100\,$mrad \\
        Solid-angle coverage & $35\,$msr \\
        Dispersion (dipole magnet) & $2.21\,\mathrm{cm}/\%$ \\
        Maximum current (dipole magnet) & $320\,$A \\
        Maximum current (quadrupole magnet) & $225\,$A \\
        \bottomrule
    \end{tabular*}
\end{table}
Of particular interest are the large solid angle acceptance of $35\,$msr and a momentum acceptance of $\pm10\,$\% essential for coincidence experiments over a broad excitation-energy range.

\subsection{Beamline} \label{Sec:Beamline}
A spatially well-tuned electron beam is crucial for $(e,e^\prime \gamma)$ measurements. 
The energy resolution of the scattered electrons measured in the QCLAM spectrometer depends strongly on the size and position of the beam spot on the target~\cite{Knirsch.10.12.1991}. 
Therefore, an active beam-stabilization system using a feedback loop has been developed and implemented~\cite{Schneider.2024}.
The purpose of this system is to stabilize the center beam position within a $200\,\upmu$m margin at the interaction point. 
The kick on the beam generated by the position control is mitigated with a secondary set of correction magnets to ensure an overall improvement of the transverse beam stability and, therefore, the spectrometer resolution.
Furthermore, beam-induced background radiation from secondary reactions of electrons scattered off the beam pipe and other materials downstream the target is strongly reduced by a well-focused electron beam. 
Thus, two quadrupole magnets for improved focusing and two steerer elements for a better tuning of the electron beam are in operation a few meters upstream the scattering chamber, compare Fig.~\ref{fig:eepg_setup}.

\subsection{Scattering Chamber} \label{Sec:Scattering_Chamber}
\begin{figure}[htbp]
    \centering
    \includegraphics[width=0.75\linewidth]{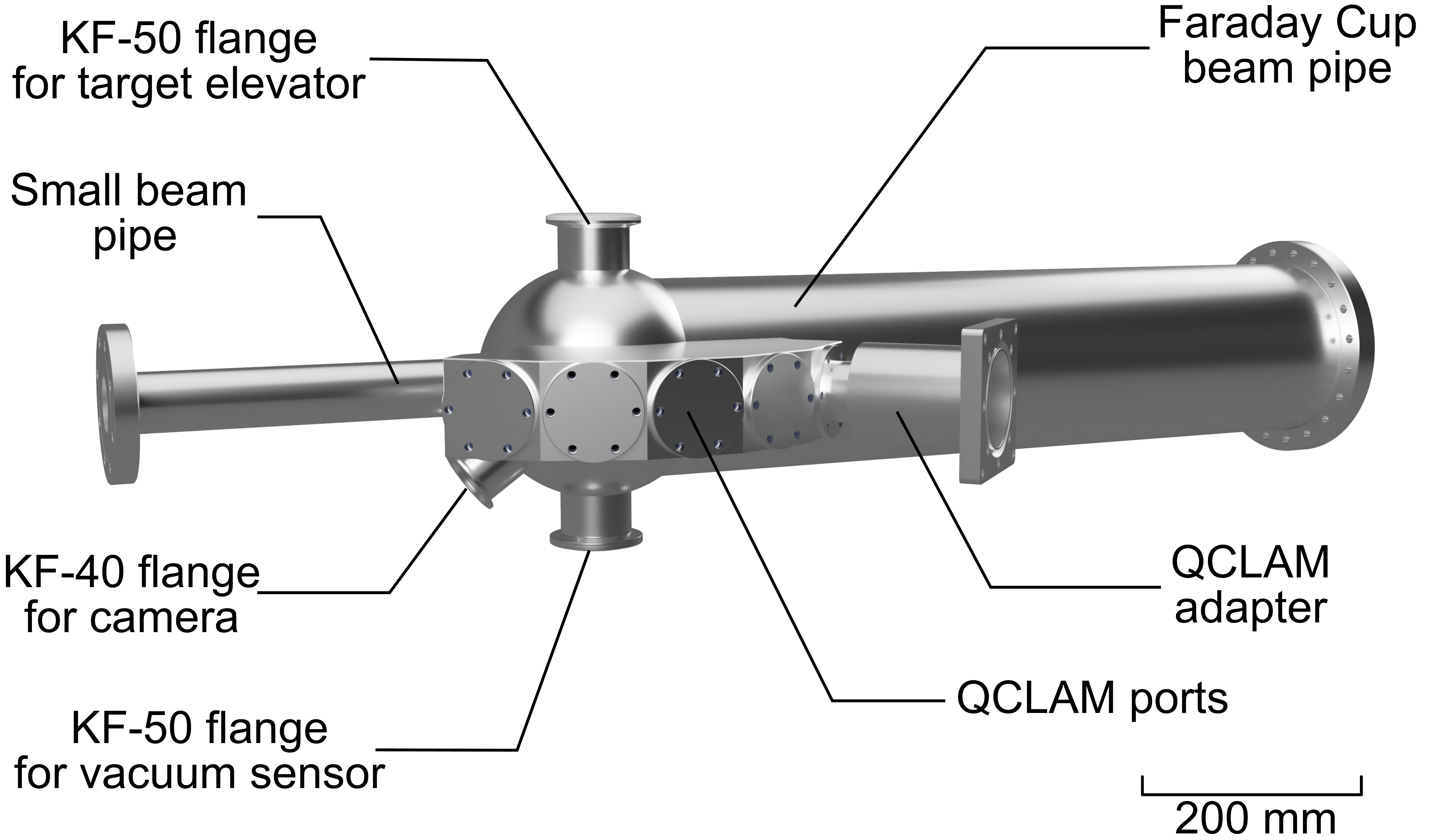}
    \caption{Design of the $(e,e^\prime \gamma)$ scattering chamber.}
    \label{fig:scattering_chamber}
\end{figure}

The scattering chamber for $(e,e^\prime \gamma)$ experiments is designed to minimize beam-induced background radiation and permit large solid angle coverage of the $\gamma$-ray detectors. 
The design of the scattering chamber is depicted in Fig.~\ref{fig:scattering_chamber}.
\\
The electron beam enters the scattering chamber from the left through a $44.5\,$mm wide beam pipe. 
The primary electron beam passes the target and exits through a beam pipe with a diameter of $154\,$mm heading towards the beam dump accounting for the divergence of the beam after interaction with the target.
The scattering chamber is made from aluminum with a wall thickness of $3\,$mm only to minimize beam-induced background. 
The flanges connecting the scattering chamber to the QCLAM spectrometer are attached to a box, which allows to select one of five QCLAM spectrometer positions in the interval from $47.5^\circ$ to $132.5^\circ$ in increments of $21.25^\circ$. 
The adapter required to connect scattering chamber and QCLAM spectrometer reduces the solid angle coverage to $25\,$msr.

\subsection{Gamma Ray Detectors and Holding Structure}
\begin{figure}[htbp]
    \centering
    \includegraphics[width=0.57\linewidth,angle=-90]{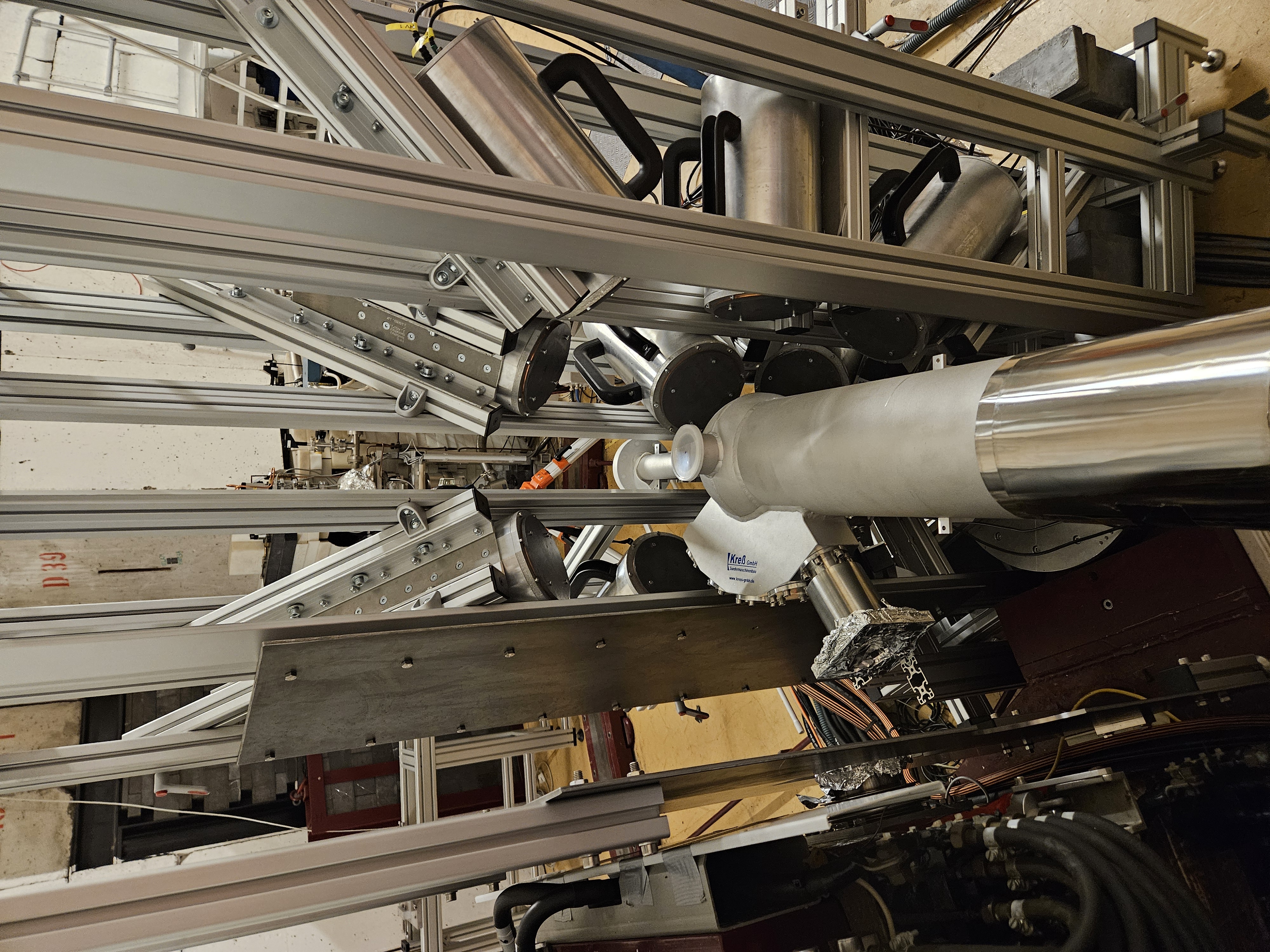}
    \caption{Photo of the DAGOBERT setup. Detector towers are placed around the scattering chamber at the center of rotation of the QCLAM spectrometer. The picture was taken from the direction of the Faraday Cup beam pipe and points to the entrance beam pipe (compare Fig.~\ref{fig:scattering_chamber}). The QCLAM spectrometer is positioned on the left side outside the image.}
    \label{fig:daqgobert_setup}
\end{figure}

DAGOBERT is composed of cerium-doped lanthanum bromide crystals (LaBr\textsubscript{3}:Ce) of the type Saint-Gobain BrilLanCe®380~\cite{SaintGobainCrystals.2021} with dimensions of $3\,'' \times 3\,''$. 
The crystals are characterized by excellent intrinsic time resolution of $0.7\,$ns important for random background suppression, good energy resolution ($2.9\,\%$ at $662\,$keV) and high count rate capabilities due to the short decay time of the scintillation signals.
Such detectors allow single count rates of several $100\,$kcps~\cite{Loher.2012, Steinhilber.2022}.
The detectors are mounted inside tower-like mechanical holding structures as illustrated in Fig.~\ref{fig:eepg_setup}.
Figure~\ref{fig:daqgobert_setup} provides a photo of the DAGOBERT array.
\\

The angle in the scattering plane relative to the beam axis is adjusted by rotating the towers around the center of the scattering chamber. 
The tilt angle of the detector within the tower relative to the scattering plane and the distance between each LaBr\textsubscript{3}:Ce scintillator and the target can be set individually. 
Due to the increased intensity of bremsstrahlung background in forward direction of the electron beam (compare Fig~\ref{fig:nuclear_bremsstrahlung}), polar angles $\theta > 180^\circ$ are favored for reduced background. 
The current design of the setup consists of three detector towers, each accommodating at least three LaBr\textsubscript{3}:Ce at variable azimuthal angles.

For further passive shielding against beam-induced background, the crystals of the LaBr\textsubscript{3}:Ce detectors are mounted inside a housing consisting of a lead tube with $3\,$cm-thick walls, while the photo-multiplier tubes (PMT) in the rear area are shielded by $2\,$cm of lead. 
Additional lead shielding in front of the active detector surfaces reduces the low-energy background originating from the target. 
An additional copper shielding is used to attenuate X-rays emitted from the lead. 
The thickness of the lead and copper shielding can be varied by stacking the $5\,$mm thick lead filters and $2\,$mm thick copper filters, respectively.

\subsection{Magnetic Field Shielding}
\begin{figure}[htbp]
	\centering
		\includegraphics[width=0.75\linewidth]{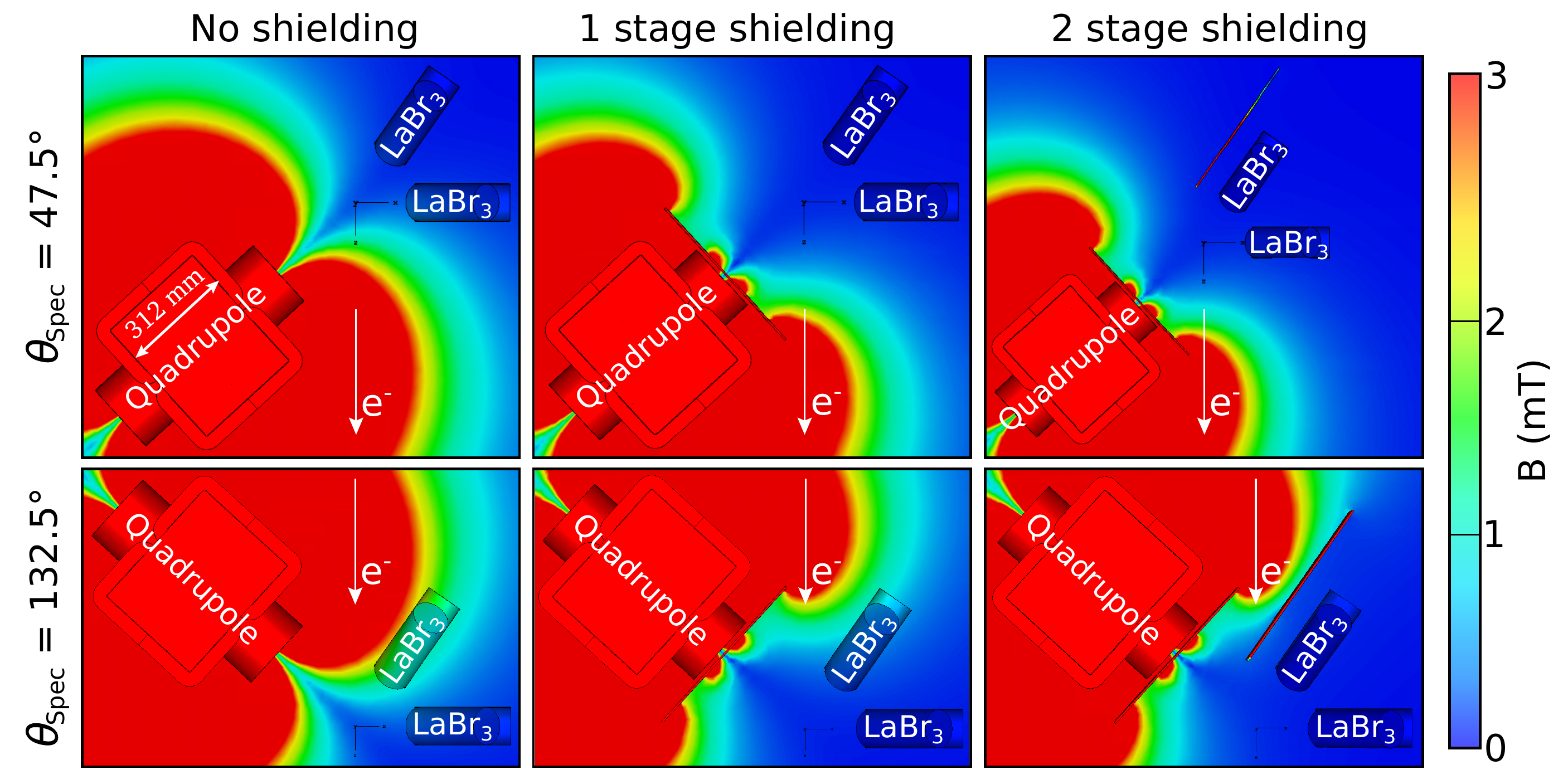}
	\caption{CST simulation of magnetic field maps for different spectrometer angles (top row: $\theta_{\textrm{Spec}} = 47.5^\circ$ and bottom row: $\theta_{\textrm{Spec}} = 132.5^\circ$) with and without magnetic field shielding. The first column represents no shielding. The second column has a magnetic field shielding in front of the QCLAM spectrometer quadrupole magnet. The third column has a shielding in front of the QCLAM spectrometer quadrupole magnet and on the side of the second $\gamma$-ray detector tower. The direction of the electron beam is indicated by the vertical arrows.}
	\label{fig:field_shielding}
\end{figure}

The PMTs of the LaBr\textsubscript{3}:Ce detectors are sensitive to external magnetic fields which impact the amplification of the signals. Thus, they must be shielded against the far-reaching magnetic field of the QCLAM spectrometer quadrupole magnet. 
Simulations with the software CST Studio Suite 2020~\cite{DassaultSystemes.2019} were performed to estimate the magnetic field strength at the position of the scintillation detectors. 
Figure~\ref{fig:field_shielding} shows simulated field maps for the most forward ($47.5^\circ$) and backward ($132.5^\circ$) electron-scattering angles without shielding and with one- or two-stage shielding. 
\\
The first stage of magnetic field shielding consists of a $5\,$mm thick iron plate in front of the quadrupole magnet, leaving the aperture of the spectrometer free, which reduces the magnetic field strength towards the scattering chamber. The second stage consists of an additional $5\,$mm thick iron plate at the side of a detector tower.
All simulations were performed with the maximum current of the QCLAM spectrometer quadrupole magnet of $225\,$A.
The two-stage shielding leads to a reduction of the magnetic field strength on the detector surface by one order of magnitude from $2.4\,$mT to $0.3\,$mT for the $132.5^\circ$ setup.

\subsection{Performance of DAGOBERT@QCLAM} \label{Sec:Performance}
The performance of the $(e,e^\prime \gamma)$ setup can be quantified by the electro-photo production coincidence resolution 
\begin{equation}
    \mathcal{R} = \frac{E_i}{\Delta E_x} \cdot \frac{E_{\gamma}}{\Delta E_{\gamma}} \cdot \frac{1}{\Gamma (T_{\textrm{coin}})},
\end{equation}
defined as the product of the energy resolutions $E/\Delta E$ for the detection of the scattered electrons and the produced photons per coincidence time resolution denoted by $\Gamma (T_{\textrm{coin}})$.
All resolutions are given as full width at half maximum (FWHM). 

In Tab.~\ref{tab:comparison_performance}, a comparison between the performance of the first $(e,e^\prime \gamma)$ experiment conducted at the University of Illinois~\cite{Papanicolas.1985, C.N.Papanicolas.1985} on the target nucleus $^{12}\textrm{C}$, a $^{12}\textrm{C}(e,e^\prime \gamma)$ experiment with a preliminary version of DAGOBERT@QCLAM and a $^{96}\textrm{Ru}(e,e^\prime\gamma)$ measurement with the full DAGOBERT setup is given. 
In the preliminary experimental setup only three LaBr\textsubscript{3}:Ce detectors, a scattering chamber not optimized for $(e,e^\prime \gamma)$ experiments and a standard beamline without the improvements described in Sec.~\ref{Sec:Beamline} were used. 
\begin{table}[htbp]
    \caption{Comparison between the performance of the first $(e,e^\prime \gamma)$ experiment at the University of Illinois~\cite{Papanicolas.1985, C.N.Papanicolas.1985} with a $^{12}\textrm{C}$ target and the $^{12}\textrm{C}(e,e^\prime \gamma)$ experimental campaign using a preliminary setup of DAGOBERT@QCLAM in 2019~\cite{Steinhilber.2022}. 
    Additionally, the performance parameters of the full DAGOBERT@QCLAM setup from the $^{96}\textrm{Ru}(e,e^\prime\gamma)$ experiment (see Sec.~\ref{chap::Data_Analysis}) are shown.} \label{tab:comparison_performance}
    \begin{tabular*}{\tblwidth}{@{} LCCC@{} }
        \toprule
        Parameters & $^{12}\textrm{C}$ Illinois~\cite{Papanicolas.1985, C.N.Papanicolas.1985} & $^{12}\textrm{C}$ pre setup~\cite{Steinhilber.2022} & $^{96}\textrm{Ru}$ this work\\
        \midrule
        Electron beam energy $E_i$ & $66.9\,$MeV & $30\,$MeV & $85\,$MeV\\
        $\Delta E_x$ for $2_1^+$ state of $^{12}\textrm{C}$ / of $^{96}\textrm{Ru}$ & $100\,$keV & $41\,$keV & $86.4(1.0)\,$keV\\
        $\Delta E_x/E_{i}$ for $2_1^+$ state of $^{12}\textrm{C}$ / of $^{96}\textrm{Ru}$ & $0.15\,\%$ & $0.14\,\%$ & $0.102(2)\,\%$\\
        Electron solid angle acceptance$/4\pi$ & $0.04\,\%$ & $0.28\,\%$ &$0.20\,\%$\\
        \midrule
        $\Delta E_{\gamma}$ for $2_1^+$ state of $^{12}\textrm{C}$ / at $4.44\,$MeV & $287\,$keV & $53.3\,$keV & $71.1(2)\,$keV\footnotemark[1]\\
        $\Delta E_{\gamma}/E_{\gamma}$ for $2_1^+$ state of $^{12}\textrm{C}$ / at $4.44\,$MeV & $6.46\,\%$ & $1.20\,\%$ & $1.60(1)\,\%$\footnotemark[2]\\
        Total photopeak efficiency at $1.33\,$MeV & $-$ & $0.21\,\%$ & $0.61(2)\,\%$\\
        \midrule
        Coincidence time resolution $\Gamma (T_{\textrm{coin}})$ & $20\,$ns & $3.1\,$ns & $1.9\,$ns\\
        Width of coincidence time gate & $43.5\,$ns & $5\,$ns & $4\,$ns\\
        \midrule
        $\mathcal{R}$ & $0.52\,\textrm{ps}^{-1}$ & $19.7\,\textrm{ps}^{-1}$ & $32.3\,\textrm{ps}^{-1}$\\
        \bottomrule
    \end{tabular*}
\end{table}\\
\footnotetext[1]{$\Delta E_{\gamma}$ at $1.33\,$MeV: $43.7(1)\,$keV.}
\footnotetext[2]{$\Delta E_{\gamma}/E_{\gamma}$ at $1.33\,$MeV: $3.28(1)\,\%$.}

Comparing the two $^{12}\textrm{C}(e,e^\prime \gamma)$ experiments the total and relative electron and $\gamma$-ray energy resolution, the coincidence time resolution and the width of the coincidence time gate are all superior for the DAGOBERT@QCLAM setup. 
This leads to an improvement of $\mathcal{R}$ by a factor of $38$ compared to the first $(e,e^\prime \gamma)$ experiment at the University of Illinois~\cite{Papanicolas.1985, C.N.Papanicolas.1985} from $0.52\,\textrm{ps}^{-1}$ to $19.7\,\textrm{ps}^{-1}$.

For the $^{96}\textrm{Ru}(e,e^\prime \gamma)$ measurement, the QCLAM spectrometer energy resolution was obtained from the excitation energy spectrum in Fig.~\ref{fig::Excitation-Energy-Spectrum}. 
The value of $\Delta E_x$ is larger compared to the $^{12}\textrm{C}$ measurement due to energy loss straggling in the target~\cite{Goldwasser.1952}. 
The $^{96}\textrm{Ru}$ target had a larger areal density and charge number compared to $^{12}\textrm{C}$ and is backed with $^{197}\textrm{Au}$ from both sides.
The $\gamma$-ray energy resolution and the total photopeak efficiency of the DAGOBERT setup was obtained from calibration measurements (compare Sec.~\ref{Sec:Gamma-Cal-energy}). 
The determination of the coincident time resolution is described in Sec.~\ref{Sec:Coin-Time-Cor} and was further improved by a factor of $1.6$ compared to the $^{12}\textrm{C}$ measurement. 
In total, the electro-photo production coincidence resolution $\mathcal{R}$ exhibits a value of $\mathcal{R} = 32.3\,\textrm{ps}^{-1}$. Along with a five times larger acceptance for scattered electrons, this results in a two-to-three orders of magnitude higher sensitivity to electro-photo production reactions as compared to the pioneering set-up at Illinois.

\section{Electronics and Data Acquisition} \label{chap::Electronics_DAQ}
The coincidence data acquisition system (DAQ) is composed of a newly developed DAQ for DAGOBERT combined with the existing DAQ for the QCLAM spectrometer~\cite{Singer.2019b}.
Both DAQs are based on the multi-branch system (MBS)~\cite{GSIHelmholzzentrumfurSchwerionenforschungGmbH.22.10.2020} developed by GSI.
The modules of the QCLAM spectrometer and DAGOBERT DAQs are based on the VME and NIM standards. 
Signals from the LaBr\textsubscript{3}:Ce detectors are processed by 14-bit SIS3316 FADCs from Struck Innovative Systems GmbH~\cite{Struck-SIS3316}. 
A key parameter of these modules are a $14\,$bit accuracy for the $2\,$V / $5\,$V range of signals and a time resolution up to $50\,$ps. 

\begin{figure}[htbp]
    \centering
    \includegraphics[width=0.75\linewidth]{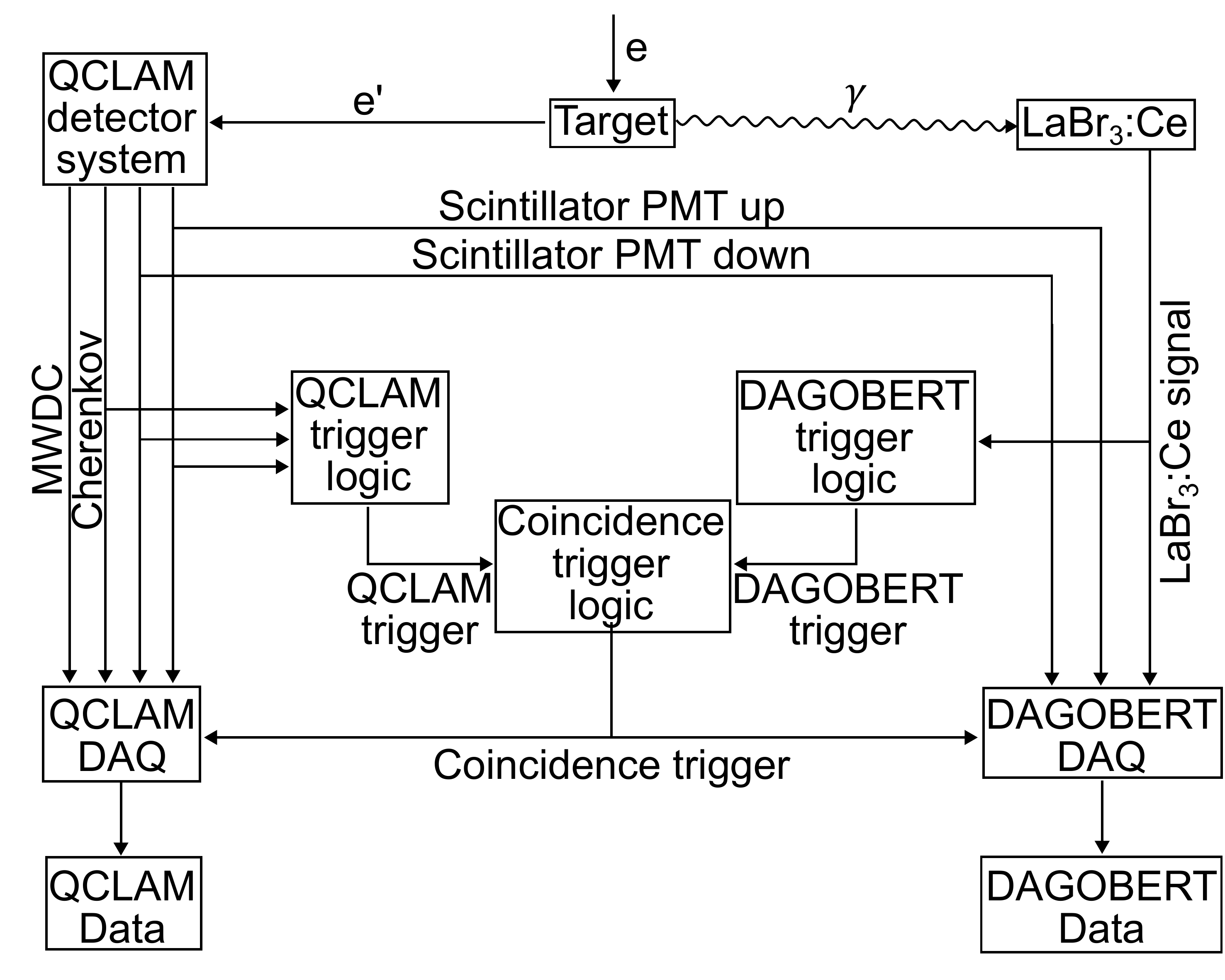}
    \caption{Concept of the electron-$\gamma$ coincidence data acquisition system.}
    \label{fig:daq_concept}
\end{figure}

For electron-$\gamma$ coincidence measurements, both DAQs are combined following the scheme given in Fig.~\ref{fig:daq_concept}. 
\\
Scattered electrons $e^\prime$ entering the QCLAM spectrometer detector system produce signals in the MWDC and PMTs of the scintillation and Cerenkov detectors. 
These signals are processed in the QCLAM spectrometer DAQ. The QCLAM trigger logic enables the combination of signals from the PMTs and the Cerenkov detector for the QCLAM trigger to suppress contributions from $\gamma$-ray background, e.g. cosmic rays. 
Photons detected with one of the LaBr\textsubscript{3}:Ce scintillation detectors produce a $\gamma$ signal in the DAGOBERT DAQ. 
The DAGOBERT trigger is built and in addition signals of the PMTs of the QCLAM spectrometer scintillation detectors are stored for a reference time stamp. 
When the LaBr and QCLAM trigger are generated within a preset time window, a coincidence trigger is built and sent back to the two data acquisition systems for activation of the writing of the corresponding two data types.

For the further analysis of the electron-$\gamma$ coincidence data, QCLAM and DAGOBERT events are combined to form coincidence events. 
This is achieved by a trigger pattern matching due to the information from the PMT signals of the QCLAM spectrometer scintillator stored in both data sets.
QCLAM spectrometer events are unpacked~\cite{Singer.2019b} and then evaluated using a custom-made software tool performing standard corrections such as energy calibrations and corrections of imaging errors of the electron-optical system. 
These corrections are described in more detail in the following Chapter~\ref{chap::Data_Analysis}. 
Raw data recorded by the DAGOBERT DAQ is unpacked with an adapted UCESB~\cite{Johansson.2010} unpacker and evaluated by a custom-made software tool.
Specific corrections such as a rate-dependent energy shift are performed.

\section{Commissioning of the DAGOBERT setup} \label{chap::Data_Analysis}
The first $(e,e^\prime \gamma)$ experiments with the full established DAGOBERT setup were performed on $^{12}\textrm{C}$ for calibration purposes and $^{96}\textrm{Ru}$ to study the $\gamma$-decay branching ratio of its mixed-symmetry $2^+_3$ state~\cite{Pietralla.2001}.
The electron beam energy was $85\,$MeV with beam currents of about $100\,$nA.
The QCLAM spectrometer was placed at a scattering angle of $46.3^\circ$. 
Its dipole and quadrupole magnetic fields were set to $I_{\textrm{Dipole}} = 112\,A$ and $I_{\textrm{Quadrupole}} = 140\,A$, respectively, to detect and momentum analyze electrons with energy losses in the range of $0 - 15\,$MeV. The beam time lasted for approximately $250\,$h.
The emitted $\gamma$ rays were detected with DAGOBERT consisting of six LaBr\textsubscript{3}:Ce detectors positioned in two towers with three detectors each. 
Their position in terms of distances from the target and angles $\theta$ and $\Phi$ defined by Fig.~\ref{fig:coordinate_system} are summarized in Table~\ref{tab:measurement_detector_position}. 
\begin{table}[htbp]
    \caption{Positioning of the six LaBr\textsubscript{3}:Ce detectors relative to the beam axis and target. The definition of $\theta$ and $\Phi$ is explained in Fig.~\ref{fig:coordinate_system}.} \label{tab:measurement_detector_position}
    \begin{tabular*}{\tblwidth}{@{} CCCC@{} }
        \toprule
         Detector & Distance (mm) & $\theta$ ($^\circ$) & $\Phi$ ($^\circ$)\\
        \midrule
        0 & $263(2)$ & $310.6(2)$ & $67.7(1)$ \\
        1 & $172(2)$ & $336.2(0)$ & $0.0(0)$ \\
        2 & $280(2)$ & $311.2(2)$ & $-67.2(1)$ \\
        3 & $264(2)$ & $277.9(2)$ & $45.5(1)$ \\
        4 & $171(2)$ & $281.2(0)$ & $0.0(0)$ \\
        5 & $263(2)$ & $278.0(2)$ & $-44.6(1)$ \\
        \bottomrule
    \end{tabular*}
\end{table}

For the measurements three targets were mounted on a remote controlled target ladder: $^{12}\textrm{C}$, BeO, and $^{96}\textrm{Ru}$. 
The $^{12}\textrm{C}$ foil with an areal density of $10\,\textrm{mg}/\textrm{cm}^2$ was used to calibrate the energy and efficiency of the QCLAM spectrometer. 
The second position was kept empty for electron beam and background optimization purposes. 
The BeO target was also used for electron beam optimization. 
The $^{96}\textrm{Ru}$ target with an areal density of $28\,\textrm{mg}/\textrm{cm}^2$ was sandwiched between two foils of $^{197}\textrm{Au}$ with a total areal density of $13.6\,\textrm{mg}/\textrm{cm}^2$.
In the following sections the necessary steps are presented to obtain an electron-$\gamma$ coincidence $E_x$-$E_\gamma$ matrix that can be further analyzed.

\subsection{Calibration: Electron Data}
Before the data of the QCLAM spectrometer can be used for analysis, corrections and calibrations on the measured raw signal need to be performed. These corrections and calibrations are discussed in detail in the following subsections.

\subsubsection{{Correction of Aberration}}
\begin{figure}[htbp]
    \centering
    \includegraphics[width=0.6\linewidth]{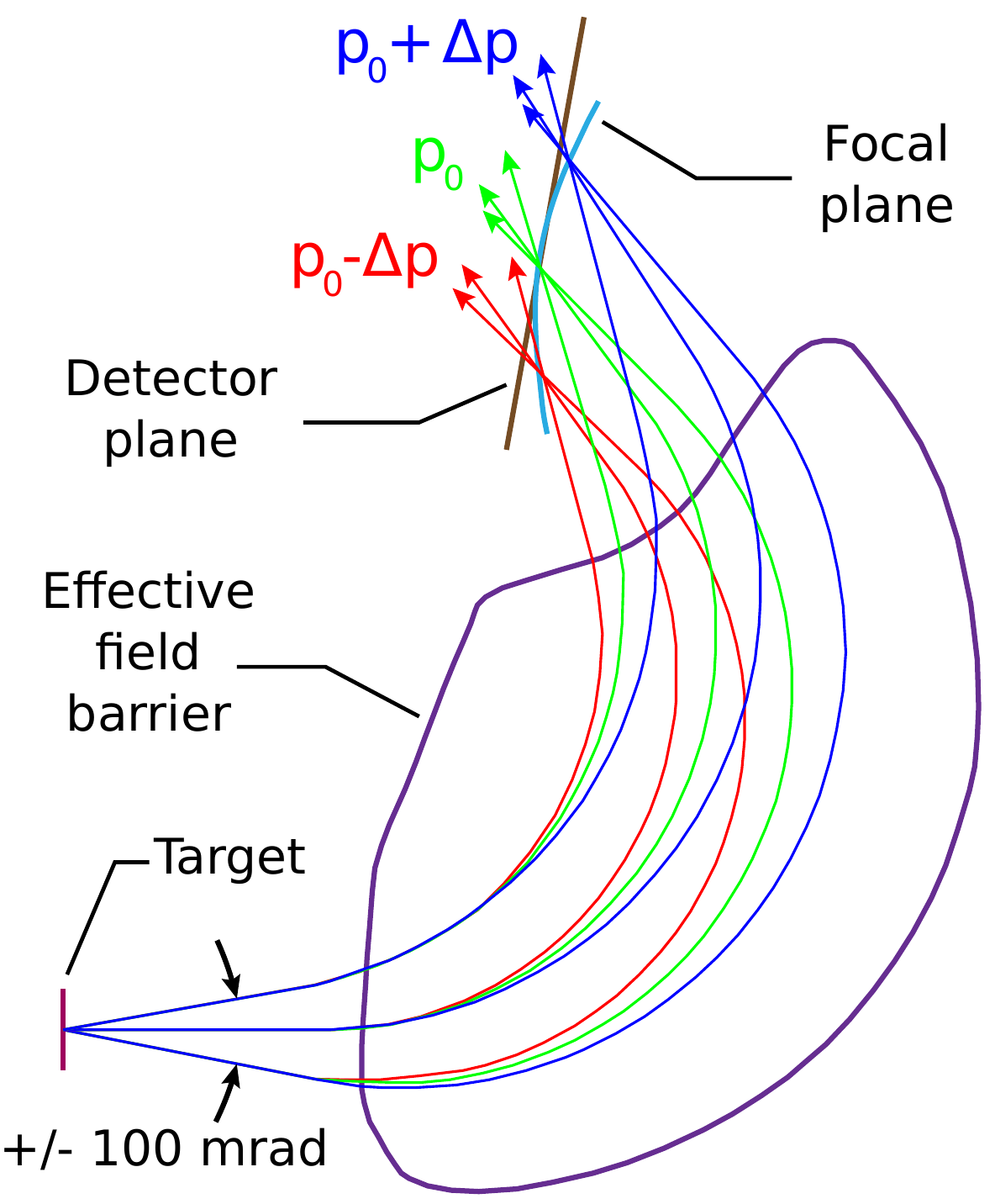}
    \caption{Deflection of the scattered electrons in the magnetic field of the QCLAM spectrometer. The magnetic system deflects electrons with the same momentum but different vertical entry angles to the same point on the curved focal plane. In the straight detector plane, the scattered electrons no longer hit the same point.}
    \label{fig::focal_plane}
\end{figure}

The focal plane of the QCLAM spectrometer is curved due to the electron-optical imaging properties of the magnet system. 
This is illustrated in Fig.~\ref{fig::focal_plane}. 
\\
The detector plane in which the drift chambers are located, on the other hand, is designed to be straight due to mechanical conditions. 
This results in a spatial mismatch between the spectrometer's focal plane and its MWDCs, which must be corrected for in the data analysis.
If scattered electrons with a different angle but the same momentum enter the QCLAM spectrometer, they hit the curved focal plane at the same point. 
This is not the case in the detector plane.
For a fixed momentum, the dispersive position $x$ and the dispersive angle $\phi$ in the detector system are coupled. 
Thus, aberration errors occur which have to be corrected prior to the energy calibration. 

\begin{figure}[htbp]
    \centering
    \begin{subfigure}[b]{0.75\textwidth}
        \includegraphics[width=\linewidth]{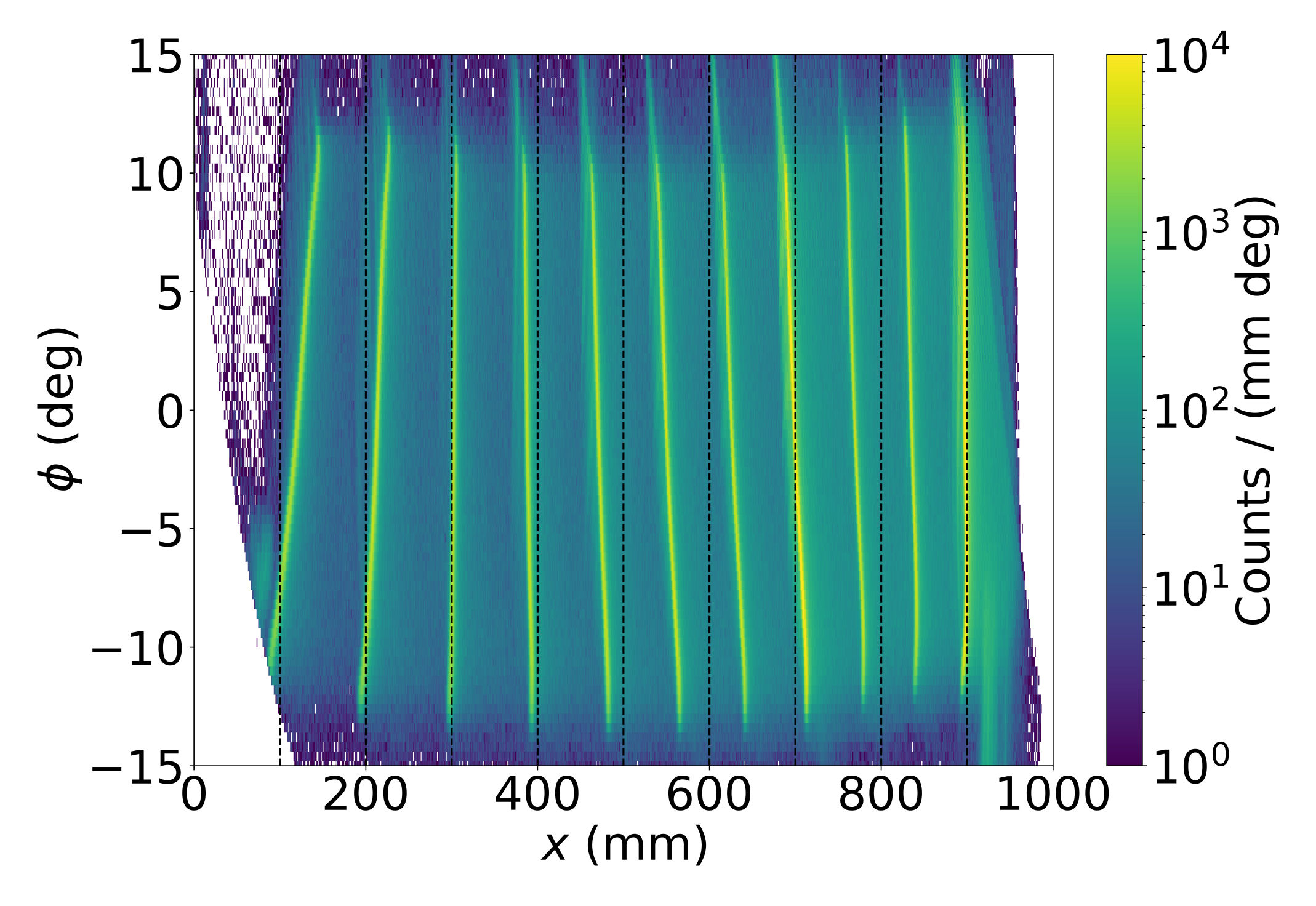}
        \caption{Uncorrected $x-\phi$ plot.}
        \label{fig:correction_of_electron_optics:A} 
    \end{subfigure}
    \begin{subfigure}[b]{0.75\textwidth}
        \includegraphics[width=\linewidth]{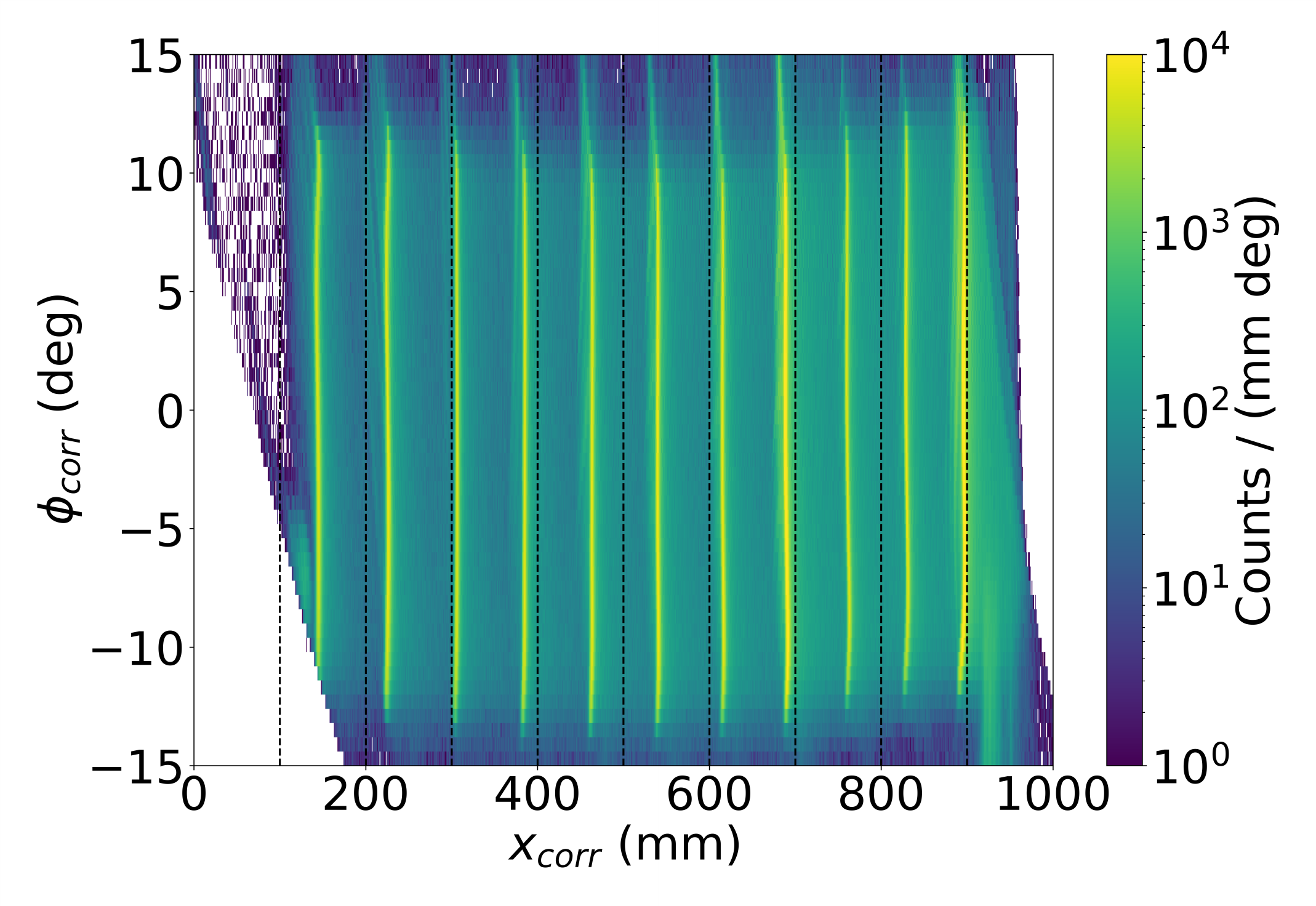}
        \caption{Corrected $x_{\textrm{corr}}-\phi_{\textrm{corr}}$ plot.}
        \label{fig:correction_of_electron_optics:B}
    \end{subfigure}
    \caption{Correction of the electron-optical aberration. The elastic line of $^{96}\textrm{Ru}$ was shifted over the focal plane by using $11$ different magnetic field settings.}
    \label{fig:correction_of_electron_optics}
\end{figure}

The aim of the correction is a mapping of the measured data to a corrected coordinate $x_{\textrm{corr}}$ using an empirical polynomial approach
\begin{equation}
    x_{\textrm{corr}}(x,\phi)=\sum_{i,j}a_{ij} x^i \phi_{\textrm{corr}}^j,
\end{equation}
where the parameters $a_{ij}$ are determined by a fit to calibration data. 
In a first step, the dependence of the dispersive angle on the dispersive coordinate is extracted using
\begin{equation}
    \phi_{\textrm{corr}}(x) = \phi - \phi(x) = \phi - (a+b\cdot x),
\end{equation}
with free parameters $a$ and $b$.
For their calibration, $x-\phi$ tuples distributed over the full acceptance of the spectrometer are required to which an $x_{\textrm{corr}}$ value is assigned.
For this purpose, a series of measurements of elastically scattered electrons at different magnetic field settings are performed covering the range from minimum to maximum dispersive positions in multiple steps.
A summed matrix of all measurements is shown in Fig.~\ref{fig:correction_of_electron_optics:A}.
Here, elastic lines of $^{96}\textrm{Ru}$ are visible for $11$ different magnetic field settings. 
These lines are curved due to the aberration of the electron spectrometer. 
After the correction of the aberration shown in Fig.~\ref{fig:correction_of_electron_optics:B}, straight vertical lines are obtained. 
This significantly improves the energy resolution of the QCLAM spectrometer, e.g for the magnetic field setting where the elastic line is positioned in the center of the focal plane from a FWHM of $1186\,$keV to $186\,$keV.

\subsubsection{{Energy Calibration}} 
\begin{figure}[htbp]
    \centering
    \includegraphics[width=0.75\linewidth]{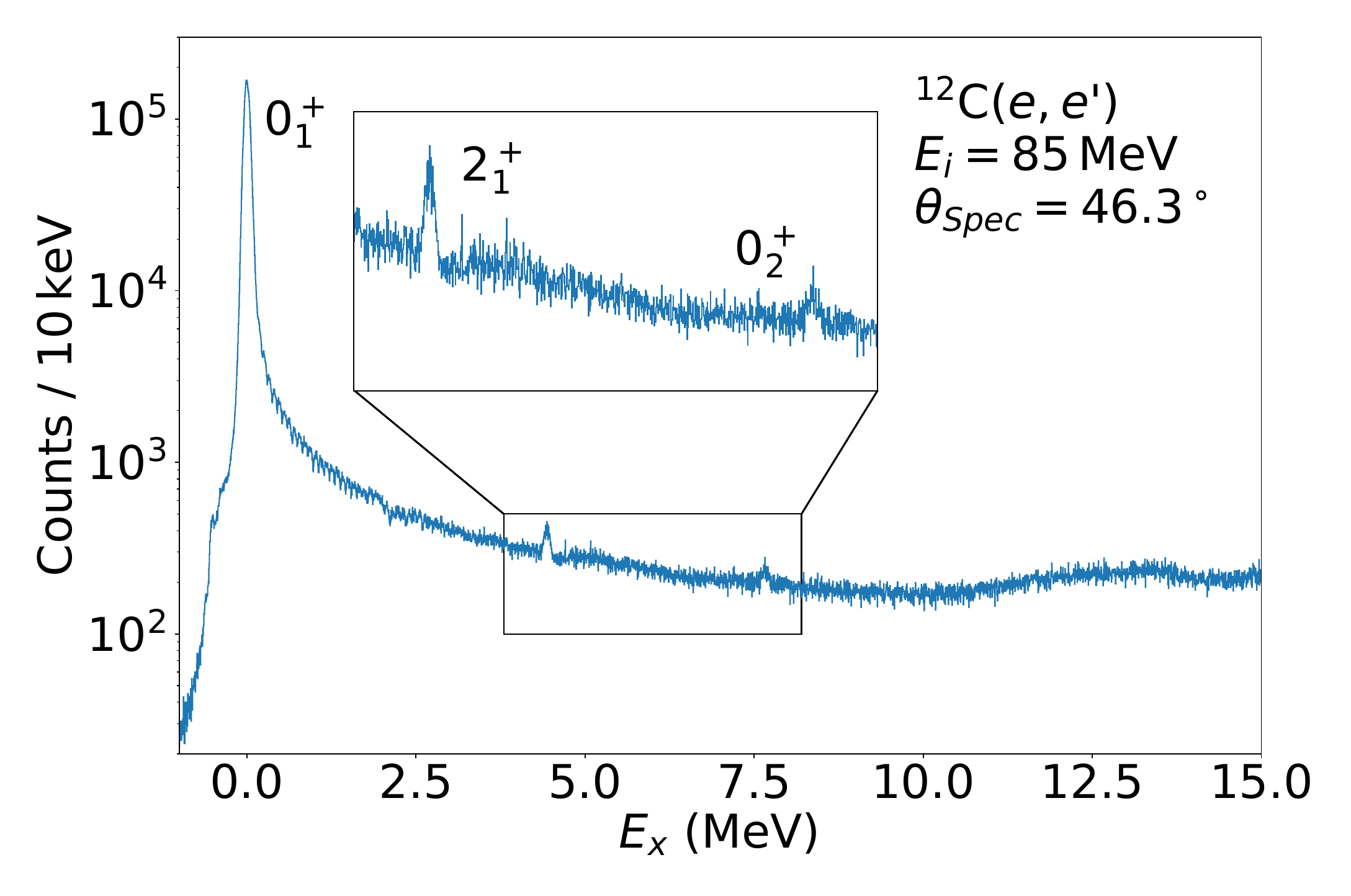}
    \caption{$^{12}\textrm{C}$ spectrum measured in inclusive electron scattering. The elastic line $0_1^+$, the $2_1^+$ and the $0_2^+$ Hoyle state~\cite{Hoyle.1954} are used for an energy calibration.}
    \label{fig:12C_electron_spectrum}
\end{figure}

An energy calibration of the QCLAM spectrometer is performed using the $x_{\textrm{corr}}$ coordinates in a $^{12}\textrm{C}(e,e^\prime)$ measurement.
The spectrum depicted in Fig.~\ref{fig:12C_electron_spectrum} exhibits three clearly visible peaks. 
\\
They originate from elastic scattering, i.e. $E_x = 0.0\,$MeV, the excitation of the $2_1^+$ state at $E_x = 4.44\,$MeV and the $0_2^+$ Hoyle state~\cite{Hoyle.1954} at $E_x = 7.65\,$MeV, respectively~\cite{Kelley.2017}. 

The peak shape in the measured spectrum is parameterized by an asymmetric Gaussian function with an exponential tail~\cite{Hofmann.2002}.
A second order polynomial is used to determine the energy calibration to account for the non-linearity of the focal plane of the spectrometer:
\begin{equation}
    E_x(x_{\textrm{corr}}) = a+b\cdot x_{\textrm{corr}} + c\cdot x^2_{\textrm{corr}}.
\end{equation}
 
To apply the energy calibration using the $^{12}\textrm{C}$ target to the $^{96}\textrm{Ru}(e,e^\prime \gamma)$ experiment, a shift in the peak positions must be taken into account. The reason for this shift is, on the one hand, the different mass of the targets, which leads to a constant shift of the energy due to different recoil factors. On the other hand, there are drifts in the peak position between the different measurement runs. These drifts are caused by different environmental conditions, such as fluctuations in the electron beam energy or the magnetic field of the QCLAM spectrometer. The shifts are determined by fitting the electron peak formula to the runs of the $^{96}\textrm{Ru}(e,e^\prime \gamma)$ measurement and then shifting the determined peak position of the elastic line to the position for the $^{12}\textrm{C}(e,e^\prime)$ measurement.

\subsubsection{Relative Efficiency Correction}
In a next step the relative efficiency of the QCLAM spectrometer is investigated using the $^{12}\textrm{C}$ target.
A white spectrum was recorded, where the energies and angles of the detected electrons are homogeneously distributed over the focal plane of the detector system. 
These requirements are fulfilled at high excitation energies due to the exponentially increasing density of excited states resulting in smooth and structureless excitation-energy spectra.
The resulting $x-\phi$ data are normalized to the total number of events providing the relative efficiency for these coordinates.

\subsection{Calibration: Gamma Data}
The required corrections and calibrations of the raw LaBr\textsubscript{3}:Ce data are discussed in detail in the following.

\subsubsection{{Count-Rate Correction}} \label{Sec:Gamma-Cal-count}
Count-rate dependent gain shifts of the LaBr\textsubscript{3}:Ce detectors are corrected for by measurements with a $^{60}\textrm{Co}$ source with an activity of $20\,$MBq using the two known lines at $1.17\,$MeV and $1.33\,$MeV~\cite{Helmer.2000}.
LaBr\textsubscript{3}:Ce spectra are recorded for different count rates achieved by varying the distance between the $^{60}\textrm{Co}$ source and the detectors.
The rate-corrected integrated detector signals are parameterized by
\begin{equation} \label{Eq:Rate-Dependency-1}
    x_{\textrm{corr},1}(r,x) = (1+a\cdot r)\cdot b \cdot x,
\end{equation}
which scales linearly with the rate $r$ in units of cps and is proportional to the integrated detector signal $x$. 
The integrated detector signal is the area of the trace of the signal, which is associated to an energy value. 
Detector 2 showed strong deviations from the equation above, thus an empirical formula
\begin{equation} \label{Eq:Rate-Dependency-2}
    x_{\textrm{corr},2}(r,x) = \frac{c\cdot x}{r^d + e\cdot r^f},
\end{equation}
was used instead. 
The parameters $a$, $b$, $c$, $d$ and $f$ are obtained by fitting the equations above to the measured rate dependence when changing the distance for the source to the detectors. The resulting parameters are summarized in Tab.~\ref{tab:rate_dependency_parameters}.
\begin{table}[htbp]
    \caption{Fit parameters of rate dependence correction given by Eqs.~(\ref{Eq:Rate-Dependency-1}) and~(\ref{Eq:Rate-Dependency-2}). 
    The parameters are determined from the rates $r$ (measured in units cps) for the $1.17\,$MeV and $1.33\,$MeV peaks of a $^{60}\textrm{Co}$ source.} \label{tab:rate_dependency_parameters}
    \begin{tabular*}{\tblwidth}{@{} CCCCCCC@{} }
        \toprule
        Detector & $a$ & $b$ & $c$ & $d$ & $e$ & $f$\\
        \midrule
        0 & $-3.91(13)\cdot 10^{-8}$ & $0.28981(8)$ & $-$ & $-$ & $-$ & $-$\\
        1 & $2.64(8)\cdot 10^{-9}$ & $0.20319(5)$ & $-$ & $-$ & $-$ & $-$\\
        2 & $-$ & $-$ & $0.3585(3)$ & $0.1623(16)$ & $0.0190(10)$ & $0.5919(53)$\\
        3 & $1.1(19)\cdot 10^{-9}$ & $0.13537(2)$ & $-$ & $-$ & $-$ & $-$\\
        4 & $1.901(96)\cdot 10^{-8}$ & $0.04087(1)$ & $-$ & $-$ & $-$ & $-$\\
        5 & $-1.161(19)\cdot 10^{-7}$ & $0.14489(5)$ & $-$ & $-$ & $-$ & $-$\\
        \bottomrule
    \end{tabular*}
\end{table}\\
By correcting for the rate dependence, the energy resolution of the LaBr\textsubscript{3}:Ce detectors is improved by up to $17\,$\%.

\subsubsection{{Energy Calibration}} \label{Sec:Gamma-Cal-energy}
\begin{figure}[htbp]
    \centering
    \includegraphics[width=0.75\linewidth]{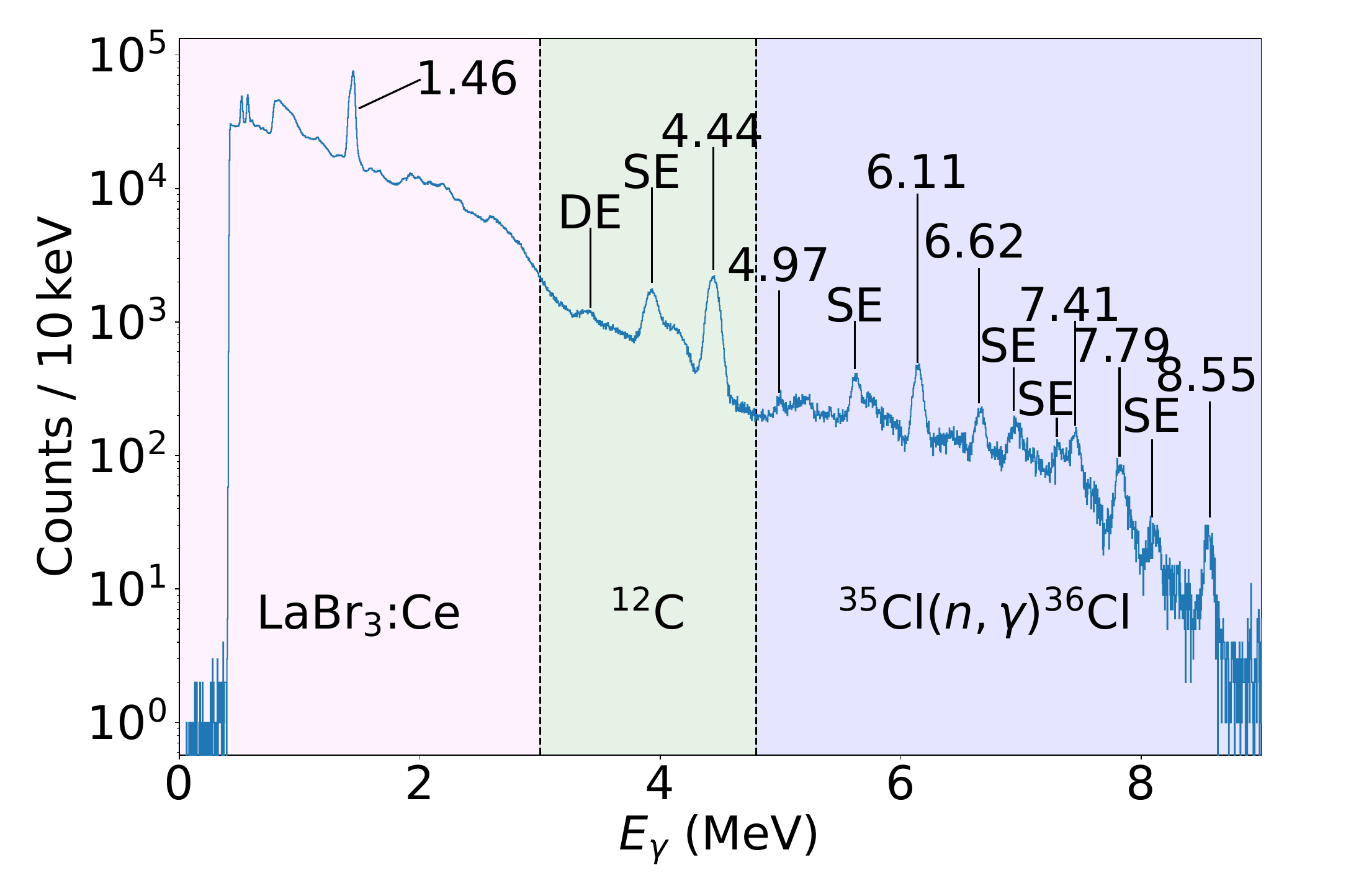}
    \caption{LaBr\textsubscript{3}:Ce spectrum of the $^{35}\textrm{Cl}(n,\gamma)^{36}\textrm{Cl}$ reaction using an $\textrm{Am} - \textrm{Be}$ source for energy calibration up to $9\,$MeV. The spectrum is divided into three parts: intrinsic radioactivity of the detector, $\textrm{Be} + \alpha \rightarrow$ $^{12}\textrm{C}$ reaction and subsequent $\gamma$ decay and $^{35}\textrm{Cl}(n,\gamma)^{36}\textrm{Cl}$ reaction. DE: Double Escape Peak, SE: Single Escape Peak.}
    \label{fig:36cl_gamma_spectrum}
\end{figure}

For the energy calibration of the $\gamma$ spectra a $^{226}\textrm{Ra}$ source for $\gamma$-ray energies $< 3\,$MeV and the $^{35}\textrm{Cl}(n,\gamma)^{36}\textrm{Cl}$ reaction for energies $> 3\,$MeV was used. 
The latter spectrum can be seen in Fig.~\ref{fig:36cl_gamma_spectrum}.
\\
The energy calibration was then performed using a second order polynomial.
An overall mean energy resolution of $43.7(1)\,$keV at $E_{\gamma} = 1.33\,$MeV resulting in a relative energy resolution of $3.28(1)\,\%$ was obtained for the six LaBr\textsubscript{3}:Ce detectors of the DAGOBERT setup. At the same $\gamma$-ray energy DAGOBERT with six LaBr\textsubscript{3}:Ce detectors in the configuration of Fig.~\ref{fig:eepg_setup} features a total photopeak efficiency of $0.61(2)\,\%$.
For the comparison of the performance in Sec.~\ref{Sec:Performance} the relative energy resolution is taken at $E_{\gamma} = 4.44\,$MeV and amounts to $1.60(1)\,\%$.

\subsection{Calibration: Coincidence Data} \label{Sec:Coin-Time-Cor}
\begin{figure}[htbp]
	\centering
	\includegraphics[width=0.75\linewidth]{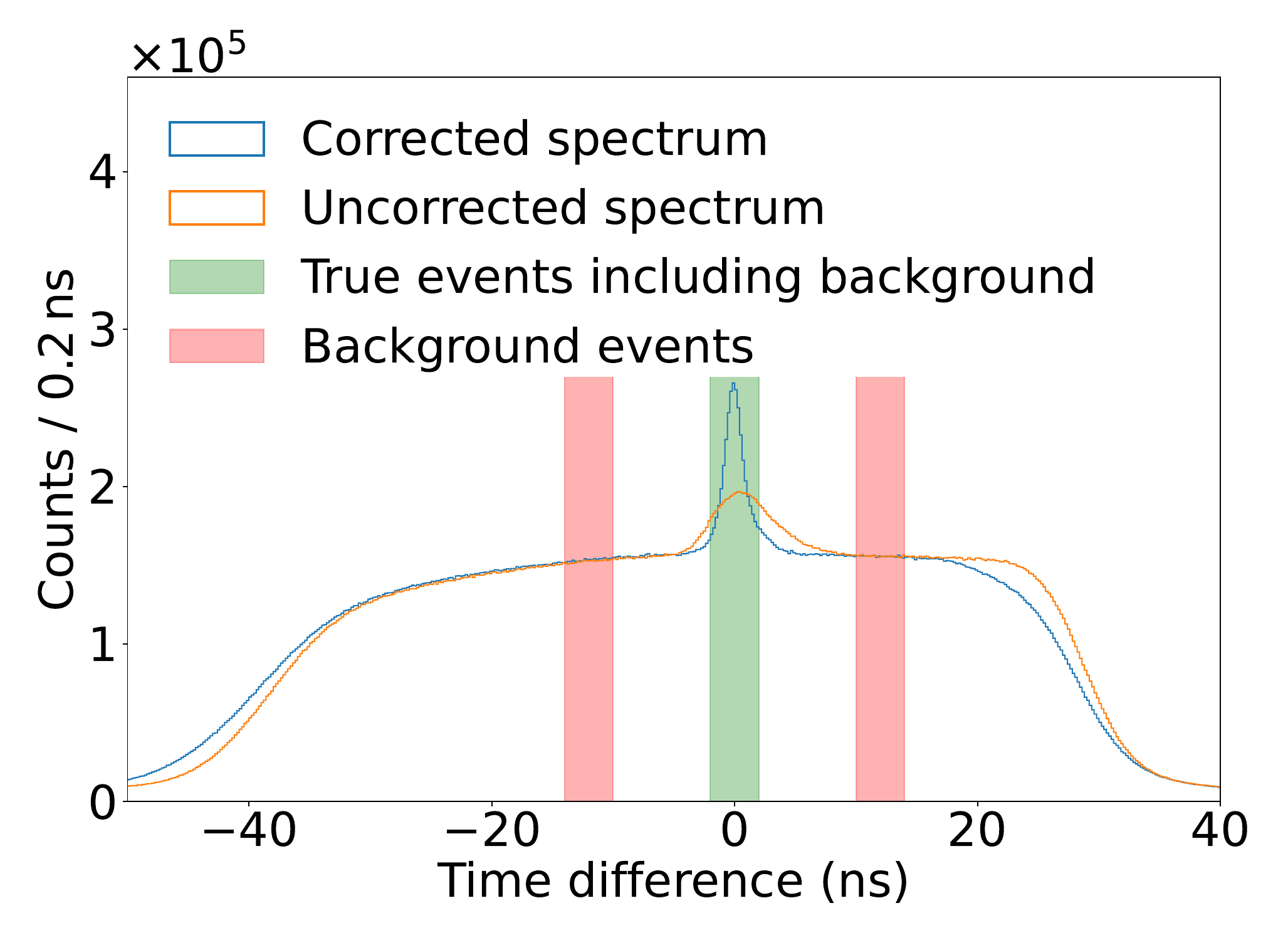}
	\caption{Time difference spectrum of QCLAM and DAGOBERT trigger timestamps. Blue: time-of-flight corrected spectrum, orange: uncorrected measured spectrum. 
    In addition, time gates for background subtraction (compare Sec.~\ref{Sec:Random_Coin}) are shown. Green: true events including random coincidences, red: random coincidences only.}
	\label{FIG:time_difference}
\end{figure}

For an enhanced signal-to-background ratio an optimized time resolution between the electron and $\gamma$-ray data is of particular importance. 
The uncorrected time difference between QCLAM ($t_e$) and DAGOBERT ($t_\gamma$) events
\begin{equation}
    \Delta t = t_\gamma - t_e,
\end{equation}
is displayed in orange in Fig.~\ref{FIG:time_difference}. 
\\
In blue the corrected time difference is shown. The correction is presented in the following.

The present prompt time-difference peak has a full width at half maximum (FWHM) of $5.6\,$ns. 
As depicted in Fig.~\ref{fig::focal_plane}, depending on their energy and scattering angle, the highly relativistic electrons ($v \approx c$) travel different path lengths from the scattering target to the detection system of the spectrometer resulting in a variation of the time-of-flight (tof) values broadening the prompt peak in the time-difference spectrum. 
A correction is performed to restore the time resolution using the parameterization 
\begin{equation}
    t_{\textrm{tof}}(x,\phi) = \sum_{i,j} a_{i,j}x^i\phi^j,
\end{equation}
with the dispersive coordinate $x$ and the dispersive angle $\phi$. 
The parameters $a_{ij}$ with $i,j = 0,1$ are determined iteratively. 
For each subsequent correction, the previous corrections are applied to the time difference. 
Starting from the correction of the $x$-dependence, the $\phi$-dependence and the $x \cdot \phi$-dependence.

\begin{figure}[htbp]
    \centering
    \includegraphics[width=0.75\linewidth]{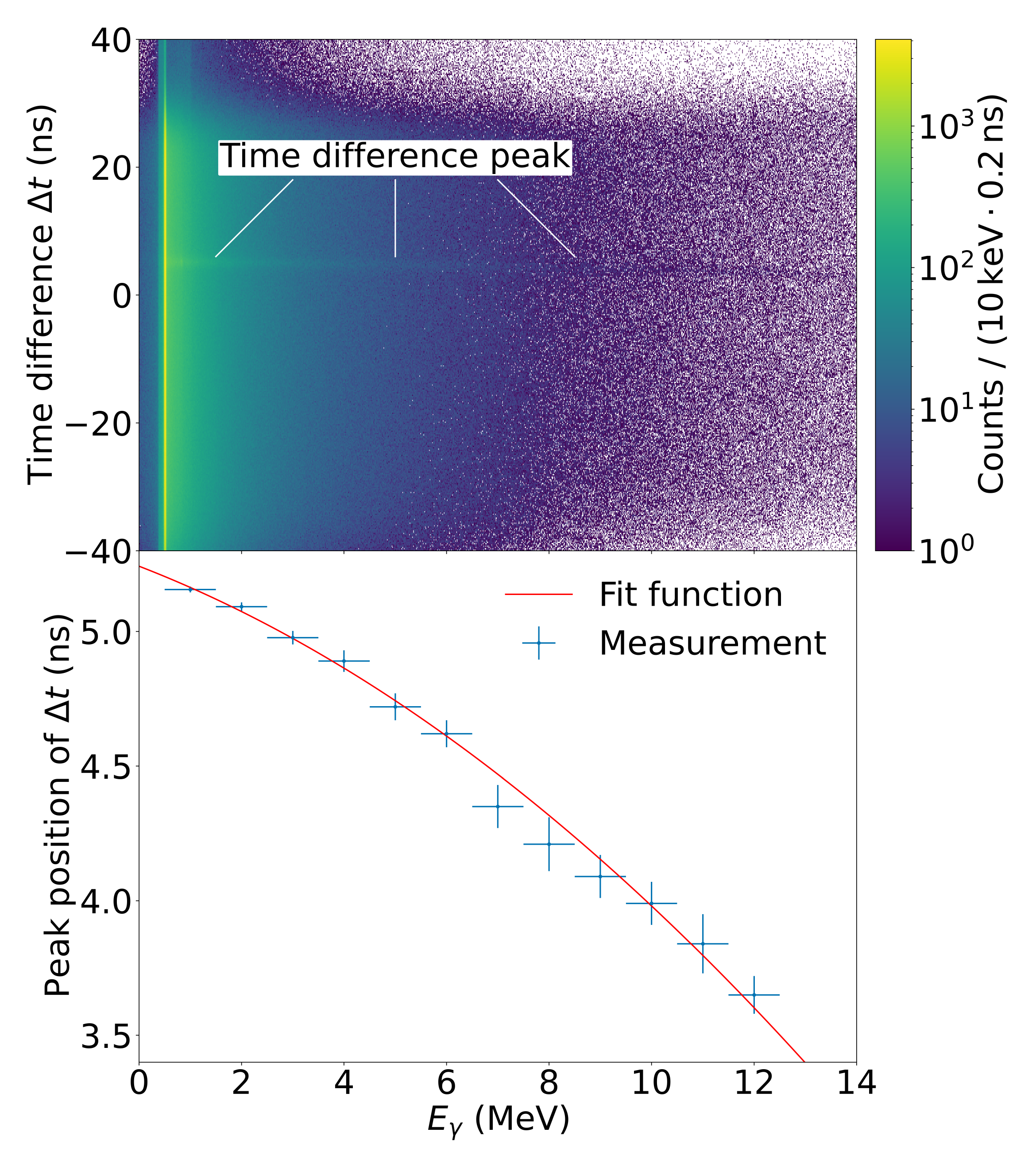}
    \caption{Time difference dependence of $E_{\gamma}$ for detector number 5. In the upper panel the photon energy $E_{\gamma}$ versus time difference $\Delta t$ is shown. An energy dependence of the time difference peak is visible. For the correction of the effect, a quadratic function is fitted to the peak position of the time difference for $E_{\gamma}$ gates of $\pm0.5\,$MeV in the lower panel.}
    \label{Fig:time_difference_Eg}
\end{figure}

Furthermore, the trigger time of the emitted $\gamma$ rays detected by DAGOBERT must be corrected because the trigger pulse depends on the waveform, which in turn depends on the amplitude and thus on the $\gamma$-ray energy.
The correction for the $\gamma$-ray energy dependence on the time difference is accomplished by the fit of a quadratic function to the prompt peak in the time difference spectrum. 
An example is shown in Figure~\ref{Fig:time_difference_Eg} for detector 5. 
\\
In the upper panel the photon energy $E_{\gamma}$ is plotted against the time difference. 
A dependence of $\Delta t$ on $E_{\gamma}$ is clearly visible. 
It can be corrected by fitting a quadratic function to the centroid of the peak in the time difference spectrum.

Finally, the corrected time difference is given by:
\begin{equation}
    \Delta t_{\textrm{corr}} = \Delta t - t_{\textrm{tof}} - \Delta t_{E_\gamma}.
\end{equation}
Applying both corrections, the time resolution of the coincidence peak was improved from $5.6\,$ns to $1.9\,$ns (FWHM). 
The corrected time-difference spectrum is shown in blue in Fig.~\ref{FIG:time_difference}.

\subsection{Correction for Random Coincidences} \label{Sec:Random_Coin}
Random coincidence events are a major source of background and therefore need to be accounted for in the analysis.
A narrow time gate of $\pm 2\,$ns is set on the events in the prompt peak of the time difference spectrum, Fig~\ref{FIG:time_difference}. 
Furthermore, random coincidences are selected by gating on the left and right from the prompt peak with the same width and subtracted from the events of the prompt peak gate.

\subsection{$E_x$-$E_\gamma$ Matrix}
\begin{figure}[htbp]
    \centering
    \begin{subfigure}[b]{0.7\linewidth}
        \includegraphics[width=\linewidth]{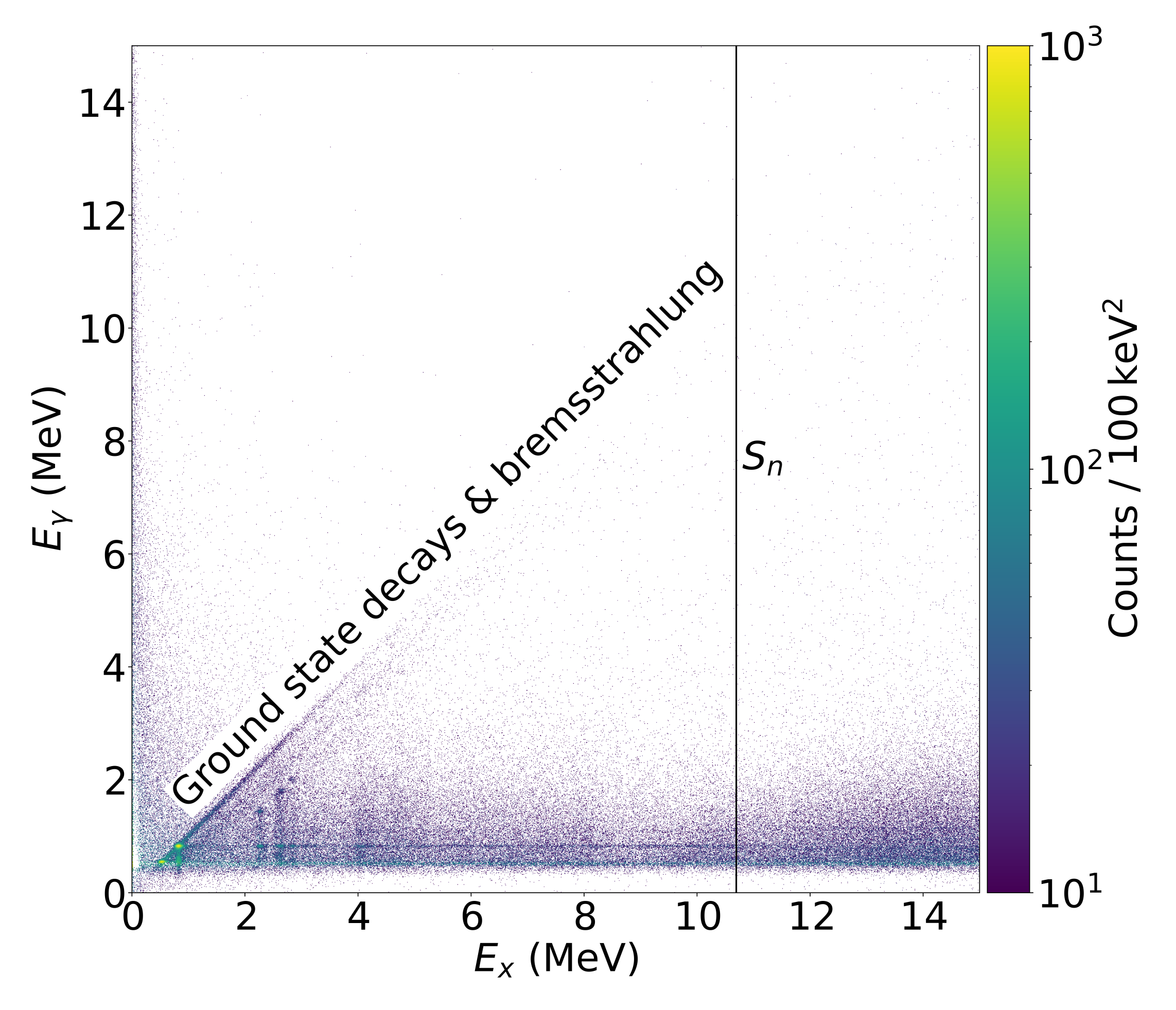}
        \caption{Full $E_x$-$E_\gamma$ matrix up to excitation energies of $15\,$MeV.}
        \label{fig::Ex-Eg-Matrix:A} 
    \end{subfigure}
    \begin{subfigure}[b]{0.7\linewidth}
        \includegraphics[width=\linewidth]{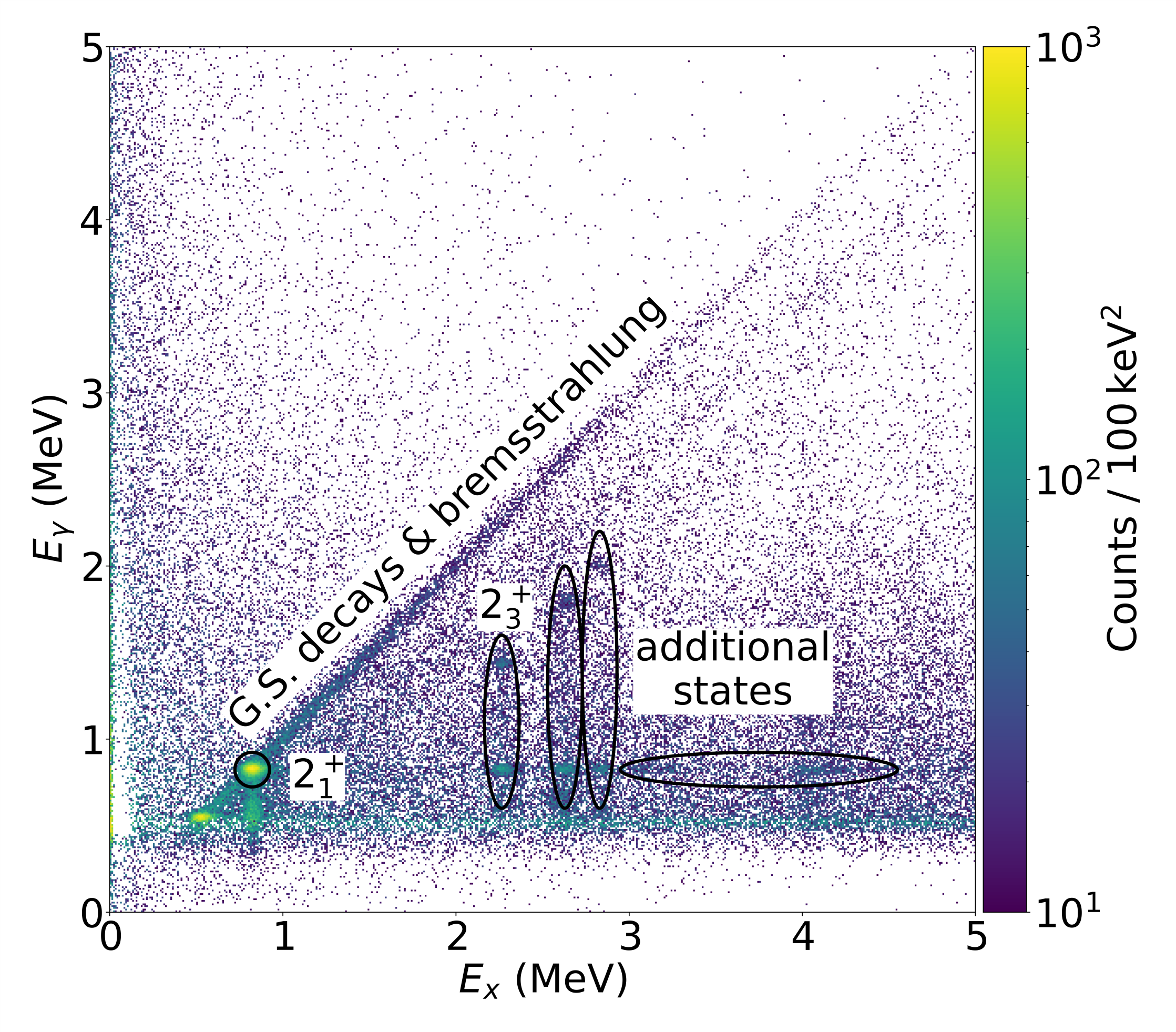}
        \caption{Zoom at low excitation energies.}
        \label{fig::Ex-Eg-Matrix:B}
    \end{subfigure}
    \caption{Background- and efficiency-corrected $E_x$-$E_\gamma$ matrix for $^{96}\textrm{Ru}$ at an electron beam energy $E_i = 85\,$MeV and a scattering angle $\theta_{\textrm{Spec}} = 46.3^\circ$.}
    \label{fig::Ex-Eg-Matrix}
\end{figure}

Figures~\ref{fig::Ex-Eg-Matrix} and~\ref{fig::Ex-Eg-Matrix-Sub} display $E_x$-$E_\gamma$ matrices corrected for the background from random coincidences and the relative efficiencies. In such a matrix, the energy loss of the electrons $E_x$ is displayed on the $x$-axis, the $\gamma$-ray energy $E_\gamma$ on the $y$-axis and the number of events per channel on the $z$-axis. 
The full measured matrix up to excitation energies of $15\,$MeV and a zoom on low excitation energies can be seen in Figs.~\ref{fig::Ex-Eg-Matrix:A} and~\ref{fig::Ex-Eg-Matrix:B}, respectively.
\\

Parts of the events along the $E_x=E_\gamma$ diagonal must be attributed to bremsstrahlung due to its indistinguishability from ground-state $\gamma$ decay. 
The bremsstrahlung contribution decreases with increasing energy loss of the electrons, i.e., with increasing excitation energy. 
At low excitation energies discrete states can be seen. One of these states is the $2_1^+$ excited state which decays directly into the ground state. Its angular distribution will be discussed in Sec.~\ref{Sec:Angular-Distribution}. 
Above the neutron separation threshold $S_{n} = 10.69\,$MeV~\cite{Wang.2012}, an increase of events with photon energies less than $2\,$MeV is observed originating from $\gamma$ decays of the $^{95}\textrm{Ru}$ daughter in the $^{96}\textrm{Ru}(e,e^\prime n \gamma)$ reaction. 
This will be further discussed in Sec.~\ref{Sec:Gamma_Decay_Above_Threshold}.
Additionally, the mixed-symmetric $2_3^+$ state can be seen whose branching ratio will be investigated in Sec.~\ref{Sec::Branching-Ratio}.

\subsection{Correction for Coincident Bremsstrahlung} \label{Sec::Bremsstrahlung-Subtraction}
The remaining background in the $E_x$-$E_\gamma$ matrices in Fig.~\ref{fig::Ex-Eg-Matrix} originates from coincident bremsstrahlung events.  
In this section two methods for the subtraction of the coincident bremsstrahlung contribution from the experimental results are discussed. 
More details can be found in~\cite{Hesbacher.2023}.

\subsubsection{{Simulation-based corrections}} \label{Sec::Bremsstrahlung-Subtraction-Sim}
The first procedure utilizes response simulations of the LaBr\textsubscript{3}:Ce detectors in combination with PWBA calculations of the bremsstrahlung process in $(e,e^\prime \gamma)$ reactions to subtract the coincident bremsstrahlung contribution. 
The angular distribution of bremsstrahlung photons produced by relativistic electrons is focused to forward angles due to the Lorentz boost.
The LaBr\textsubscript{3}:Ce detectors are positioned at backward angles relative to the incoming electron beam, where bremsstrahlung is suppressed by $4$ to $5$ orders of magnitude compared to forward angles, see Fig.~\ref{fig:nuclear_bremsstrahlung}. 
As a consequence, it is computationally not feasible to simulate the bremsstrahlung process for such kinematics. 
Therefore, a different approach is used in this work. 

In order to take the detector response into account, photons emitted isotropically from the target position and detected with DAGOBERT were simulated using GEANT4~\cite{GEANT4-1, GEANT4-2, GEANT4-3} leading to homogeneously distributed events in all LaBr\textsubscript{3}:Ce detectors. 
The differential cross section of the bremsstrahlung contribution is determined for the same kinematical conditions as in the experiment by PWBA calculations with the Bethe-Heitler formula, Eq.~(\ref{Bethe-Heitler-Formel}).
The simulation results are combined for each photon with the PWBA differential cross section calculations to generate bremsstrahlung spectra in each LaBr\textsubscript{3}:Ce detector.

\begin{figure}[htbp]
    \centering
    \begin{subfigure}[b]{0.7\linewidth}
        \includegraphics[width=\linewidth]{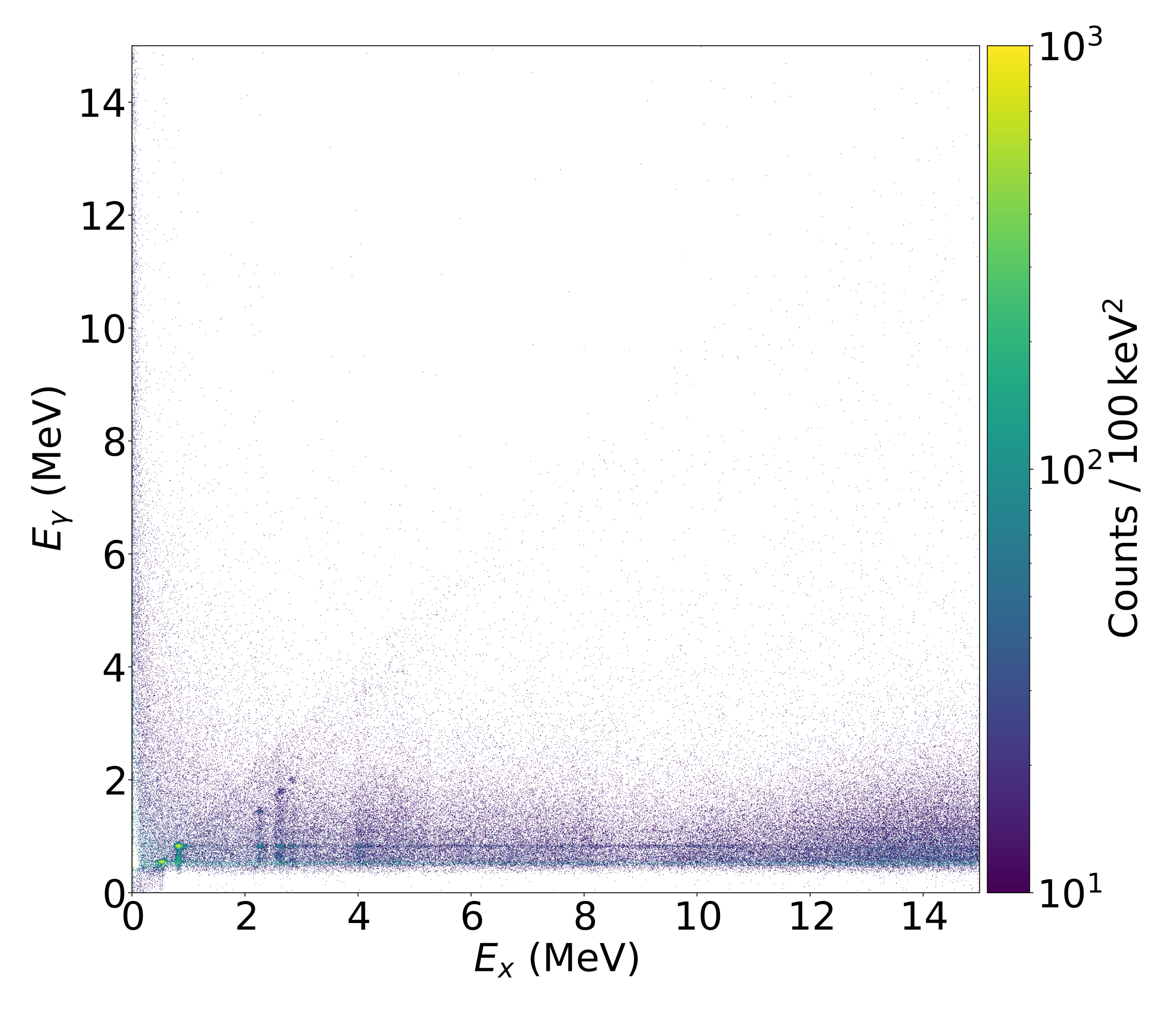}
        \caption{Full $E_x$-$E_\gamma$ matrix up to excitation energies of $15\,$MeV.}
        \label{fig::Ex-Eg-Matrix-Sub:A} 
    \end{subfigure}
    \begin{subfigure}[b]{0.7\linewidth}
        \includegraphics[width=\linewidth]{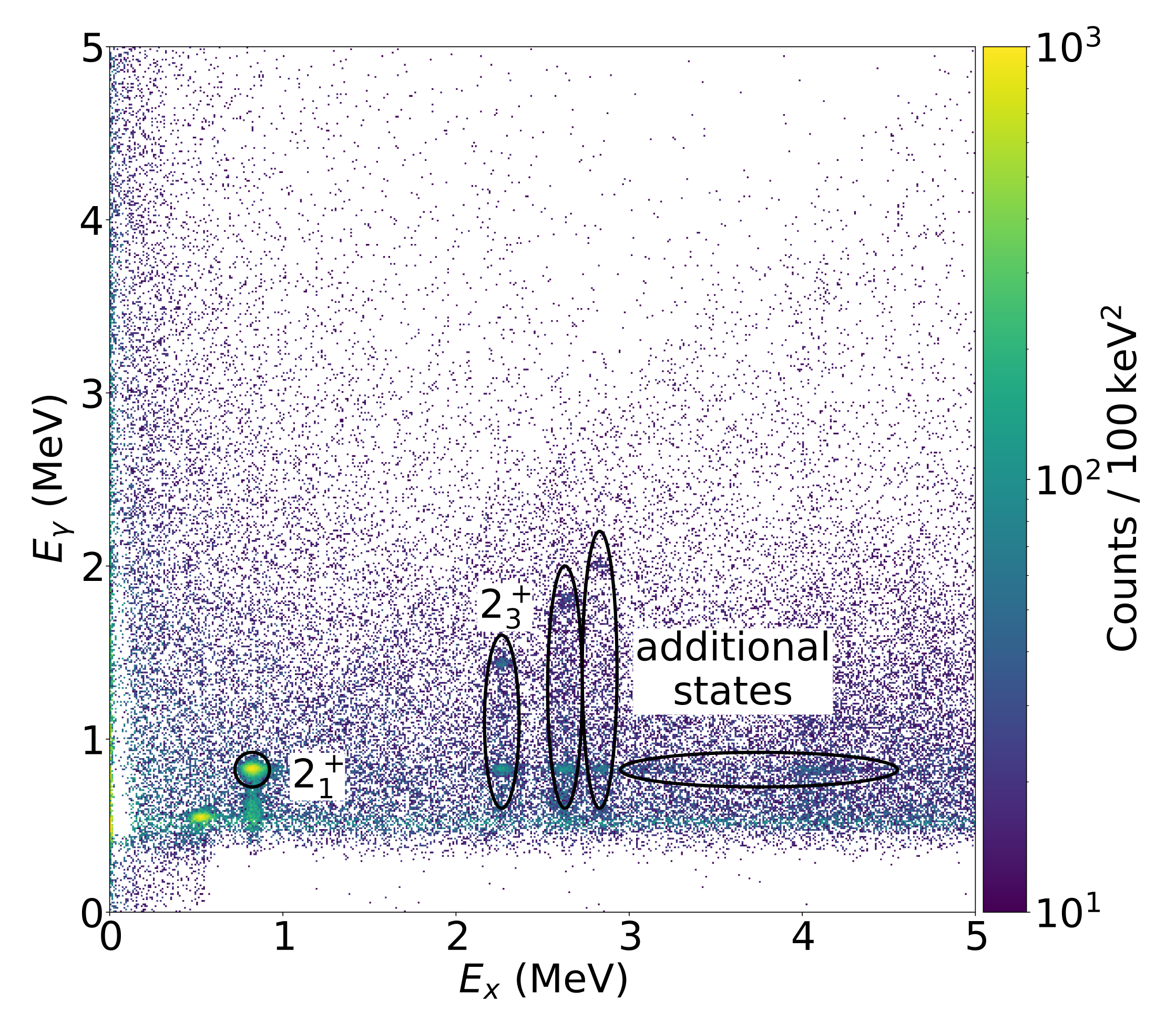}
        \caption{Zoom at low excitation energies.}
        \label{fig::Ex-Eg-Matrix-Sub:B}
    \end{subfigure}
    \caption{Same as Fig.~\ref{fig::Ex-Eg-Matrix}, but with additional subtraction of the bremsstrahlung background.}
    \label{fig::Ex-Eg-Matrix-Sub}
\end{figure}

With this procedure, $\gamma$ energy spectra of pure brems\-strahlung can be generated for discrete energies of the bremsstrahlung photons. 
A continuous $E_x$-$E_\gamma$ matrix is created from the individual $\gamma$ energy spectra of the brems\-strahlung by interpolation. 
Thereby, the experimental energy resolution of the LaBr\textsubscript{3}:Ce detectors and the QCLAM spectrometer determined from the measurements are taken into account.
Finally, the resulting $E_x$-$E_\gamma$ matrix of the simulated coincident bremsstrahlung in $^{96}\textrm{Ru}(e,e^\prime \gamma)$ reactions is scaled to the measured data and subtracted generating a bremsstrahlung-corrected $E_x$-$E_\gamma$ matrix shown in Fig.~\ref{fig::Ex-Eg-Matrix-Sub}.
\\
A significant reduction of the background is clearly visible in comparison to Fig~\ref{fig::Ex-Eg-Matrix}.

\subsubsection{{Experiment-based correction}} \label{Sec::Bremsstrahlung-Subtraction-Meas}
\begin{figure}[htbp]
    \begin{minipage}[b]{.49\linewidth}
    		\centering
    		\includegraphics[width=\linewidth]{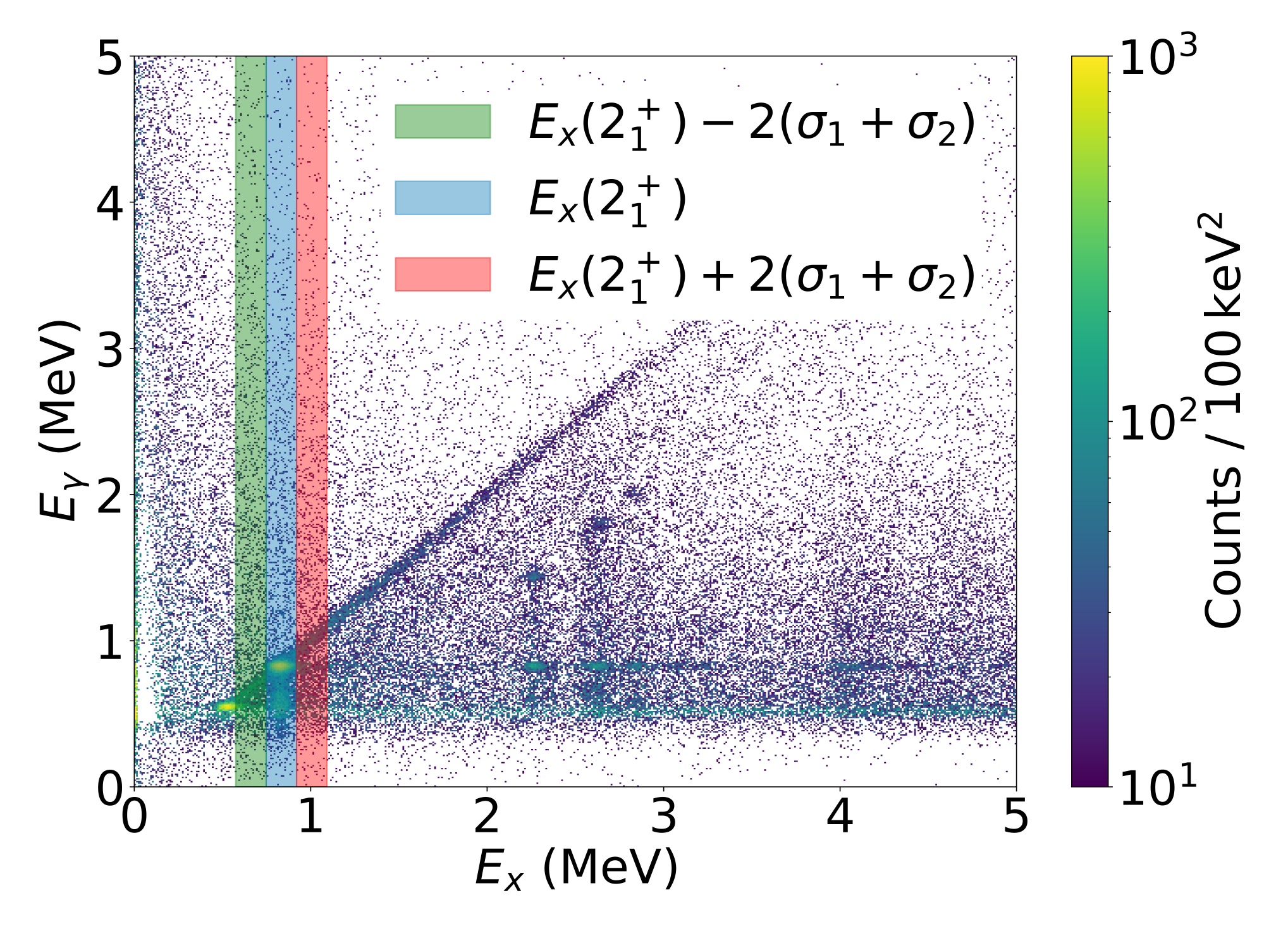}
            \subcaption{Three gates with a width of $-2\sigma_1$ and $+2\sigma_2$ of the electron energy resolution (Tab.~\ref{tab::electron_peak_fit_parameters}) in the measured $E_x$-$E_{\gamma}$ matrix. 
        Blue: Bremsstrahlung plus nuclear decay of interest. Green and red: Bremsstrahlung plus background nuclear decay.}
        \label{fig::Bremsstrahlung-Sub-Measurement:A} 
    \end{minipage}
    \hfill
    \begin{minipage}[b]{.49\linewidth}
    		\centering
    		\includegraphics[width=\linewidth]{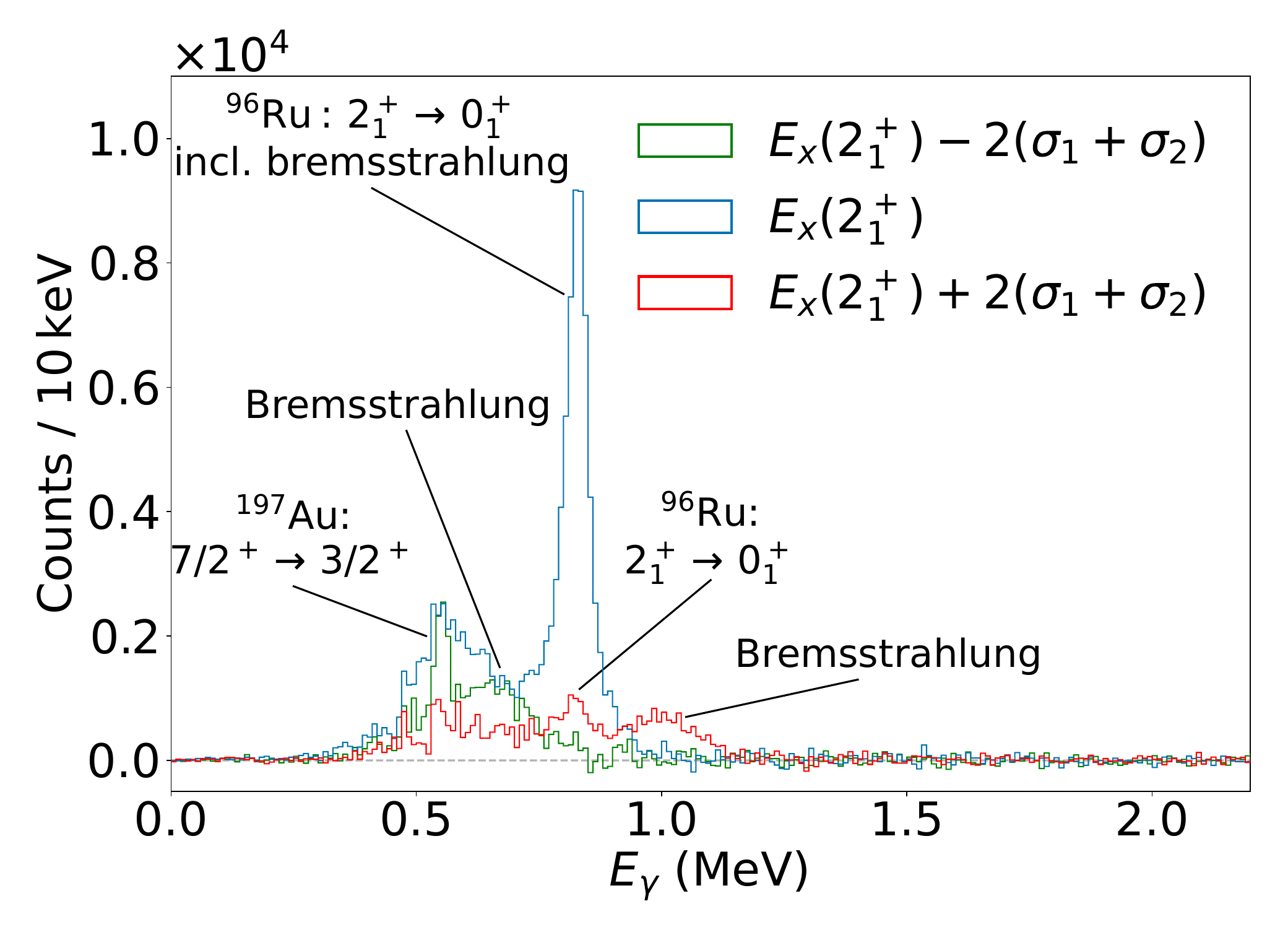}
            \subcaption{Projected $\gamma$ energy spectra of the gates defined in (a).\vspace*{3\baselineskip}}
        \label{fig::Bremsstrahlung-Sub-Measurement:B}
    \end{minipage}
    \begin{minipage}[b]{.49\linewidth}
    		\centering
    		\includegraphics[width=\linewidth]{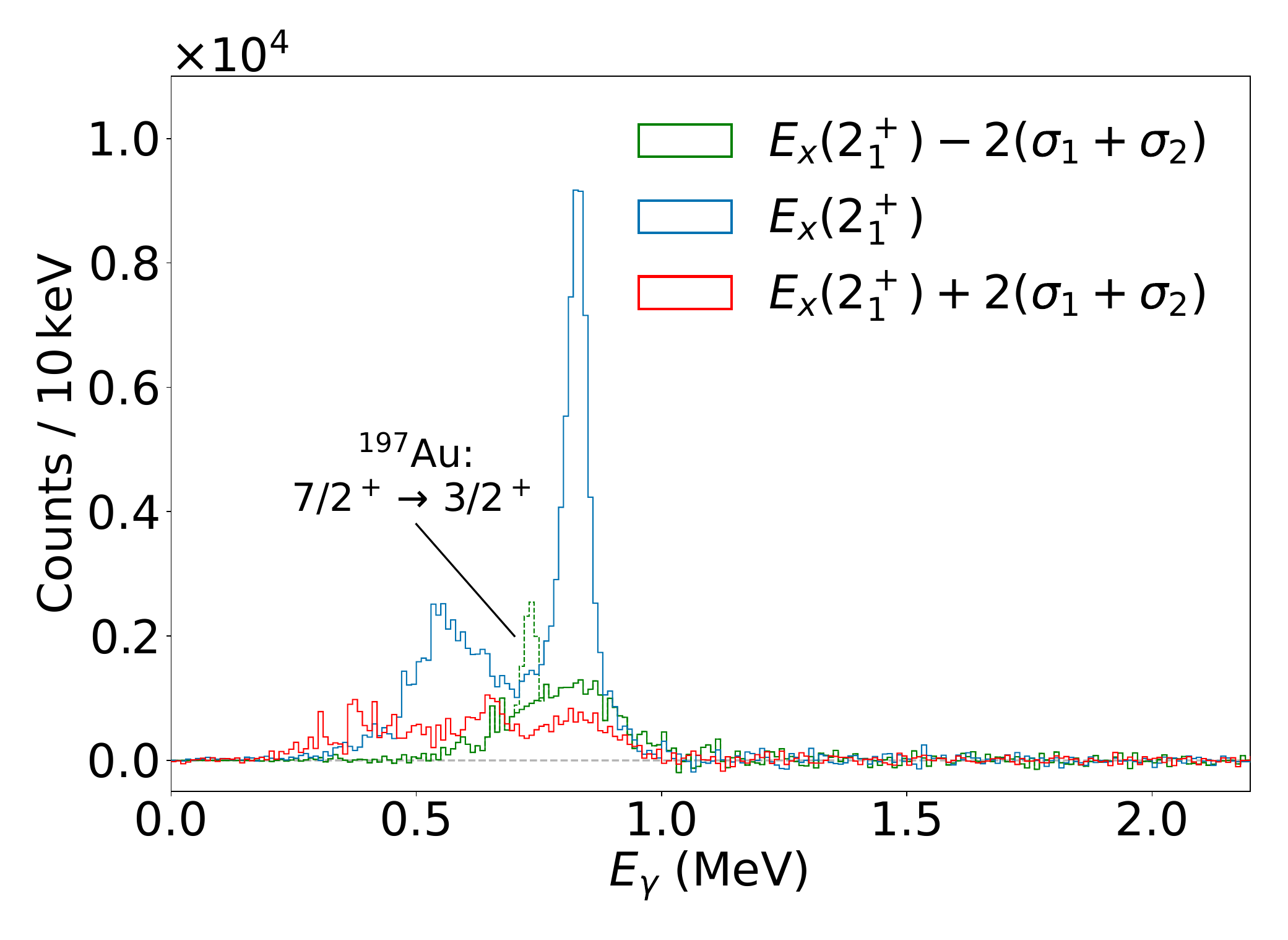}
            \subcaption{Background spectra shifted to the energy of the excited state. The bremsstrahlung contribution under the $\gamma$ decay of $^{197}\textrm{Au}$ was approximated by a second order polynomial.}
            \label{fig::Bremsstrahlung-Sub-Measurement:C}
    \end{minipage}
    \hfill
    \begin{minipage}[b]{.49\linewidth}
    	\centering
    	\includegraphics[width=\linewidth]{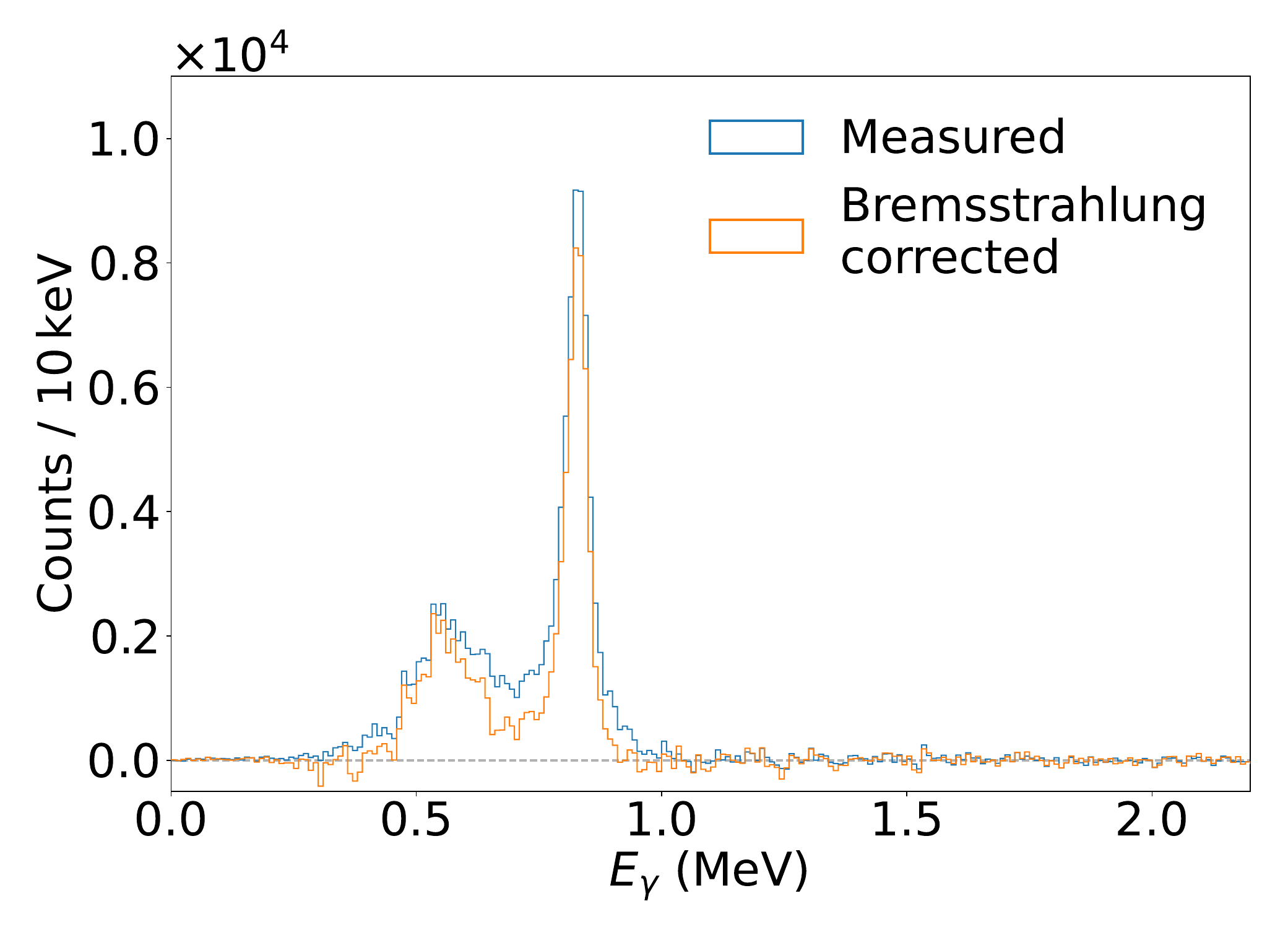}
        \subcaption{Comparison of measured $\gamma$ spectrum with bremsstrahlung-subtracted spectrum. Below the ground state $\gamma$-decay the Compton edge and continuum is visible down to the detector threshold.}
        \label{fig::Bremsstrahlung-Sub-Measurement:D}
    \end{minipage}
    \caption{Procedure for subtracting bremsstrahlung from the measured $(e,e^\prime \gamma)$ spectra for individually resolved states using the decay of the $2_1^+$ state of $^{96}\textrm{Ru}$ as example.}
    \label{fig::Bremsstrahlung-Sub-Measurement}
\end{figure}

Another method for the correction of coincident brems\-strahlung makes use of the measured data. 
The procedure is illustrated by way of example for the $2_1^+$ state of $^{96}\textrm{Ru}$ in Fig.~\ref{fig::Bremsstrahlung-Sub-Measurement}.
\\
Due to the low density of states, regions in the $E_x$-$E_\gamma$ matrix at small excitation energies can be selected where no excited states are present.
Hence, all electron-$\gamma$ coincidence events in those regions originate solely from bremsstrahlung.
Sections below (green colored band) and above (red colored band) the excitation energy of the $2^+_1$ state (blue colored band) selected for the analysis are illustrated in Fig.~\ref{fig::Bremsstrahlung-Sub-Measurement:A}. 
The projected low-energy $\gamma$-ray spectrum is shown in Fig.~\ref{fig::Bremsstrahlung-Sub-Measurement:B}. 
The spectrum of the ground state decay of the $2_1^+$ state corresponding to the blue band overlaps with coincident bremsstrahlung.
The detector response of the $2^+_1 \rightarrow 0^+_1$ transition is visible down to the electronically set threshold of about $450\,$keV. 
The green spectrum shows a bremsstrahlung peak at the corresponding excitation energy and also the decay from a transition in $^{197}\textrm{Au}$ backing material. 
In the $\gamma$ energy spectrum in red the bremsstrahlung peak is visible too. 
In addition, contributions from the radiative tail of the $2_1^+ \rightarrow 0_1^+$ decay can be seen.
Before subtraction, the $\gamma$-ray spectra obtained from the low-energy gate at ${\sim}650$~keV (green) and the high-energy gate at ${\sim}1000$~keV (red) have to be shifted to the excitation energy of the $2_1^+$ state as illustrated in Fig.~\ref{fig::Bremsstrahlung-Sub-Measurement:C}. 
Within a small excitation-energy region, the bremsstrahlung cross section is expected to be a smooth function of the excitation energy. 
Thus, the mean value of the two background $\gamma$-ray spectra is used to subtract the bremsstrahlung contribution. In addition, the peak from the $\gamma$ decay of $^{197}\textrm{Au}$ has been removed assuming a second-order polynomial background below the peak structure. 
Figure~\ref{fig::Bremsstrahlung-Sub-Measurement:D} shows the final bremsstrahlung-subtracted $\gamma$-ray spectrum of the $2_1^+$ state.
This method is restricted to isolated excited states in regions of low level density, since the background gates need to be free from ground-state $\gamma$-ray decays of excited states.

\section{Results of $^{96}\textrm{Ru}(e,e^\prime \gamma)$ Measurement}\label{chap::Results}
This chapter deals with the analysis of the first $(e,e^\prime \gamma)$ experiment at the S-DALINAC conducted on $^{96}\textrm{Ru}$. 
The corrected $E_x$-$E_\gamma$ matrix is used.

\subsection{Excited States}
\begin{figure}[htbp]
	\centering
	\includegraphics[width=0.75\linewidth]{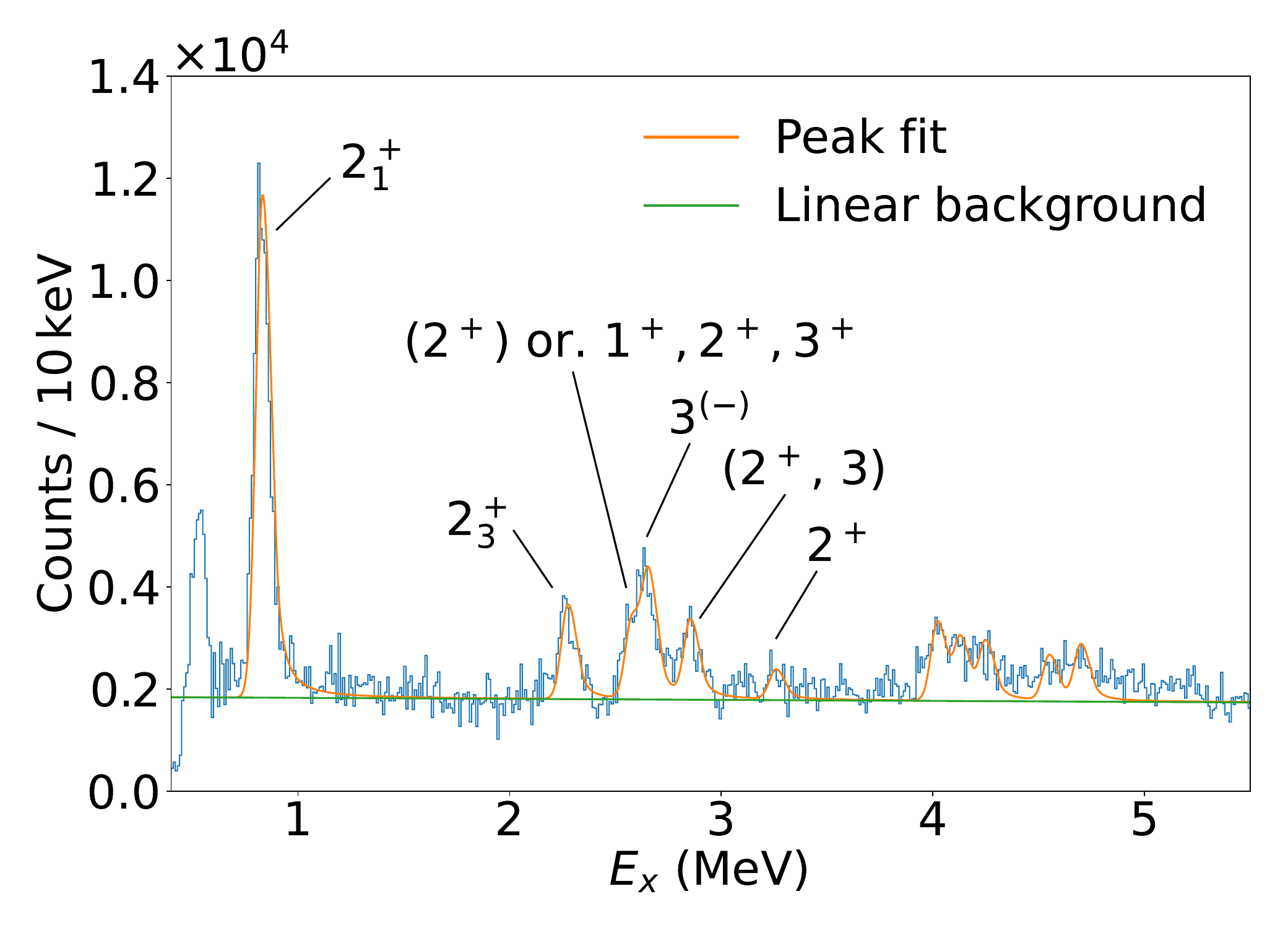}
	\caption{Projection of the $E_x$-$E_{\gamma}$ matrix on the $E_x$ axis. Fits of asymmetric Gaussian functions with radiative tail and linear background to the recognizable peaks are shown in orange.}
	\label{fig::Excitation-Energy-Spectrum}
\end{figure}

The excitation energy spectrum taken from a projection of the $E_x$-$E_\gamma$ matrix onto the $E_x$-axis is investigated up to $5.5\,$MeV. 
This allows to analyze the excitation of individual states. The spectrum is displayed in Fig.~\ref{fig::Excitation-Energy-Spectrum}.
\\
Numerous peaks are observed corresponding to nuclear excitations of $^{96}\textrm{Ru}$ or $^{197}\textrm{Au}$. For the determination of excitation energies $E_{x,\textrm{fit}}$ and the number of counts present in each peak, an asymmetric Gaussian function with a radiative high-energy tail is used to approximate the line shape~\cite{Hofmann.2002}:
\begin{equation} \label{Eq::electron peak shape-HDTV}
	y = y_0 \cdot \begin{cases}
		\exp(-\ln 2 \cdot (x-x_0)^2/\sigma_1^2) & , x \le x_0, \\
		\exp(-\ln 2 \cdot (x-x_0)^2/\sigma_2^2) & , x_0 < x \le x_0 + \eta\sigma_2, \\
		\frac{A}{(B + x - x_0)^\delta}  & , x > x_0 + \eta\sigma_2.
	\end{cases}
\end{equation}
Here, $y_0$ indicates the amplitude of the peak and $x_0$ the position of the maximum of the peak. The parameters $\sigma_{1,2}$ describe the widths of the two half Gaussian functions, $\delta$ the power-law decay and $\eta$ specifies the attachment point, with $\eta > 1$. 
The parameters $A$ and $B$ are defined by:
\begin{align}
    B &= \frac{\sigma_2\delta - 2\sigma_2 \eta^2 \ln 2}{2 \eta \ln 2}, \\
    A &= 2^{-\eta^2} (\sigma_2\eta + B)^\delta.
\end{align}
To determine the parameters, a fit of Eq.~(\ref{Eq::electron peak shape-HDTV}) to clearly identifiable peaks in the electron energy spectrum was performed by assuming a linear background. 
The line shape parameters are assumed to be independent of excitation energy.
The results are summarized in Tab.~\ref{tab::electron_peak_fit_parameters}.
\begin{table}[htbp]
    \caption{Fit parameters for the electron energy spectrum according to Eq.~(\ref{Eq::electron peak shape-HDTV}) with linear background.} 
    \label{tab::electron_peak_fit_parameters}
    \begin{tabular*}{\tblwidth}{@{} CCCC@{} }
        \toprule
        $\sigma_1$ (keV) & $\sigma_2$ (keV) & $\eta$ & $\delta$\\
        \midrule
        $36.7(4)$&$49.7(5)$&$1.36(2)$&$1.77(8)$\\
        \bottomrule
    \end{tabular*}
\end{table}\\
The energy resolution of the QCLAM spectrometer was determined as $86.4(1.0)\,$keV (FWHM).

From the first five fitted excited states an energy recalibration to literature values~\cite{Hennig.2015} using a first order polynomial, with a resulting slope of $1.0013(13)$ and an offset of $10.1(15)\,$keV, was carried out. 
After the recalibration, the fitted excited states are provided in Table~\ref{tab::Excitation-Energy-Spectrum}.
\begin{table}[htbp]
    \caption{Table of fitted excitations $E_{x,\textrm{fit}}$ in the electron energy spectrum compared with known excitation energies $E_{x,\textrm{Lit}}$~\cite{Hennig.2015} and their associated states $J_i^{\pi}$.} \label{tab::Excitation-Energy-Spectrum}
    \begin{tabular*}{\tblwidth}{@{} CCC@{} }
        \toprule
        $E_{x,\textrm{fit}}$ (keV) & $E_{x,\textrm{Lit}}$ (keV) & $J_i^{\pi}$\\
        \midrule
        \eqmakebox[E][l]{$832.6(23)$} & {$832.5(1)$} & {$2_1^+$}\\
        \eqmakebox[E][l]{$2276(5)$} & {$2283.3(2)$} & {$2_3^+$}\\
        \eqmakebox[E][l]{$2578(7)$} & {$2575.7(4)$ or. $2578.5(5)$} & {$2^+$ or. $1^+, 2^+, 3^+$}\\
        \eqmakebox[E][l]{$2658(5)$} & {$2650(2)$} & {$3^{(-)}$}\\
        \eqmakebox[E][l]{$2855(6)$} & {$2851.4(2)$} & {$(2^+, 3)$}\\
        \eqmakebox[E][l]{$3258(9)$} & {$3261.5(8)$} & {$2^+$}\\
        \eqmakebox[E][l]{$4022(6)$} & {-} & {-}\\
        \eqmakebox[E][l]{$4131(8)$} & {-} & {-}\\
        \eqmakebox[E][l]{$4251(8)$} & {-} & {-}\\
        \eqmakebox[E][l]{$4551(9)$} & {-} & {-}\\
        \eqmakebox[E][l]{$4700(9)$} & {-} & {-}\\
        \bottomrule
    \end{tabular*}
\end{table}\\

11 excited states can be clearly identified below $5.5\,$MeV. 
The first excited states are well known from the literature~\cite{Hennig.2015}. 
The most pronounced transition corresponds to the excitation of the $2_1^+$ state of $^{96}\textrm{Ru}$ at $833\,$keV, which is analyzed in more detail in the subsequent subsection. 
Other prominent excited states, which could be identified by comparison with the literature values, are, for example, the mixed symmetric $2_3^+$ state at $2283.3\,$keV. 
This special excited state will be analyzed further in Sec.~\ref{Sec::Branching-Ratio}. 
The double peak structure between $2.5\,$MeV and $2.7\,$MeV corresponds to a $2^+$ or rather $1^+, 2^+, 3^+$ excited state at the lower edge and the $3^{(-)}$ state at the higher edge. 
Furthermore, another $2^+$ and spin-$3$ state could be associated with known excited states.
Starting at $4\,$MeV additional excited states could be identified.
Furthermore, at $539\,$keV the first excited $7/2^+$ state of $^{197}$Au~\cite{Bolotin.1979} is observed.

\subsection{Angular Distribution of the $2_1^+ \rightarrow 0_1^+$ Decay} \label{Sec:Angular-Distribution}
Photons emitted in the $^{96}\textrm{Ru}(e,e^\prime \gamma)$ reaction were detected with DAGOBERT using the six LaBr\textsubscript{3}:Ce scintillation detectors positioned at different angles. 
This allows for the measurement of angular distributions of $\gamma$-rays emitted in the decays. 
For a comparison with theoretical results the angle $\theta$ between the momentum transfer vector of the electron $\vec{q}$ and the momentum vector of the emitted photon $\vec{k}$ (compare Fig.~\ref{fig:coordinate_system}) is used. 
The corresponding $\theta$ values of the experiment can be found in Tab.~\ref{tab:measurement_detector_position}. 

In PWBA, the angular dependence of the nuclear decay on $\theta$ and $\Phi$ is encoded in Eq.~(\ref{Eq:Acker-Rose}). Writing out the kinematic factors $V$ one obtains $V_I \propto \sin(\theta) \cos(\Phi)$ and $V_S \propto \sin^2(\theta) \cos(2\Phi)$. 
In addition, the generalized form factors $W$ show a dependence on the Legendre polynomials $P(\cos(\theta))$ and their derivatives. 
The detailed expressions are given in Appendix~\hyperref[Sec:Appendix_A]{A}.

\begin{figure}[htbp]
	\centering
	\includegraphics[width=0.75\linewidth]{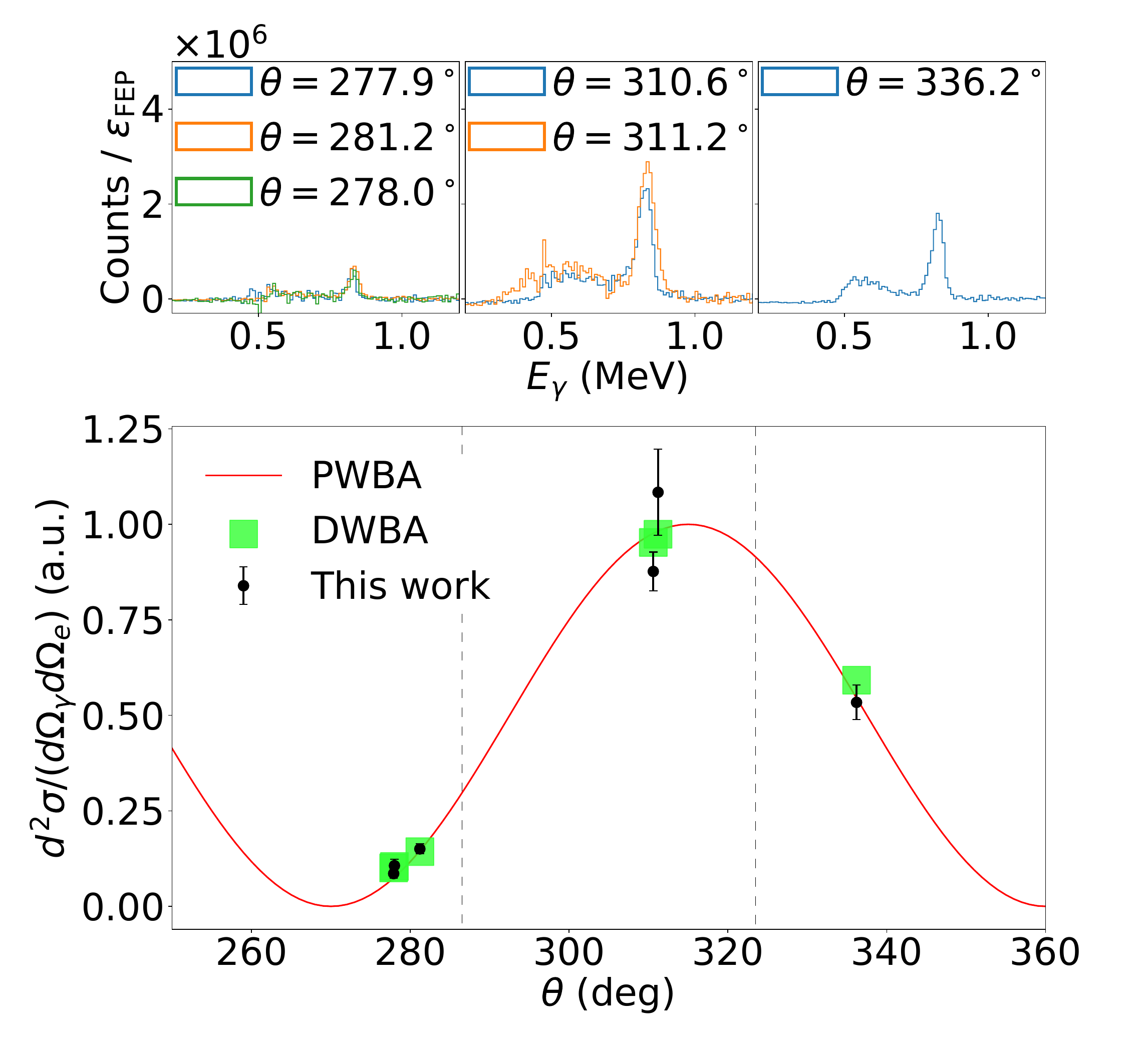}
	\caption{Angular distribution of the $2_1^+ \rightarrow 0_1^+$ decay in the $^{96}\textrm{Ru}(e,e^\prime \gamma)$ reaction. The results are compared to PWBA and DWBA calculations. The data is normalized to the PWBA calculations and given in arbitrary units. The corresponding absolute FEP efficiency $\epsilon_{\mathrm{FEP}}$ corrected $\gamma$ energy spectra of the six LaBr\textsubscript{3}:Ce detectors can be seen on top of the angular distribution. The left figure features spectra of the three detectors with the smallest angles $\theta$. The middle figure shows spectra of the two detectors near the maximum of the angular distribution. The right figure depicts the spectrum of the detector with the highest angle $\theta$.}
	\label{fig::Angular-Distribution-Gamma-Decay}
\end{figure}

By gating on the excitation of the $2_1^+$ state, the characteristic quadrupole pattern of the $E2$ transition to the ground state is observed.
PWBA calculations based on Eq.~(\ref{Eq:Acker-Rose}) are performed for a comparison with the experimental results and shown in Fig.~\ref{fig::Angular-Distribution-Gamma-Decay}. 
In addition, calculations based on the distorted-wave Born approximation (DWBA) discussed below are displayed.

For the $^{96}\textrm{Ru}$ target ($Z=44$), Coulomb distortion effects are non-negligible. 
A distorted-wave Born approximation theory for the $(e,e^{\prime}\gamma)$ process was presented in Ref.~\cite{JakubassaAmundsen.2017}, extending a previous model where only the charge contribution to the nuclear excitation was formulated~\cite{Ravenhall.1987}. 
In this theory, the triple differential cross section is given by
\begin{equation}\label{Eq:Jakubassa1}
    \begin{split}
        \frac{d^3\sigma}{d\Omega_e d\Omega_\gamma d\omega}=\frac{4\pi^2 \omega^2 E_iE_f p_f}{p_i\,c^7\,f_{\rm rec}}\;\frac12
        \cdot \sum_{\sigma_i \sigma_f} \sum_\lambda \left| M_{fi}^{(1)} + M_{fi}^{\rm brems}\right|^2,
    \end{split}
\end{equation}
where it is averaged over the spin projection $\sigma_i$ of the incoming electron, summed over the spin projection $\sigma_f$ of the scattered electron and summed over the photon polarization $\lambda$.
The matrix element $M_{fi}^{(1)}$ accounts for the excitation of the nucleus, followed by radiative decay
\begin{equation}\label{Eq:Jakubassa2}
    \begin{split}
        M_{fi}^{(1)} = i\;\frac{Zc^2}{4\pi \sqrt{\omega}} \;\frac{1}{\omega -E_x+i\Gamma_J/2}\;\sqrt{\frac{\Gamma_J^{\rm rad}}{\Gamma_J}}
        \cdot \sum_{M=-J}^{J} A_{fx}^{\rm dec}(M)\;A_{xi}^{\rm exc}(M),
    \end{split}
\end{equation}
where $J=2$ for a $2_1^+$ excited state. 
Since for collision energies above $30\,$MeV an exact (one-photon exchange) bremsstrahlung theory is not feasible, a relativistic modification of the Bethe-Heitler formula for the bremsstrahlung amplitude $M_{fi}^{\rm brems}$ is used, including a reduction factor by the ground-state charge form factor ${F_0(\vec{q}-\vec{k})}$~\cite{Hubbard.1966} as well as a recoil factor $f_{\rm rec}$ in the cross sections. Explicit expressions for the excitation and decay amplitudes $A_{xi}^{\rm ex}$ and $A_{fx}^{\rm dec}$ are given in Ref.~\cite{JakubassaAmundsen.2017}.

Since $^{96}\textrm{Ru}$ has a $0^+$ ground state, the charge transition density $\varrho_2(r)$ and the magnetic transition densities $J_{21}(r)$ and $J_{23}(r)$ contribute to $A_{xi}^{\rm ex}$, while only the two magnetic densities determine $A_{fx}^{\rm dec}$ (see Eq.~(2.7) in Ref.~\cite{JakubassaAmundsen.2017}, where a factor $(4\pi)^{-1/2}$ is missing). 
For the ground-state charge distribution a Fermi parametrization~\cite{Visscher.1997} with $c=5.017\,$fm and $a=0.55\,$fm is used in order to reproduce the rms radius, and the code RADIAL~\cite{Salvat.1995} is applied for the calculation of the electronic scattering states.
The nuclear transition densities are obtained by means of the self-consistent Hartree-Fock plus Random Phase Approximation~\cite{Colo.2013,Colo.2021} calculations based on the Skyrme parametrization SkP~\cite{Dobaczewski.1984}, which provides a reasonable description of low-lying collective excited states. 

\begin{figure}[htbp]
	\centering
	\includegraphics[width=0.75\linewidth]{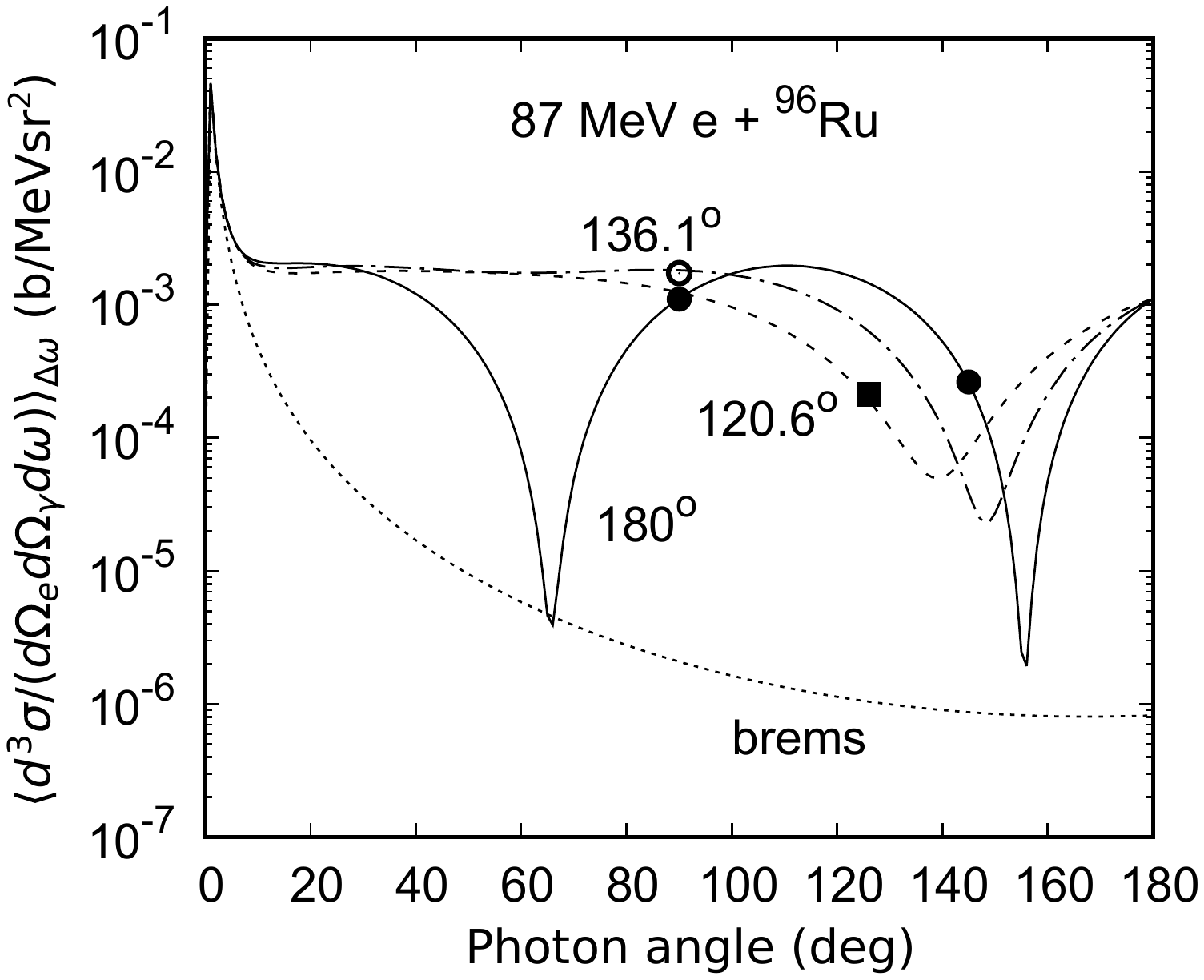}
	\caption{Averaged triple differential cross section $\langle \frac{d^3\sigma}{d\Omega_e d\Omega_{\gamma} d\omega}\rangle _{\Delta \omega}$ for $87\,$MeV electrons scattering from $^{96}\textrm{Ru}$ at an angle of $47.5^\circ$. Shown is the dependence on the photon angle $\theta_k$ for $\varphi_k=180^\circ$ (------, $\bullet$), $\varphi_k=120.6^\circ \;(---, \;\blacksquare$) and $\varphi_k=136.1^\circ \;(-\cdot -\cdot -, \; \odot$), including bremsstrahlung. The lines refer to theoretical calculations and the symbols to measurement data points from the LaBr\textsubscript{3}:Ce detectors. The bremsstrahlung angular distribution (for $\varphi_k=120.6^\circ$) is also shown $(\cdots)$, being slightly higher than the ones for the other two azimuthal angles.}
	\label{fig::Jakubassa}
\end{figure}

Figure~\ref{fig::Jakubassa} shows DWBA results for a scattering angle of $47.5^\circ$. 
The cross section is averaged over the detector resolution $\Delta \omega/\omega=3\%$, corresponding to $50\,$keV (FWHM).
Since the line width is negligibly small ($\Gamma_J^{\rm rad}=\Gamma_J=0.162(7)\,$meV~\cite{Klein.2002}) compared to the resolution, this averaging modifies the resonance factor in Eq.~(\ref{Eq:Jakubassa2}).
\\
For higher-order theories, the $z$-axis is usually chosen along the beam direction, such that the photon angle $\theta_k$ (the abscissa in Fig.~\ref{fig::Jakubassa}) is the angle between $\vec{p}_i $ and $\vec{k}$. 
The azimuthal angle $\varphi_k$ between $\vec{k}$ and $\vec{p}_f$ is chosen such that $\varphi_k=0^\circ$ or $180^\circ$ corresponds to in-plane photon emission. 
The relation to the angles $\theta$ and $\Phi$ of Fig.~\ref{fig:coordinate_system} and Tab.~\ref{tab:measurement_detector_position} are given by:
\begin{equation}
    \varphi_k=\arctan \;\frac{\sin \theta \;\sin \Phi}{\sin \theta \cos \delta \cos \Phi - \sin \delta \cos \theta},
\end{equation}
\begin{equation}\label{25}
    \theta_k=\arccos \;(\cos \delta \cos \theta + \sin \delta \sin \theta \cos \Phi).
\end{equation}
Angular distributions for three azimuthal angles $\varphi_k$ are shown in Fig.~\ref{fig::Jakubassa} together with the corresponding experimental data normalized at the datum point of detector 1 ($\theta_k=90^\circ,\varphi_k=180^\circ$). 
The bremsstrahlung contribution for $\varphi_k = 120.6^\circ$ is shown as dashed line.
In general it is suppressed by more than two orders of magnitude for angles $90^\circ \leq \theta_k \leq 145^\circ$. 
The suppression increases with improved detector resolution $\Delta \omega$.

The difference between PWBA and DWBA was studied in detail for the $(e,e^\prime \gamma)$ process on the $2_1^+$ state of a $^{92}\textrm{Zr}$ target~\cite{JakubassaAmundsen.2017}, which has a similar nuclear charge number. For scattering angles in the forward hemisphere, the shift of the minima in the photon angular distribution is negligibly small, while the intensity increases by about a factor of $1.5$.
A similar factor enhancing DWBA over PWBA is expected for $^{96}\textrm{Ru}$.

Excellent agreement between the data and the calculations of the angular distribution of the $2_1^+ \to 0_1^+$ transitions of $^{96}\textrm{Ru}$ is visible in Figs.~\ref{fig::Angular-Distribution-Gamma-Decay} and~\ref{fig::Jakubassa}.
In this context it should be pointed out that in Fig.~\ref{fig::Angular-Distribution-Gamma-Decay} the experimental data and the DWBA calculations are normalized to the maximum of the PWBA curve and the cross sections are given in arbitrary units. Hence, the mentioned difference in the calculated intensities between DWBA and PWBA are not shown.

\subsection{Gamma Decays above Neutron Separation Threshold} \label{Sec:Gamma_Decay_Above_Threshold}
\begin{figure}[htbp]
	\centering
	\includegraphics[width=0.75\linewidth]{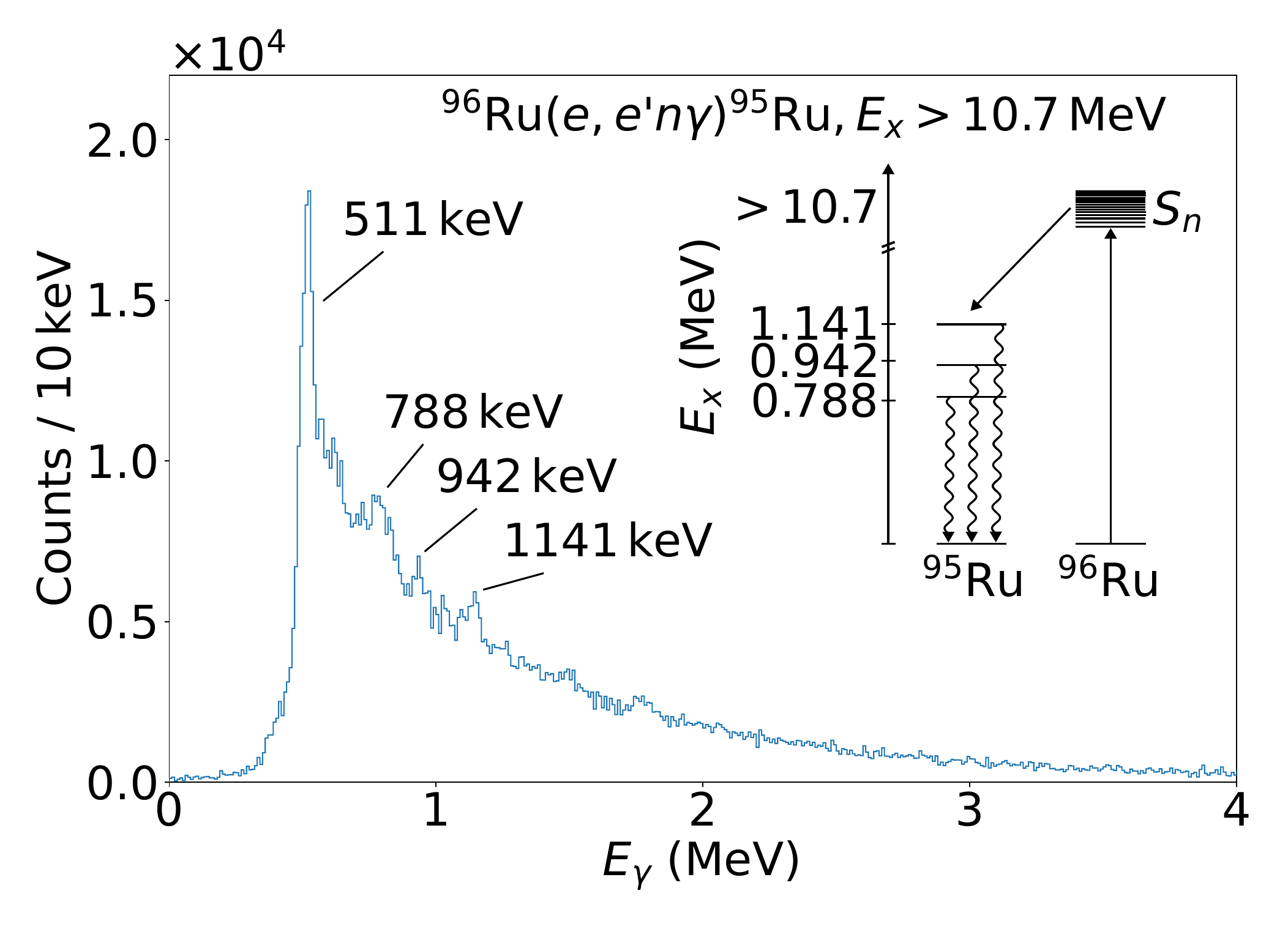}
	\caption{Depopulation of low-lying states of $^{95}\textrm{Ru}$ excited in the $^{96}\textrm{Ru}(e,e^\prime n\gamma)$ reaction above the neutron separation threshold.}
	\label{fig::Neutron-Separation-Threshold}
\end{figure}

A further potential of the newly established $(e,e^\prime \gamma)$ setup is the study of $\gamma$-ray decay following $(e,e^\prime n)$ reactions above the neutron separation threshold. 
In the $E_x$-$E_{\gamma}$ matrix (compare Fig.~\ref{fig::Ex-Eg-Matrix}), an increase in events with $\gamma$-ray energies less than $2\,$MeV is observed for excitation energies above the neutron separation threshold $S_n = 10.69\,$MeV~\cite{Wang.2012}. 
Figure~\ref{fig::Neutron-Separation-Threshold} presents a $\gamma$-ray spectrum obtained from a projection of the $E_x$-$E_{\gamma}$ matrix limiting the excitation energy to values above $S_n$. 
\\ 
Three peaks at energies of $788\,$keV, $942\,$keV, and $1141\,$keV are visible, which can be assigned to the depopulation of the first three excited states of $^{95}\textrm{Ru}$~\cite{Ball.1971}. The $1141\,$keV $\gamma$ transition has been observed for the first time.

\subsection{Branching Ratio of $2_3^+$ Excited State} \label{Sec::Branching-Ratio}
\begin{figure}[htbp]
    \centering
    \begin{subfigure}[b]{0.75\textwidth}
        \includegraphics[width=\linewidth]{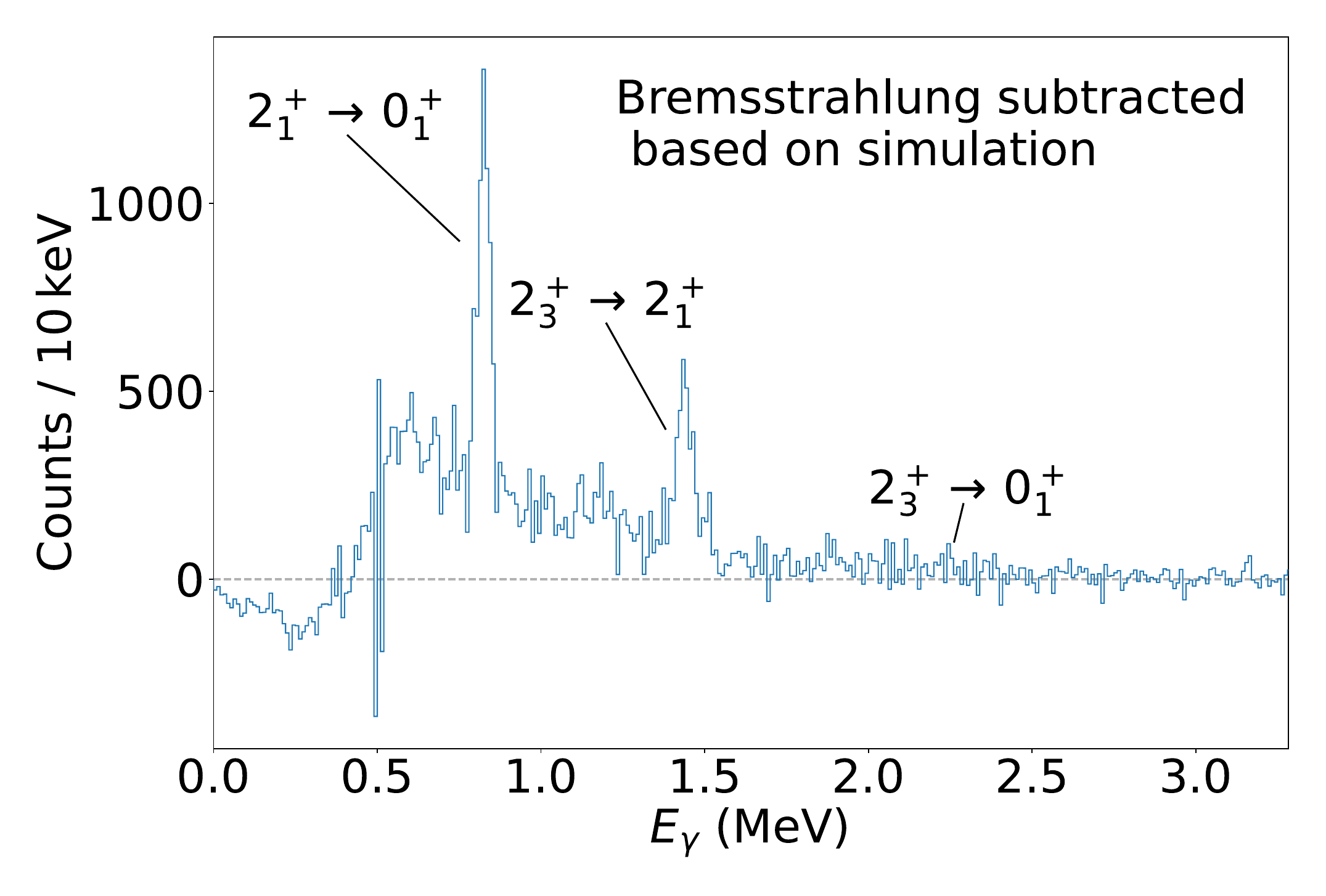}
        \caption{Bremsstrahlung subtraction based on the simulation method, Sec.~\ref{Sec::Bremsstrahlung-Subtraction-Sim}.}
        \label{fig::Branching-Ratio:A} 
    \end{subfigure}
    \begin{subfigure}[b]{0.75\textwidth}
        \includegraphics[width=\linewidth]{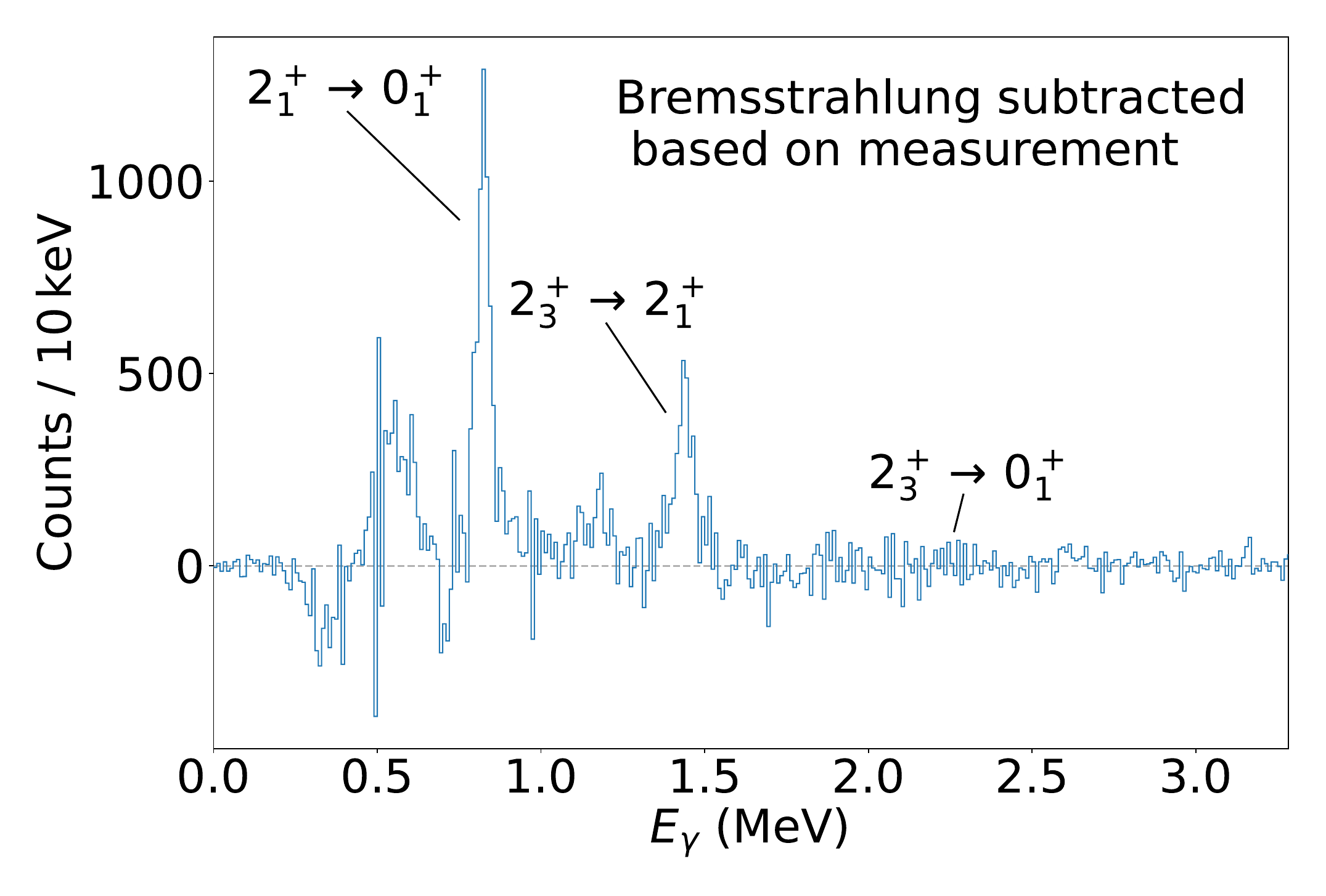}
        \caption{Bremsstrahlung subtraction based on the measurement method, Sec.~\ref{Sec::Bremsstrahlung-Subtraction-Meas}.}
        \label{fig::Branching-Ratio:B}
    \end{subfigure}
    \caption{Comparison of the $\gamma$-energy spectra of the $2_3^+$ excited state of $^{96}\textrm{Ru}$.}
    \label{fig::Branching-Ratio}
\end{figure}

The first analysis of an off-yrast state with $(e,e^\prime \gamma)$ reaction was possible in the $^{96}\textrm{Ru}$ experiment. 
In Fig.~\ref{fig::Excitation-Energy-Spectrum} the excitation of the $2_3^+$ state at $2.283\,$MeV is observed, 
which is known to be a mixed-symmetric state~\cite{Hennig.2015, Pietralla.2001, Klein.2002}.
Its signature is a strong $M1$ transition to the $2_1^+$ state and a weak $E2$ transition to the ground state.
The ground state decay can be used to test the methods for subtraction of coherent bremsstrahlung discussed in Sec.~\ref{Sec::Bremsstrahlung-Subtraction}.
Figure~\ref{fig::Branching-Ratio} presents the $\gamma$-ray spectra gated on the excitation of the $2_3^+$ state after the bremsstrahlung correction based on simulation and measurement, respectively.
\\
In Fig.~\ref{fig::Branching-Ratio:A} the $\gamma$-ray spectrum after bremsstrahlung correction based on the simulation method, Section~\ref{Sec::Bremsstrahlung-Subtraction-Sim}, can be seen. The decay of the $2_3^+$ state into the first excited $2_1^+$ state and its subsequent decay into the ground state ($2_1^+ \rightarrow 0_1^+$) are clearly visible on top of the Compton continuum of the two decays. 
Below $400\,$keV, an oversubtraction of the bremsstrahlung background is recognizable due to the threshold of the LaBr\textsubscript{3}:Ce scintillation detectors. 
After subtraction of the coincident bremsstrahlung, it is difficult to recognize a ground state decay of the $2_3^+$ state. Hence, an upper limit can be determined, only. 
A similar result is obtained by subtracting the bremsstrahlung with the measurement method, Section~\ref{Sec::Bremsstrahlung-Subtraction-Meas}, as shown in Fig.~\ref{fig::Branching-Ratio:B}. 

From these spectra the branching ratios are extracted and compared with values from the literature~\cite{Hennig.2015} in Table~\ref{tab::Branching-Ratio}.
\begin{table}[htbp]
    \caption{Comparison of the literature values for the branching ratios $BR_{\textrm{Lit}}$ of the $2_3^+$ state~\cite{Hennig.2015} with the results from the present work with subtraction of the bremsstrahlung by the simulation $BR_{\textrm{Sub, Sim}}$, Section~\ref{Sec::Bremsstrahlung-Subtraction-Sim}, and measurement $BR_{\textrm{Sub, Meas}}$, Section~\ref{Sec::Bremsstrahlung-Subtraction-Meas}.} \label{tab::Branching-Ratio}
    \begin{tabular*}{\tblwidth}{@{} CCCC@{} }
        \toprule
        $\gamma$ decay & $BR_{\textrm{Lit}}$ $(\%)$ & $BR_{\textrm{Sub, Sim}}$ $(\%)$ & $BR_{\textrm{Sub, Meas}}$ $(\%)$\\
        \midrule
        $2_3^+ \rightarrow 0_1^+$ & $6.7(8)$ & $6.1^{+1.1}_{-6.1}$ & $6.0^{+2.0}_{-6.0}$\\
        $2_3^+ \rightarrow 2_1^+$ & $93(10)$ & $90(11)$ & $90(10)$\\
        \bottomrule
    \end{tabular*}
\end{table}\\
The two methods presented for bremsstrahlung subtraction are compatible with the literature values~\cite{Hennig.2015}. 
This confirms the functionality of the subtraction.

\section{Summary and Outlook} \label{chap::Summary_Outlook}
The $(e,e^\prime \gamma)$ experimental setup DAGOBERT@QCLAM at the S-DALINAC of the Technische Universit\"at Darmstadt represents a unique device for nuclear structure studies with electron scattering.
Compared to the pioneering experiments of Ref.~\cite{Papanicolas.1985, C.N.Papanicolas.1985}, the electro-photo production coincidence resolution $\mathcal{R}$ of the DAGOBERT@QCLAM set-up could be increased by a factor of about $60$.
DAGOBERT@QCLAM is capable of measuring angular distributions of nuclear transitions as demonstrated for the $2_1^+$ state in $^{96}\textrm{Ru}$.
Also, branching ratios can be determined as illustrated  for the mixed-symmetric $2_3^+$ state. This represents the first $(e,e^\prime \gamma)$ data on an off-yrast nuclear state.
Furthermore, $(e,e^\prime n\gamma)$ data above $S_n$ were extracted for $^{96}\textrm{Ru}$, where three $\gamma$ decays of $^{95}\textrm{Ru}$ could be resolved. 
The first successful $(e,e^\prime \gamma)$ study of an $A \geq 90$ nucleus indicates that experiments are now possible for heavy nuclei as well.
A decisive step for increased sensitivity is the subtraction of the coincident bremsstrahlung contributions. 
Two possible subtraction methods were presented and demonstrated in the analysis of the mixed-symmetric $2_3^+$ state of $^{96}\textrm{Ru}$.

Overall, DAGOBERT@QCLAM is now ready for science production.
Projects underway with the setup include the following. Near the neutron separation threshold, the low-energy dipole response can be investigated with $(e,e^\prime \gamma)$ experiments analog to particle-$\gamma$ coincidence experiments~\cite{Savran2006, Savran2018} to select dipole transitions to the ground state, to study the Pygmy Dipole Resonance (PDR)~\cite{Savran.2013, Lanza2023} and the toroidal electric dipole mode~\cite{Semenko.1981, Nesterenko.2018,Nesterenko.2019,Repko.2019,vonNeumannCosel.2024}. 
Furthermore, a complete separation of $M1$ and $E1$ excitation strength distributions will be feasible by comparing results of nuclear resonance fluorescence (NRF) experiments with polarized photon beams and $(e,e^\prime \gamma)$ experiments.
Finally, energy-resolved studies of the neutron-decay branches of the Isovector Giant Dipole Resonance (GDR) with $(e,e^\prime n\gamma)$ experiments as function of the excitation energy similar to~\cite{Kleemann.2025} are planned as future projects exploiting the DAGOBERT setup.

\printcredits

\section*{Declaration of competing interest}
The authors declare that they have no known competing financial interests or personal relationships that could have appeared to influence the work reported in this paper.

\section*{Acknowledgments}
We thank the accelerator group of the S-DALINAC and the technical staff at the Institute for Nuclear Physics at Technische Universit{\"a}t Darmstadt for their support.
This work is funded by the Deutsche Forschungsgemeinschaft (DFG, German Research Foundation) under  project-ID No. 279384907 - SFB 1245 and project-ID No. 499256822 - GRK 2891 ''Nuclear Photonics''. 
The authors acknowledge the support by the State of Hesse within the Research Cluster ELEMENTS (Project ID 500/10.006).
X.~Roca-Maza acknowledges support by MICIU/AEI/10.13039/501100011033 and by FEDER UE through grants PID2023-147112NB-C22; and through the “Unit of Excellence María de Maeztu 2020-2023” award to the Institute of Cosmos Sciences, grant CEX2019-000918-M. 
Additional support is provided by the Generalitat de Catalunya (AGAUR) through grant 2021SGR01095.

\section*{Data availability}
The processed data underlying this work is openly available at the TUdatalib repository of Technische Universit{\"a}t Darmstadt~\cite{TUdatalib.2025}.

\newpage
\onecolumn
\appendix
\section*{Appendix A. Nuclear part of differential cross section of the $(e,e^\prime \gamma)$ reaction in PWBA} \label{Sec:Appendix_A}
The differential $(e,e^\prime\gamma)$ cross section for a nuclear excitation of a spin-zero ground state with subsequent $\gamma$ decay is described in Plane-Wave-Born-Approximation (PWBA) by~\cite{H.L.Acker.1967}
\begin{equation}
        \frac{\mathrm{d}^3\sigma_{\mathrm{nucl}}}{\mathrm{d}\Omega_e \mathrm{d}\Omega_\gamma \mathrm{d}\omega} = 2 \left(\frac{Z\alpha \hbar c}{4\pi}\right)^2 (2J+1) \cdot \frac{p_f}{p_i} \frac{\Gamma_J^{\mathrm{rad}}}{\Gamma_J} \frac{\Gamma_J}{(E_x-\omega)^2+\frac{\Gamma_J^2}{4}} 
        \cdot \left( V_LW_L+V_TW_T+V_IW_I+V_SW_S\right),\tag{A.1}
\end{equation}
using the charge number of the excited nucleus $Z$, the spin of the excited state $J$, the momentum of the scattered electron $p_f$, the initial electron momentum $p_i$, the partial width for the photon decay of the excited state $\Gamma_J^{\mathrm{rad}}$, the total width of the excited state $\Gamma_J$, the energy of the excited state $E_x$, the energy loss of the electron $\omega$ and a sum of products of kinematic factors $V$ and generalized form factors $W$. 
The meanings of the indices are $L$ for longitudinal, $T$ for transversal, $I$ for interference and $S$ for spin.

The kinematic factors $V$ can be expressed by:
\begin{align}
    V_L &= \frac{{\left(E_i+E_f\right)}^2 - \tilde{q}^2c^2}{q^4c^4},\tag{A.2}\\
    V_T &= \frac{2{\left(\vec{p}_i \times \vec{p}_f\right)}^2 + q^2\tilde{q}^2}{q^2\tilde{q}^4c^2} = \frac{2{\left(p_i p_f \sin(\theta_{\textrm{Spec}})\right)}^2 + q^2\tilde{q}^2}{q^2\tilde{q}^4c^2},\tag{A.3}\\
    V_I &= \frac{{4\left(E_i+E_f\right)}{\left(\vec{k} \times \vec{q}\right)}{\left(\vec{p}_f \times \vec{p}_i\right)}}{\omega q^4 \tilde{q}^2 c^2} = \frac{{4\left(E_i+E_f\right)}p_i p_f \sin(\theta) \sin(\theta_{\textrm{Spec}}) \cos(\Phi)}{q^3 \tilde{q}^2 c^2},\tag{A.4}\\
    V_S &= \frac{2{\left(2{\left({\left(\vec{k} \times \vec{q}\right)}{\left(\vec{p}_f \times \vec{p}_i\right)}\right)}^2 - {\left(\vec{k} \times \vec{q}\right)}^2{\left(\vec{p}_f \times \vec{p}_i\right)}^2\right)}}{\omega^2 q^4 \tilde{q}^4 c^4} = \frac{2\left(p_i p_f \sin(\theta) \sin(\theta_{\textrm{Spec}}) \right)^2 \cos(2\Phi)}{q^2 \tilde{q}^4 c^4}.\tag{A.5}
\end{align}
Here $E_i$ represents the energy of the inital electron, $E_f$ the energy of the scattered electron, $q$ the momentum transfer of the electron and $k$ the momentum of the emitted photon. 
The underlying coordinate system described by angles $\theta$, $\theta_{\textrm{Spec}}$ and $\Phi$ is displayed in Fig.~\ref{fig:coordinate_system}. 
Four vectors are denoted by $\tilde{x}$, three vectors by $\vec{x}$ and the magnitude of the three vectors by $x$.

The generalized form factors $W$ are given by:
\begin{align}
    W_L &= - |F_L(q)|^2 \cdot \sum_{l=0,2,...} C_{1,-1}^{J,J,l} C_{0,0}^{J,J,l} P_l(\cos(\theta)),\tag{A.6}\\
    W_T &= |F_T(q)|^2 \cdot \sum_{l=0,2,...} (C_{1,-1}^{J,J,l})^2 P_l(\cos(\theta)),\tag{A.7}\\
    W_I &= - F_L(q) \cdot F_T(q) \cdot \sum_{l=2,4,...} C_{1,-1}^{J,J,l} C_{0,1}^{J,J,l} (l(l+1))^{1/2} P_l^{\prime}(\cos(\theta)),\tag{A.8}\\
    W_S &= - s(\Pi) \cdot |F_T(q)|^2 \cdot \sum_{l=2,4,...} C_{1,-1}^{J,J,l} C_{1,1}^{J,J,l} {\left(\frac{(l-2)!}{(l+1)}\right)}^{1/2} P_l^{\prime\prime}(\cos(\theta)).\tag{A.9}
\end{align}
Here, $F_L$ and $F_T$ are the longitudinal and transverse form factors of the nucleus. 
$P_l(x)$, $P_l^{\prime}(x)$ and $P_l^{\prime\prime}(x)$ represent Legendre polynomials and their first and second derivatives. 
The Clebsch-Gordan coefficient for the coupling of $|\,j_1,m_1\rangle$ and $|\,j_2,m_2\rangle$ to $|\,J,M\rangle$ is given by $C_{m_1,m_2}^{j_1,j_2,J}$. The function $s(\Pi)$ is $+1$ for an electric transition and $-1$ for a magnetic transition.

\clearpage

%\twocolumn
\bibliographystyle{model1-num-names}

\bibliography{eepg_literatur}

\end{document}